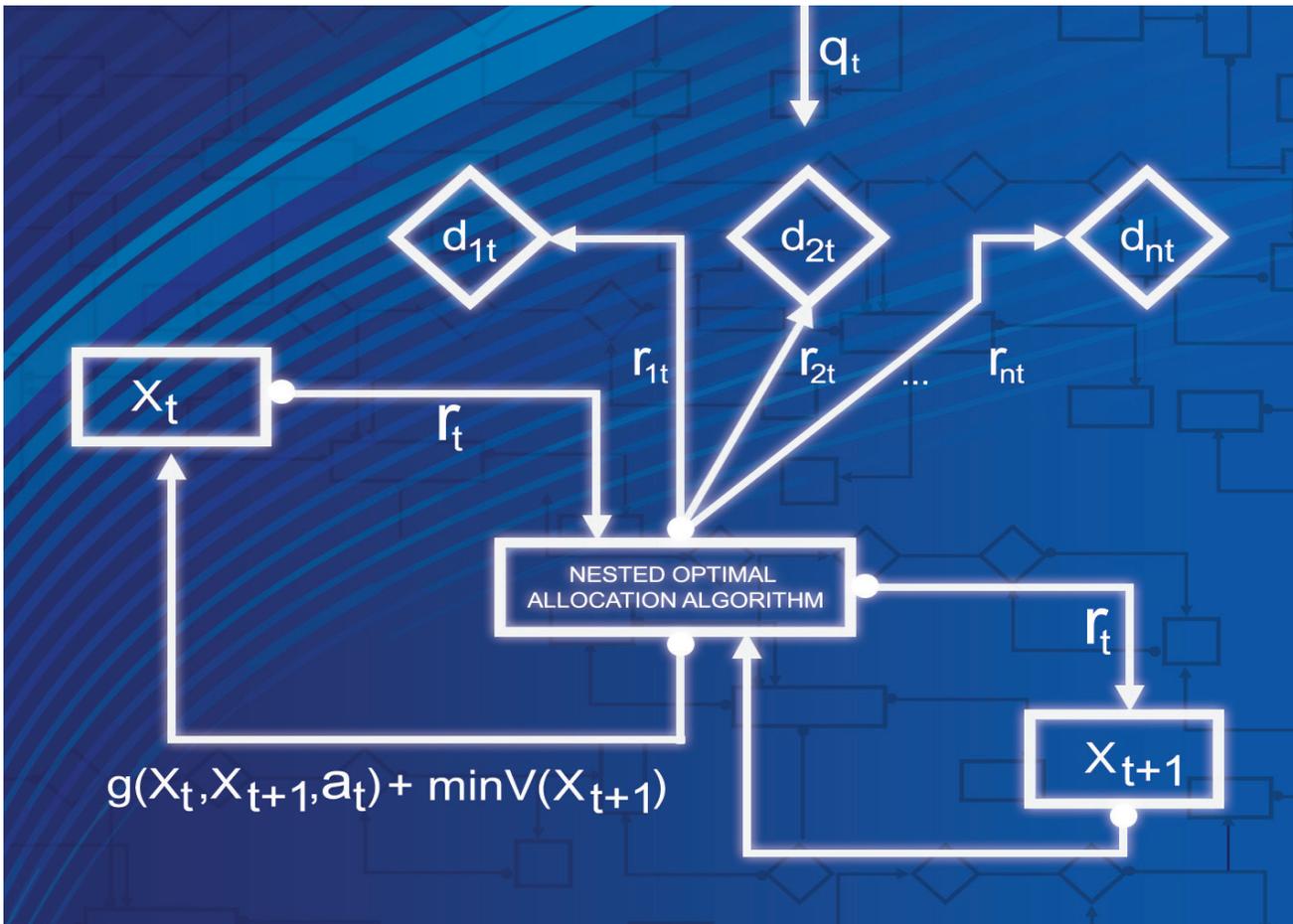

# Nested Algorithms for Optimal Reservoir Operation and Their Embedding in a Decision Support Platform

Blagoj Delipetrev

# NESTED ALGORITHMS FOR OPTIMAL RESERVOIR OPERATION AND THEIR EMBEDDING IN A DECISION SUPPORT PLATFORM

# NESTED ALGORITHMS FOR OPTIMAL RESERVOIR OPERATION AND THEIR EMBEDDING IN A DECISION SUPPORT PLATFORM

DISSERTATION

by

**Blagoj DELIPETREV**

Master of Science in Information Technology,
University "Ss Cyril and Methodius", Skopje
born in Shtip, Republic of Macedonia

*To my family & friends*

# SUMMARY

Population growth, development imperatives, and possible climate change impacts are putting continuously increasing pressures on water resources worldwide. This presents challenges for design and operation of water resources systems, which frequently need to satisfy multiple purposes, such as drinking water supply, industrial water supply, irrigation water for agricultural production, hydropower, etc. Optimal operation of such systems, particularly optimal reservoir operation (ORO), is therefore increasingly required for realising efficient and equitable water allocation across multiple users and functions. This is a known multi-objective optimisation problem with competing and sometimes conflicting objective functions. Over the last few decades, this problem has been subject of extensive scientific research aimed at development and implementation of improved and more efficient reservoir operation (policy) algorithms. Operational practice, on the other hand, requires that such improved optimal reservoir operation algorithms become integral part of decision support systems used in the design and operation of water resources systems.

Pressures on water resources are also evident in the Republic of Macedonia. The demand for clean water in the country is continuously growing, following the increasing living standards of the population, development of new industries and agriculture. Macedonia is located in a zone of continental climate, characterised with wet and cold winter season and hot and dry summer season. Water shortages are sometimes severe during summer, and providing water to all users in these periods may become an issue of very high importance in future. This, in turn, requires improved operation of existing water resources systems, and planning and design of new water resources infrastructure. These processes would benefit from developments of better reservoir optimisation algorithms and implementation of adequate decision support systems.

This situation brings the main motivation for this PhD research, which spans across the two areas of hydroinformatics: 1) methods and tools for model-based optimization of water resources, and 2) decision support systems.

In this work the multi-objective (MO) ORO problem is described by a single aggregated weighted objective function where each of the individual objectives has user-assigned weights. Existing solutions to this problem are provided by methods such as Dynamic Programming (DP), Stochastic Dynamic Programming (SDP) and, more recently, Reinforcement Learning (RL). The DP and SDP methods are well known and established, but suffer from the so-called 'dual curse': 1) curse of dimensionality and 2) curse of modelling. The increased number of variables in the state-action space of the MO ORO problem provokes the curse of dimensionality. This is especially noticeable when multiple water demand objectives are concerned, which is often the case in many optimal reservoir operation problems. This brings the first main research question addressed in this work: How to include multiple water users in the optimal reservoir operation problem while reducing the problems associated with the curse of dimensionality?

To address the issue this research proposes to use an idea of "nesting", i.e. solving the optimal water allocation problem among several water users inside each transition step of DP, SDP and RL, while maintaining the single aggregated weighted objective function that needs to be optimised. The approach allows inclusion of additional variables (corresponding to the various water users) in DP, SDP and RL without provoking the curse of dimensionality. The "nesting" idea was implemented in DP, SDP and RL, and, correspondingly, three new algorithms have been developed named nested DP (nDP), nested SDP (nSDP) and nested RL (nRL). These algorithms are in fact composed of two algorithms: 1) DP, SDP or RL as main algorithms, and 2) nested optimisation algorithm for water allocation implemented with the Simplex algorithm for linear problem formulations, and the quadratic optimisation algorithm for non-linear formulations. Nesting lowers the problem dimension and alleviates the curse of dimensionality.

The nested algorithms have been developed and tested for single aggregated weighted objective function. However, by employing the MOSS approach (multi-objective optimization by a sequence of single-objective optimization searches), these algorithms acquire the multi-objective properties. This approach has also been tested in this research and the corresponding algorithms have been denoted as multi-objective nDP (MOnDP), multi-objective nSDP (MOnSDP) and multi-objective nRL (MOnRL).

The developed algorithms were implemented and tested in the Zletovica hydro system case study, located in the eastern part of Macedonia, within the larger Bregalnica river basin. The optimisation problem formulated for this single-reservoir hydro system has eight objectives, of which two relate to the (soft) constraints on the reservoir level (minimum and maximum), five for water demand users, and one for hydropower. The problem has six decision variables, of which five are releases for the water demand users (also used for hydropower) and one is the reservoir state at the next time step. The Zletovica hydro system case study is in fact somewhat more complex than classical single reservoir case, with spatially distributed users that can partly be satisfied with incremental flows from the catchment downstream of the reservoir. Therefore, the nPD, nSDP and nRL algorithms were modified to fit the case study. The implementation with the needed modifications and the subsequent testing showed both the limitations and the capabilities of the developed algorithms.

The nested algorithms were tested using 55 years (1951-2005) of monthly and weekly data from the Zletovica hydro system. The nDP algorithm was tested on 55 years' monthly data demonstrating that it is more capable than the classical DP. Further analyses indicated that it is also more versatile when compared to the so-called aggregated water demand DP algorithm (a simplified approach in which water demands for all users are aggregated into one demand, which is then used in DP; the distribution to individual users is done separately using the DP results) The nSDP and nRL algorithms trained/learned the optimal reservoir policy using 45 years (1951-1995) of weekly data. These optimal reservoir policies were tested using 10 years (1995-2005) of weekly data. The ORO solution calculated by nDP over the same 10 years' period was set as target for the nSDP and nRL policies. The results showed that the nRL produces better optimal reservoir policy that the nSDP. All three nested algorithms (nDP, nSDP and nRL) can solve a problem with multiple water users without significant increase in algorithm complexity and computational expenses. Computationally, the algorithms are very efficient and can handle dense and irregular variable discretization. The nDP algorithm can handle multiple model and decision variables, while nSDP is limited in accepting additional model variables. The nRL algorithm is more capable than the nSDP in handling additional variables related to multiple users, but it requires quite a lot of tuning and has a relatively complex implementation.

The case study problem was also solved by using the multi-objective nested optimization algorithms MOnDP, MOnSDP, and MOnRL. The found solutions form the Pareto optimal set in eight dimensional objective functions space (since the eight different objectives were considered). The MOnDP was used as a scanning algorithm with 10 sets of varying weights that can identify the most desirable MO solutions. The MOnDP was selected because it is much quicker than MOnSDP and MOnRL. From the 10 sets of weights and their MOnDP results, the three sets were selected to be used by MOnSDP and MOnRL. (The results also confirmed the previous conclusions about the relative performance of various algorithms.) The solutions generated by the MOnRL were found to be much better than those of the MOnSDP.

The "nested" algorithms need to be included in a platform (application) so they are accessible and available to multiple users. Commonly this is done in desktop application that may include these algorithms (and possibly offer other functionalities). This approach, however, has drawbacks regarding support for multiple users' collaboration, limited portability, constraints related to software versioning, and important limitations on software scalability. Decision support applications are therefore increasingly being developed as web and cloud applications, which can overcome many of the drawbacks of desktop applications. This was also the main motivation for the second main research question addressed in this thesis: How to develop a water-resources decision support application that is available 24/7, accessible from everywhere, that is scalable and interoperable, and can support collaboration from concurrent multiple users?

This consideration has led to the development of a cloud decision support platform for water resources. This platform embedded the previously developed algorithms nDP, nSDP and nRL. It was developed using the open source software, open standards, web services and web GIS. The cloud platform is comprised of the four services for: (1) data infrastructure, (2) support of water resources modelling (3) water resources optimisation and (4) user management. The cloud platform was developed using several programming languages (PHP, Ajax, JavaScript, and Java), libraries (OpenLayers, JQuery), and open source software components (GeoServer, PostgreSQL, PostGIS).

The cloud decision support platform was developed and tested with the data from the Zletovica hydro system. The web service for supporting water resources modelling enables creation, editing and management of geospatial objects representing the system, such as the reservoir Knezevo, Zletovica River and its tributaries, derivation canals, water users, tributary inflow points and agriculture areas. This service can be seen as a customised web GIS application for water resources, providing online GIS capabilities. The web service for water resources optimisation provides web interface for the nDP, nSDP and nRL algorithms. It provides web-based user interface with forms for entering algorithms input data, buttons for executing the nested algorithms, and charts and tables for presentation of optimisation results.

The concurrent usage of the developed web services was tested by a group of students imitating the decision procedures in water resources. These tests showed that multiple users can jointly collaborate and manage the geospatial objects representing the water resources system, execute optimisation runs and view results. The developed cloud platform was deployed in a distributed computer environment running on two separate virtual machines (VM) and the testing demonstrated its advantages in terms of being available all the time, accessible from everywhere and serving as collaboration platform for multiple users. Using latest technologies and standards in development of its components, it also provides interoperability and flexibility for including additional components and services, potentially without scalability issues.

The case study area of the Zletovica hydro system has a number of water resources issues that need to be addressed, especially related to water shortages during the summer period. There are ongoing developments in the country for creating river basin management plans, adjusting operations of the available water infrastructure and designing new infrastructure elements. This research and the developed hydroinformatics technologies and systems can contribute to the efforts aimed at improving water resources system optimisation, planning, and management in the Republic of Macedonia.

# Contents

# Chapter 1    Introduction

*"Think globally, act locally"*

The introduction chapter begins by highlighting the water resources importance, with accent on construction and management of reservoirs supported by examples from different states and periods. Macedonia is a country that does not utilize enough its water resources, which is the main motivation for developing this PhD. The PhD research focuses on two principal problems: 1) ORO and 2) building of a decision support platform. The research objectives are established regarding these two main problems, followed by the PhD thesis outline.

___________________________________________________________________

## 1.1    Motivation

Water is a valuable resource. When we see the oceans, rivers, glaciers, we get the impression that there is enough water for everything, but often this is not the case. The first notice is that not all water is usable, at least in the way we need it. Of all water in the world freshwater makes up less than 3%, and over two-thirds of this is locked up in glacial ice caps and glaciers. Fresh water lakes and rivers make up only 0.009% of water on Earth and ground water makes up 0.28% (Gleick 2001).

The importance of water resources becomes crucial with continuous increase in human population, living standard, food and energy demands, since recently combined with possible effects of climate change (Vörösmarty et al. 2000). At present 50% of the world population live in cities, while in 1900 this was only 10%. In addition, individual cities are growing to unprecedented sizes, now known as megacities (Grimm et al. 2008). The population growth is expected to reach over 10.1 billion until 2100 (Jackson 2011).

The recent Fukushima disaster made a vast impact on the nuclear energy future and shifted many countries towards closing down their existing nuclear power plants, or postponing / cancelling plans for building new ones (Joskow and Parsons 2012). This event contributed to seeking alternative energy sources and especially focusing on renewable sources. One of the principal renewable energy sources is hydropower.

Water is a part of the energy-water-food nexus. These three are probably most important human resources, which are extremely interconnected and dependable on each other. The food production depends on water. Water is used to produce energy in hydropower plants, and it is a key resource in coal and nuclear plants (Feeley et al. 2008). It also can be the other way around, when energy produces water, often done by desalination plants. Desalination plants produce water with substantial cost, and are the only alternative in many parts of the world. Bio-fuels are an agricultural product that utilizes water, and produce energy. These show the energy-water-food nexus complexity, and why water (resource) is very important.

Different countries, depending on their circumstances, developed their own water resources systems and strategy. In many cases, the solution has been to build reservoirs and supporting infrastructure. The general development in building reservoirs and utilizing them for various purposes, including water supply, irrigation, hydropower production, flood protection, food, and recreation has taken place in different periods in different countries. The USA built most of its infrastructure, including many reservoirs, at the time of the great depression during 1930–1940. Brazil and Paraguay built the Itaipu dam and reservoir in the 1970s, which in 2008 supplied 90% of Paraguay energy demands and 20 % of Brazil (Barros et al. 2009). The Aswan reservoir in Egypt, constructed in the period 1960-1970 is crucial for controlling the river Nile, providing water for irrigation, hydropower, and delivers an enormous impact on the economy and culture of Egypt. More recently, Turkey made more than $30 billion investment in what is named the Great Anatolia project, which is a complex of 22 reservoirs and 19 hydropower plants. The project will increase the amount of irrigated land in Turkey by 40% and provide 25% of the country power needs (Loucks and Van Beek 2005). Finally, yet importantly, China is presently acquiring new water resources infrastructure that is unprecedented and unseen in human history. They have constructed the Three Gorges dam, which is the biggest reservoir in the world and led to the displacement of two million people (Heming et al. 2001). Another massive project in this country is the waterway from the wet south to the dry north that will alleviate water scarcity for 300-325 million people (Berkoff 2003). From these examples, it is clear that countries' development progress is closely associated with and dependent on the development of water resources, including the construction of large dams and reservoirs.

In Macedonia, most of the reservoirs have been built in the period after the Second World War, especially between 1960-1975. There has been a master plan (MIT 1978) for development of Macedonian water resources up to 2025 that was followed somewhere until 1990, and afterwards it was put aside, likely because of the tough economic state of affairs and wars in the region. In the period after 1990, until recently, there were very limited investments in building new reservoirs. The reservoirs are however quite important because of Macedonian geography and climate. Macedonia is generally a mountainous country with 11 peaks over 2000 m amsl, many rivers, and three main natural lakes. The region of Macedonia is in the zone of continental climate, characterized with wet and cold winter season and long dry summer season. Most of the precipitation takes place during the fall, winter, and early spring, while the summer season has significantly smaller precipitation compared to the remaining part of the year. The reservoirs store water in the wet periods and use it mostly for agriculture during the summer period. Providing water for irrigation in the summer period is very important, without jeopardizing the satisfaction of water requirements for other water users (urban and municipal water supply, industry, etc.). The irrigation systems in Macedonia have been constructed in the same period as the reservoirs, but nowadays they suffer from poor maintenance. The building/restoring of the irrigation systems would substantially increase food yield and the country's economic prosperity. This is likely one of the best way to solve the high unemployment problem in the country, i.e. by drawing people to farming. Additionally, Macedonia is electricity importer. Investments in building new reservoirs will contribute to higher power production and less import, again contributing to the country's economic prosperity.

Currently, there are plans for substantial investments in the Macedonian water resources. Two reservoirs have been recently constructed (Knezevo and Kozjak) while the construction of two additional reservoirs (Lukovo pole and Boskov Most) is to be initiated in the near future. The Macedonian government has made a plan to make huge investments in restoring the existing and creating new irrigation infrastructure (approximately around 200 million Euros). There are also new policies and establishment of a centralized state government body to control and manage the country water resources. Until now water resources management was divided between several government ministries and municipalities and was often lacking funds and a clear strategy. This shows the Government's dedication to invest in the water resources sector.

Considering the previously stated conditions that indicate how development of Macedonia depends critically on its water resources, I have decided to carry out my PhD research in the field of hydroinformatics. The PhD thesis main topic is the research and development of reservoir optimization algorithms. The most widely applied ORO optimization algorithms are dynamic programming (DP), stochastic dynamic programming (SDP), and reinforcement learning, (RL). These algorithms in their standard formulations lack the possibility to include several water users as decision variables in the optimization, because this can lead to the "curse of dimensionality". This particular feature is very valuable in optimizing water resources, and in the case study used in this PhD research. The motivation was to investigate whether it is possible to include additional decision variables in the previously mentioned algorithms, without significantly increasing the computation cost and provoking the curse of dimensionality. This led to the development of novel optimization algorithms for ORO.

Optimization algorithms are often part of decision support systems that can provide water resources modelling, scenario analyses, and optimization capabilities. The required water resources modelling and optimization tasks are performed using different software applications. These applications often work on a desktop computer with limited processing and storage power, with constrains in data and model portability. They are frequently dependent on software vendors and versions and they lack multi-user support. The software developers recognize these limitations and research solutions that will shift the applications to the web and cloud. This was also the motivation in this study, namely to research how to develop a state-of-the-art cloud decision support platform that deals with most of the limitations and constraints described above and embeds the novel optimization algorithms.

The novel reservoir optimization algorithms and the cloud decision support platform are implemented in the Zletovica hydro system case study, located in the north-eastern part of the Republic of Macedonia. The hydro system Zletovica is in the driest part of Macedonia, and because of its complexity presents an implementation challenge. The Zletovica hydro system is a proof of concept that developed algorithms and decision support platform can be used as a foundation for other Macedonian hydro systems. My hope is that further research and development will be in implementing this PhD research in the government institutions.

## 1.2   Problem description

### 1.2.1   Optimal reservoir operation

The ORO problem deals with the derivation of a policy for operating water reservoirs (determining dynamically changing releases and storages) in which all objectives, including water users, hydropower, reservoir levels, etc., are satisfied as much as possible. Frequently these objectives are in direct conflict, e.g. water releases are limited and need to be distributed among several conflicting water demand users.

Historically, the two most widely practiced methods for ORO have been dynamic programming (DP) and stochastic dynamic programming (SDP). These two methods suffer from the so-called "dual curse" which forbids them to be employed in reasonably complex water systems. The first one is the "curse of dimensionality" that is characterised with an exponential growth of the computational complexity with the state – decision space dimension (Bellman 1957). The second one is the "curse of modelling" that requires an explicit model of each component of the water system (Bertsekas and Tsitsiklis 1995) to calculate the effect of each system's transition. The application of various DP and SDP methods in optimal reservoir operation are reviewed in (Yeh 1985) and for multireservoir systems in (Labadie 2004).

Typically, in a single ORO problem there is only one decision variable at each time step to be identified - the reservoir release. This problem, if posed in the dynamic programming setup, uses the Bellman equation (Bellman 1957):

$$V(x_t) = \min\{g(x_t, x_{t+1}, a_t) + V_{t+1}(x_{t+1})\} \qquad (1.1)$$

(for stages *t=T-1, T-2…1*)

where $x_t$ is the state vector at the beginning of the period *t*; *T* is the number of stages in the sequential decision process; $V(x_t)$ is the state value function; $a_t = \{a_{1t}, a_{2t}…a_{nt}\}$ is the actions or decision variables vector during period *t*; $g(x_t, x_{t+1}, a_t)$ is the reward from period *t* when the current state is $x_t$, the action $a_t$ is executed and the resulting state is $x_{t+1}$. This is in fact a general formulation for any system that needs to be optimised in a multi stage decision process. For reservoir operation, the state transition is calculated with a reservoir model based on the mass balance equation:

$$s_{t+1} = s_t + q_t - r_t - e_t \tag{1.2}$$

where $q_t$ is the reservoir inflow, $e_t$ is the evaporation loss and $r_t$ is the total reservoir release and $s_t$ is the reservoir volume. Often the state vector $x_t$ is described by discrete reservoir storage volume and the reservoir inflow $x_t = \{s_t, q_t\}$.

The solution of such problems is obtained by iteratively solving Equation (1.1) as a backward looking solution process over the period *T-1, T-2 … 1* and repeating the cycle until a suitable termination test is satisfied, say after *k* cycles. Then, the last *V*-functions are the optimal $V^*$-functions from which the optimal operating rule at any time is derived as:

$$p^*(x_t) = \arg\max_{a_t} V^*(x_t) \tag{1.3}$$

and

$$a_t^* = p^*(x_t) \tag{1.4}$$

where *p\** is the optimal policy (decision rule).

To determine the right hand side of Equation (1.3), the domain of $S_x$ state and $S_a$ actions needs to be discretized and explored exhaustively at each iteration step of the resolution process. The choice of the domain discretization is essential as it reflects on the algorithm complexity, which is combinatorial in the number of states, release decision, and in their domain discretization. Let's assign $N_x$ and $N_a$ as the number of elements in the discredited state and action sets. The recursive function usually needs $kT$ iteration steps (where $k$ is usually lower than ten) to evaluate the entire state – action space.

$$kT \cdot (N_x \cdot N_a) \tag{1.5}$$

Equation (1.5) shows the so–called curse of dimensionality (Bellman 1957), i.e., an exponential growth of computational complexity with the state and decision (action) dimension. The curse of dimensionality prevents DP, SDP and RL to be applied to design operating policies with too many accounted state or decision variables.

<u>The main research here is focused on this problem</u>: how to overcome the curse of dimensionality in ORO? The specific characteristic of the problem considered in this PhD thesis is that the reservoir release $r_t$ is to be allocated to *n* competing users $r_{1t}, r_{2t}, \ldots r_{nt}$ and this multiplies the total number of decision variables. The main research question is how to include these additional decision variables and other objectives in the optimization algorithms?

The latter chapters demonstrate that it is possible to alleviate the curse of dimensionality and decrease the Equation (1.5) to:

$$kT \cdot (N_x \cdot C) \qquad (1.6)$$

Where the action space is decreased to C, which is constant and in our case is the number of reservoir level discretization, using a novel method called "nesting." The "nesting" method can include additional multiple water demand users and objectives, dense state and action variables discretization. The "nested" method is applied in the three algorithms DP, SDP and RL, creating novel optimization algorithms nDP, nSDP, and nRL (the small n stands for nested).

The ORO is a MO problem by its nature because often different objectives (water demands, hydropower, and reservoir levels) are concerned. In the reservoir operation problem there are constraints (reservoir volume, dead storage, etc.) that need to be taken into consideration. There are several possibilities to deal with the MO ORO problem. In this research, it is first reduced to the single objective optimization problem by employing the single-objective aggregated weighted sum (SOAWS) function. Then the single-objective optimization algorithms are executed multiple times with the several weight sets, i.e. the multi-objective optimization by a sequence of single-objective optimization searches (MOSS). The MOSS method is applied to nDP, nSDP and nRL and creates MOnDP, MOnSDP and MOnRL algorithms.

The Zletovica hydro system is a relatively complex water resource system, including one reservoir - Knezevo, significant tributary inflow downstream of the reservoir, several intake points, several water supply and irrigation users, and hydropower. The specific problem addressed here is how to operate the Knezevo reservoir, to satisfy as much as possible water users and other objectives. The main issue is to include five water users, two towns and two agricultural users, ecological demand, minimum and maximum reservoir critical levels, and hydropower, creating an optimization problem with in total eight objectives and six decision variables.

### 1.2.2   Development of a cloud decision support platform

Water resources planning and management task can be facilitated with a decision support platform that integrates several components, including water resources models, optimization algorithms, geospatial databases, etc. In addition to tasks related to data and model management, optimisation algorithms should be part of a decision support platform where then they can be accessed and utilized. If this is achieved, the decision support platform can provide additional functionality in storing and presenting algorithms optimization data and results. Current ICT and web GIS standards provide tools to develop such a cloud decision support platform.

Most existing water resources applications are desktop-based, designed to work on one computer and without multi user support, which limits their accessibility and availability. Data and model portability is restrained within the version, or the software vendor. Sharing of data and models between multiple users in real time is hardly possible. The classical desktop applications often lack support to connect to other applications and components and there are rigid limits of available memory, storage, and processing power, or the application scalability. These issues associated with classic desktop applications need to be addressed and resolved. Currently the only viable solution for this lies in developing web and cloud applications.

While web/cloud orientation is by now clearly recognized and elaborated in research, this is not yet reflected in the practice for varying reasons. The established practices of using software products in a traditional way seem still to be convenient for consumers and profitable for software producers. The lack of clearly formulated business models together with the investment for changing the existing software is additional constraint. Additional reasons can be in the continuous and rapid change of many web technologies, often not followed by adequate standardization efforts, which also discourages their adoption.

Recently, however, a vast body of research focused on migrating applications to the web has emerged (Choi et al. 2005, Delipetrev et al. 2014, Delipetrev et al. 2008, Horak et al. 2008, Rao et al. 2007). With the explosive growth of the Internet and the enabled web access through diverse devices (computers, mobile phones, tablets) many organizations are turning to the internet as a platform for the provision of their software solutions. Applications are now offered mainly as services accessible via the web rather than as products to be obtained, installed and run as stand-alone applications. Recently, researchers have been dealing with the development of web GIS application (Gkatzoflias et al. 2012) based on web services, cloud computing platform (Bürger et al. 2012) and mobile application that depends critically on the same web orientation (Jonoski et al. 2012). Frequently, all implementation details are hidden from the end-users and the only software that they need is the familiar web-browser. No knowledge is required about the location of storage or computing elements, platforms (operating system) on which they run or their interconnections. Obviously, such approaches may significantly increase the number and diversity of users of such services.

This PhD research continues in this direction with a creation of cloud decision support platform as a demonstrator application. The cloud platform is implemented in the Zletovica hydro system representing its complex network of rivers, canals, water users, agricultural land, etc., together with the embedded nested optimization algorithms.

## 1.3   Research objectives

The primary research objectives of this thesis are to 1) develop novel algorithms for ORO, 2) create a cloud decision support platform, and 3) implement them both in the Zletovica case study.

The specific objectives are as follows:

   a)   Develop ORO solutions capable of handling multiple decision variables without provoking the curse of dimensionality. Afterwards, implement these findings in DP, SDP and RL algorithms, developing novel optimization algorithms named nDP, nSDP, and nRL. Finally, design and implement MO solutions with MOSS and previously developed algorithms as foundation, creating MOnDP, MOnSDP and MOnRL algorithms.

b) Analyse the current state of affairs in the Zletovica hydro system concerning all facets of the system (reservoirs, irrigation channels, irrigation studies, water resources, water demands, water distribution, hydropower, etc.) and create the Zletovica model, define constraints, OFs, and optimization problem.

c) Implement the developed optimization algorithms (nDP, nSDP, and nRL) on the Zletovica river basin, and explore their capabilities and limitations.

d) Compare and discuss the nDP with classical DP and aggregated water demand DP algorithm, demonstrating the nDP advantages and features.

e) Identify a set of Pareto optimal solutions with MOnDP, MOnSDP, and MOnRL algorithms.

f) Build up a cloud decision support platform that embeds the previously developed algorithms and provides additional web services.

g) Deploy the cloud platform on two virtual machines including Amazon web services (AWS). Demonstrate cloud platform scalability, distributed computer environment, availability, accessibility, real-time multiuser collaboration environment, flexibility to add additional components and connect to other desktop software, and its advantages over classical desktop application.

## 1.4   Outline of the thesis

This dissertation is organized in eight chapters.

Chapter 2 describes in details the ORO problem, its mathematical formulation, and main solution approaches using SDP and RL. The multi-objective and multi-agent RL methods are introduced as potential future approaches for ORO.

Chapter 3 presents the nested optimization algorithms nDP, nSDP, nRL, MOnDP, MOnSDP, and MOnRL, their designs, and pseudo codes.

Chapter 4 describes the Zletovica river basin case study in detail, including the case study requirements. In this chapter, the optimization problem is formulated containing decision variables, constraints, and the objective function.

Chapter 5 presents the nDP, nSDP, and nRL algorithms implementation issues. The nDP is compared with classical DP and AWD DP, and corresponding discussion and conclusions are drawn. This chapter presents the nSDP limitation in including additional stochastic variables and the nRL complex settings in parameters, initial state, boundary condition, action list, and convergence criteria.

Chapter 6 presents experiments, results, and discussion in employing the nDP, nSDP, and nRL on the case study. The nDP experiments with monthly data over a 55-year horizon (1951-2005) demonstrate how weights changes influence optimal reservoir policy and the overall results. The nDP is also tested with variable storage discretization. The nDP experiments on weekly data over a 55-year horizon (1951-2005) comply with the case study requirements. The nSDP and nRL ORO policies are derived by training on weekly data (1951-1994). These ORO polices are then compared to the nDP ORO (used as target) on the testing data (1994-2004). Finally, yet importantly, multi-objective solutions are obtained by MOnDP, MOnSDP, and MOnRL.

Chapter 7 presents the cloud decision support platform describing its architecture and a four web services. The cloud platform is implemented in the Zletovica river basin. Some results from tests of the cloud platform web services by multiple concurrent users are also presented. The web service for water resources optimization embodies the previously developed algorithms the nDP, nSDP, and nRL.

Chapter 8 provides the conclusions from this research and recommendations for further research.

# Chapter 2 Optimal reservoir operation: the main approaches relevant for this study

*Tell me and I forget. Teach me and I remember. Involve me and I learn*

*Benjamin Franklin*

This chapter presents the mathematical formulation of the ORO problem and reviews its possible solutions. A short introduction and literature review is presented on the DP, SDP, RL, and their applications to MO reservoir optimization.

___________________________________________________________________

## 2.1 Mathematical formulation of reservoir optimization problem

The MO ORO problem with a period $T$ equal to one year (Castelletti et al. 2007) can be schematized with a feedback control framework as shown in Figure 2.1. For each time step $t$ of the planning horizon, given the storage volume $s_t$ available in the reservoir and other information, the operating policy $p$ returns the decisions (releases) $a_t = \{a_{1t}, a_{2t}...a_{nt}\}$ to be released over the time interval $[t, t+1]$. The other information can include additional meteorological information (precipitation, temperature) and/or hydrological information (previous period inflow, soil moisture, evaporation) $I_t = \{I_t^1, I_t^2...I_t^F\}$. $F$ shows how many additional factors are considered in the reservoir operation.

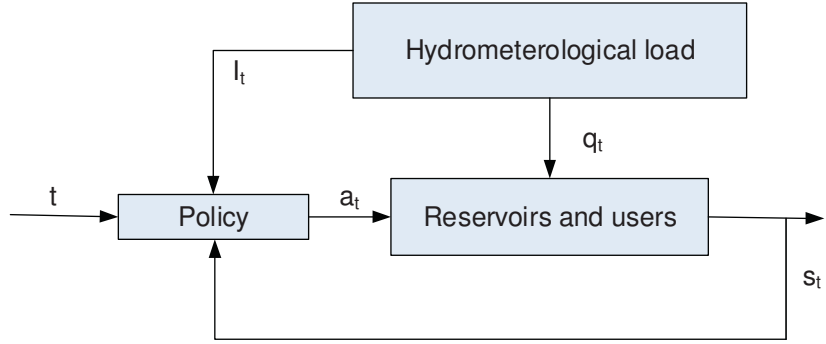

Figure 2.1 Feedback control framework for optimal operation of a reservoir

The following basic mass conservation equation governs the reservoir dynamics:

$$s_{t+1} = s_t + q_t - r_t - e_t \qquad (2.1)$$

where $s_t$ is the reservoir storage volume, $q_t$ is the reservoir inflow volume in the time $[t,t+1]$, $e_t$ is reservoir evaporation; and $r_t$ is the release over the same period, which is a function of the release decision $a_t$ made at time $t$, the storage $s_t$ and the inflow $q_t$.

The following vector equation compactly represents the typically used model of a water system, composed of the catchment and the reservoir:

$$a_t = p(x_t) \qquad (2.2)$$

where the $x_t$ is the state vector that often include the reservoir storage $s_t$ and the $I_t$ hydro-meteorological information; in our case the state vector $x_t = \{s_t, q_t\}$ is described by the reservoir storage $s_t$ and the reservoir inflow $q_t$; $a_t$ is the decision vector including releases for multiple users, and $p$ represents the policy.

The MO ORO problem is often described with $n$ (multiple) objectives, corresponding to different water users and other social and environmental interests, which are in conflict with each other. The MO ORO solution is represented by a set of Pareto-optimal release vectors. Alternatively, the objective functions (OFs) can be aggregated into a single-objective aggregated weighted sum (SOAWS) function as shown in Equation (2.3):

$$g_t(x_t, x_{t+1}, a_t) = \sum_{i=1}^{n} w_{it} \cdot g_{it}(x_t, x_{t+1}, a_t) \qquad (2.3)$$

where $g_t(x_t, x_{t+1}, a_t)$ is the aggregated reward of $n$ objectives at time step $t$, $w_{it}$ is the objective weight at time step $t$ and $g_{it}(x_t, x_{t+1}, a_t)$ is the step reward of each objective at time step $t$. The problem time horizon $T$ can be finite or infinite. The finite time horizon requires establishment of boundary conditions, or a definition of the final state penalty function. On the other hand, when an infinite time horizon is considered a discount factor must be included to ensure convergence of the policy design algorithm. For a given value of the weights $w_{it}$, the total reward function associated with the state value function $V(x_t)$ over a time horizon can be defined as:

$$V(x_t) = \lim_{t \to \infty} \left[ \sum_{t=1}^{\infty} \gamma^t \cdot g_t(x_t, x_{t+1}, a_t) \right] \quad (2.4)$$

where $\gamma$ is a discount factor and $0 < \gamma < 1$. The $\gamma^t$ value is decreasing with every time step and at infinity is zero. Often in optimization problems the state value function $V(x_t)$ needs to be maximized or minimized, depending on the objectives and reward functions. Further in this thesis minimization of the state value function is used as default, except if not denoted differently. The solution of the following optimal control problem produces the optimal state value function $V^*$:

$$V^* = \arg \min \ V(x_t) \quad (2.5)$$

subject to model equations. The Equation (2.4) on a finite horizon $T$ is:

$$V(x_t) = \left[ \sum_{t=1}^{T-1} \gamma^t \cdot g_t(x_t, x_{t+1}, a_t) + \gamma^T \cdot V(x_T) \right] \quad (2.6)$$

where $V_T(x_T)$ is a penalty function that expresses the total expected reward one would incur in starting from $x_T$ and applying optimal release decision over the period $[T, \infty]$.

Since $\gamma^t$ vanishes for $t$ going to infinity the solution to the Equation (2.6) is equivalent to the limit of the following sequence of policies for the horizon $T$ going to infinity denoted in Equation (2.7).

$$V^*(x_1) = \arg\min \left[ \sum_{t=1}^{T} \gamma^t \cdot g_t(x_t, x_{t+1}, a_t) \right] \quad (2.7)$$

$$x_{t+1} = f(x_t, a_t, q_t) \quad (2.7a)$$

$x_1$ is given  $\quad (2.7b)$

$$V^* = \{V_0(\cdot), V_1(\cdot), \ldots V_T(\cdot)\} \quad (2.7c)$$

$t = 1, 2 \ldots T \quad (2.7d)$

The ORO mathematical formulation is described in Equations (2.1) – (2.7d). Equation (2.7a) governs the state transition function $f$, where the next state $x_t$ can be calculated by the current state $x_t$, and the hydro-meteorological information here denoted with the reservoir inflow $q_t$, and the set of actions $a_t$. Often the starting state $x_0$ is given, shown in Equation (2.7b). The goal is to minimize the objective $V^*(x_1)$, defined by equation (2.7).

If $n$ different objectives (in Equation (2.3)) are considered, it is possible to describe the ORO in multi-objective contexts and calculate Pareto optimal solutions. This can be achieved with setting $m$ multiple weights $\{w_1^i...w_n^i\}$, where ($i$ starts from 1 to $m$) that are applied as a SOAWS for different weight sets (MOSS). Each of the $m$ weights sets will produce its own ORO policy and create the Pareto layer. The Pareto layer gives opportunity to analyse different MO solutions and if necessary, select one of $m$ possible alternatives as the final solution.

## 2.2 Dynamic programming

The ORO problem formulation shown in Equations (2.7) a-d) makes an assumption that all model variables are known at each time step $t$, and all state transitions can be calculated. The reservoir model consists of the mass balance equation, the state $x_t = \{s_t, q_t\}$ and the step reward function are known, i.e., $g(x_t, x_{t+1}, a_t)$ only depends on variables defined for the time interval $[t, t+1]$. The solution of the problem shown in Equations (2.7) a-d) is computed recursively from the following Bellman equation:

$$V(x_t) = \min \left[ g_t(x_t, x_{t+1}, a_t) + \gamma \cdot V(x_{t+1}) \right] \qquad (2.8)$$

(for stages $t=T-1, T-2,...1$)

where $V(x_t)$ is the so-called state value function, i.e., the cumulative expected reward resulting from applying the release decision $a_t$ at time $t$ in state $x_t$ and assuming optimal decision (i.e. a greedy policy) in any subsequent system transition, and $\gamma$ is the discount factor.

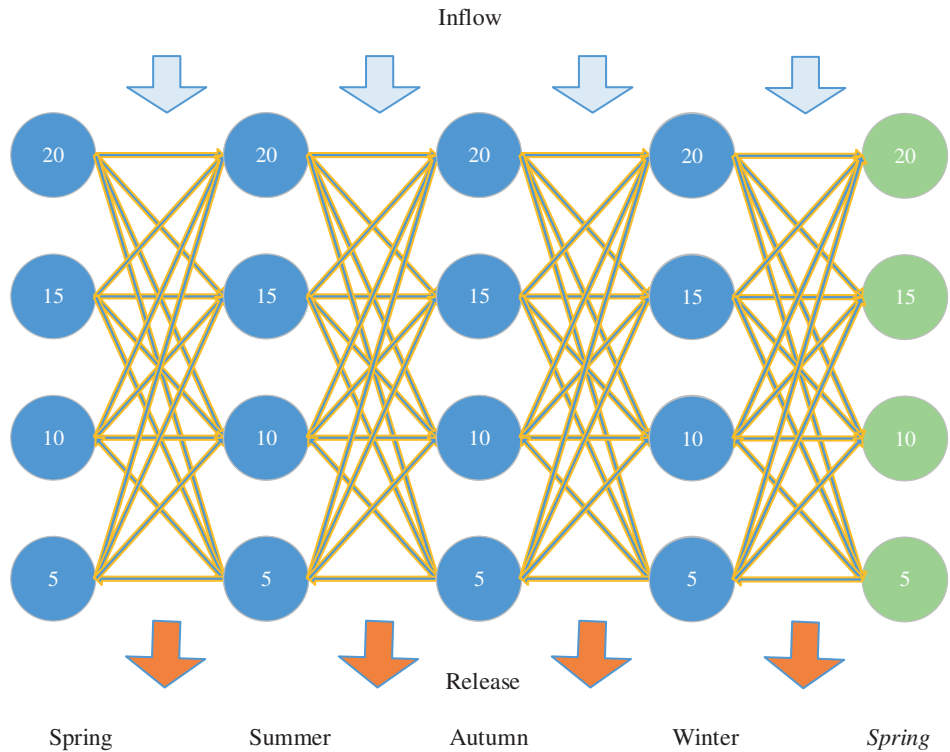

Figure 2.2 A simple representation of reservoir transitions

There is a very good explanation of dynamic programming (DP) with a numerical example in (Loucks and Van Beek 2005) pp.103-113. To explain the DP algorithm application on a reservoir operation let us consider a simple example presented in Figure 2.2 with the four time steps. The blue lines from above represent the reservoir inflow $q_t$ that is changing at each time step. The reservoir is discretized in four reservoir volumes $s_t$ (5, 10, 15 and 20 arbitrary volume units). The state $x_t$ is defined as a reservoir storage volume $s_t$ and the reservoir inflow $q_t$. The orange lines below are representing the reservoir release $r_t$ and the yellow lines are representing the reservoir state transitions. Each of the state transitions has a reward function $g(x_t, x_{t+1}, a_t)$. Some of the state transitions are impossible because they do not satisfy the mass balance equation shown in Equation (2.1). These four transitions can be viewed as four seasons (spring, summer, autumn and winter).

The presented figure characterizes a multistage decision-making problem. At the beginning at each time step $t$, the reservoir storage volume $s_t$ can be in any of the four discretized states (5, 10, 15 and 20). The solution is to find the path thought the network nodes shown in Figure 2.2 that minimizes the sum of reward function $g(x_t, x_{t+1}, a_t)$ or solve the Bellman equation shown in Equation (2.8).

The path can be found with the backward-moving solution procedure. The backward-moving procedure begins at any arbitrarily selected time period or season when the reservoir presumably produces no further benefits and proceeds backward, from right to left one stage at a time, towards the present. At each node (representing a state $x_t$), the state transition $x_{t+1}$ and the reward $g(x_t, x_{t+1}, a_t)$ is calculated. The optimal release $a_t$ completely describes the next state $x_t$ and vice versa (mass balance equation). The generalized DP algorithm pseudo code is as follows:

Algorithm 1. DP pseudo code.

1. Discretize storage $s_t$ and $s_{t+1}$ in $m$ intervals, i.e., $s_{it}$ ($i = 1, 2, …, m$), $s_{j,t+1}$ ($j = 1, 2, …, m$) and set $k=0$.
2. Set time at $t=T-1$ and $k=k+1$.
3. Set reservoir level $i=1$ (for time step $t$)
4. Set reservoir level $j = 1$ (for time step $t+1$)
5. Calculate the total release $r_t$ using mass balance Equation (2.1) ($s_{it}$, $s_{jt+1}$, $q_{it}$ are known)
6. Calculate the $g(x_t, x_{t+1}, a_t)$ and update $V(x_t)$.
7. $j=j+1$.
8. If $j \leq m$, go to step 5.
9. Select the optimal actions (decision variables) $\{a_{1t}, a_{2t}…a_{nt}\}_{opt}$, which consist of the optimal transition $\{x_{t+1}\}_{opt}$ and the users releases $\{r_{1t}, r_{2t}…r_{nt}\}_{opt}$ that give minimal value of $V(x_t)$.
10. $i = i +1$.
11. If $i \leq m$, go to step 4.
12. $t = t -1$.
13. If $t > 0$, go to step 3.
14. If $t = 0$, Check if the optimal actions (decision variables) $\{a_{1t}, a_{2t}…a_{nt}\}_{opt}$ are changed from the previous episode (or in the last three consecutive episodes)? If they are changed, go to step 2, otherwise stop.

One of the main issues is the boundary condition or the ending state value functions denoted with $V_T(x_T)$. A typical approach that solves the boundary condition problem is to connect the first state values, with the last one, making a cycle, as indicated in Figure 2.2. If one year is considered, this is a natural cycle, meaning there is a transition between spring and winter (time step 1 and 4). In Figure 2.2 the cycle is shown in green colour representing storage volume discretization that is in fact the spring of time step 1. With this approach, there is no need to establish boundary condition state value functions $V_T(x_T)$. Several cycles (denoted in the text by $k$) are needed to converge to the optimal state value functions $V^*$. The DP pseudo code steps from 2 to 14 are one cycle (episode) over all possible states and actions. If the optimal actions (decision variables) $\{a_{1t}, a_{2t}…a_{nt}\}_{opt}$ stay the same in the two or three consecutive episodes then the DP stops. This also means that the state value functions $V$ has converged to $V^*$. The $V^*$ describes the optimal policy (decision):

$$p^*(x_t) = \arg \min_{a_t} V^*(x_t) \tag{2.9}$$

The reached steady-state policy is calculated when reservoir operates with the same objectives for a very long time ($k$ cycles). An annual yearly policy $p$ is produced that defines the actions (decisions releases) $a_t = \{a_{1t}, a_{2t}...a_{nt}\}$ at each state $x_t$ to be released/executed over the time interval $[t, t+1]$.

The presented DP algorithm state space is described with the time step $t$, reservoir storage $s_t$ and the inflow $q_t$. The action space in the current settings is represented by the next reservoir level $s_{t+1}$ that describes the next state $x_{t+1}$ and the reservoir release $a_t$. Often in reservoir operation there are many different objectives to consider, for example: releases for specific user (municipal water supply, agriculture, ecology, etc.), minimum and maximum reservoir critical levels, hydropower production, etc. Including these objectives in the DP pseudo code would mean discretization of these objectives variables (similar to reservoir storage volume) and exponential growth of state and actions (decision) dimension. This growth of state and (action) decision dimensions and computational complexity is referred to as "curse of dimensionality" (Bellman 1957). Figure 2.3 shows graphically how different releases $r_{it}$ for a specific water user's demands $d_{it}$ can be included in a classical DP algorithm. The curse of dimensionality limits the number of state/action variables and prevents DP to be used in complex reservoir optimization problems.

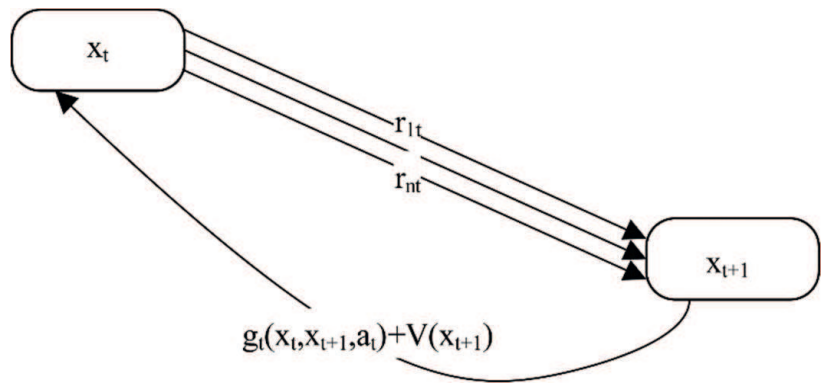

Figure 2.3 Classical DP model of multiple water users

There are various attempts to overcome the curses (Anvari et al. 2014, Castelletti et al. 2012, Li et al. 2013), or earlier DP-based on Successive Approximations (Bellman and Dreyfus 1962), Incremental DP (Larson and Larson 1968), and Differential DP (Jacobson and Mayne 1970). The Differential DP (DDP) starts with an initial guess of values and policies for the goal and continues with improving the policy using different techniques (Atkeson and Stephens 2007). The Incremental DP (IDP) attempts to find a global solution to a DP problem by incrementally improving local constraint satisfaction properties as experience is gained thought interaction with the environment (Bradtke 1994). A number of authors propose decomposition/ aggregation methods for reducing the system to a smaller, computationally tractable one. Most of these methods, however, exploit some particular topological features of the system and are thus are problem-specific.

## 2.3   **Stochastic dynamic programming**

DP assumes perfect knowledge of the input parameters. However, real word problems often have some unknown (uncertain) parameters that are described by their probability distributions. In ORO problems, one of the uncertain parameters is the reservoir inflow $q_t$ that is part of hydro-meteorological information $I_t$. The reservoir inflow $q_t$ uncertainty affects the state value function $V(x_t)$ that consequently is described with the probability distribution as in Equation (2.10). Equation (2.10) replaces Equation (2.4) in describing SDP, while other equations are the same (2.7) a-d).

$$V(x_1) = \lim_{t \to \infty} E\left[\sum_{t=1}^{\infty} \gamma \cdot g_t(x_t, x_{t+1}, a_t)\right] \qquad (2.10)$$

The solution of the SDP problem shown in Equation (2.10) are computed recursively solving the following Bellman equation.

$$V(x_t) = \min \; E[g(x_t, x_{t+1}, a_t) + V(x_{t+1})] \qquad (2.11)$$

(for stages $t=T-1, T-2,…1$)

The SDP algorithm works with the transition probabilities (Loucks and Van Beek 2005, pp. 236-240) that describe the state $x_t$ transitions and their probabilities. As described previously the state $x_t$ consist of reservoir storage volume $s_t$ and reservoir inflow $q_t$. The transition matrices require the reservoir inflow discretization. There are several options in discretizing the reservoir inflow, and one of them is to discretize the reservoir inflow $q_t$ into equal intervals.

$$P_{ij}^t = P\{\, q_{t+1}^j \text{ in interval } j \mid q_t^i \text{ in interval } i \,\} \qquad (2.12)$$

The transition probability matrix $TM$ describe the probability $P_{ij}^t$ for a reservoir inflow $q_t$ that is in interval $i$ in time step $t$, to become $q_{t+1}^j$ that is in the interval $j$ in time step $t+1$. If the reservoir inflow is discretized in regular intervals, then the interval's middle value is taken as the representative. The SDP needs long historical time series reservoir inflow data to derive reasonably accurate transition matrices.

$$\sum_j P_{ij}^t = 1 \qquad (2.13)$$

Equation (2.13) shows that summation of transition probabilities of all intervals $i$ in time step $t$ is one. When Equation (2.12) is included into Bellman's Equation (2.8), the resulting equation is:

Chapter 2  Optimal reservoir operation: review of main approaches                                    19$$V(x_t) = \min\left\{g(x_t, x_{t+1}, a_t) + \gamma \cdot \sum_j p_{q_{t+1}|q_t} \cdot V(x_{t+1})\right\} \quad (2.14)$$

A SDP ORO numerical example is presented in (Loucks and Van Beek 2005, pp. 244-251). The SDP is quite similar to the DP, where multiple years and $k$ cycles are need to obtain the steady-state optimal reservoir policy. The SDP pseudo code is presented below:

Algorithm 2. SDP pseudo code.

---

1. Discretize the reservoir inflow $q_l$ into $L$ intervals i.e., $q_{lt}$ ($l=1, 2..., L$)
2. Create the transition matrices *TM* that describe the transition probabilities $p_{q_{t+1}|q_t}$
3. Discretize storage $s_t$ and $s_{t+1}$ in $m$ intervals, i.e., $s_{i,t}$ ($i = 1, 2, ..., m$), $s_{j,t+1}$ ($j = 1, 2, ..., m$) and set $k=0$.
4. Set time $t=T$-1 and $k=k+1$.
5. Set reservoir level $i=1$ (for time step $t$)
6. Set reservoir level $j = 1$ (for time step $t+1$)
7. Set reservoir inflow interval centres $l=1$ (for time step $t$)
8. Calculate the total release $r_t$ using Equation (2.1) ($s_{it}$, $s_{jt+1}$, $q_{tt}$ are known).
9. Calculate the $g$ ($x_t$, $x_{t+1}$, $a_t$) and update $V(x_t)$.
10. $l=l+1$.
11. If $l \leq L$, go to step 8.
12. $j=j+1$.
13. If $j \leq m$, go to step 7.
14. Select the optimal actions (decision variables) $\{a_{1t}, a_{2t}... a_{nt}\}_{opt}$, which consist of the optimal transition $\{x_{t+1}\}_{opt}$ and the users releases $\{r_{1t}, r_{2t}... r_{nt}\}_{opt}$ that give minimal value of $V(x_t)$.
15. $i = i + 1$.
16. If $i \leq m$, go to step 6.
17. If t>0
18. $t=t$-1.
19. Go to step 4.
20. If t = 0, Check if the optimal actions (decision variables) $\{a_{1t}, a_{2t}...a_{nt}\}_{opt}$ are changed from the previous episode (or in the last three consecutive episodes)? If they are changed, go to step 4, otherwise stop.

---

As in DP, several cycles $k$ are needed to derive the ORO policy. The difference with the DP algorithm presented above is that there is an additional reservoir inflow discretization $L$, and *TM* that are used in the calculation of the state value function $V(x_t)$ from Equation (2.14).

Because the DP and SDP are quite similar, the methods for alleviating the curses mentioned earlier can be applied to both of them.

## 2.4  Reinforcement learning

Reinforcement learning (RL) is a machine learning method that maps situation and actions to maximize the cumulative reward signal. The RL components are an agent, an environment, and a reward function. The environment is observable to the agent through state $x_t$ (state variables). Agent observes the state $x_t$ and takes action $a_t$. The environment reacts to this action, and based on the changes in the environment, gives a reward $g\ (x_t, x_{t+1}, a_t)$ to the agent. In principle, RL has two possible types of action: exploration and exploitation. The exploration action is when the agent makes a random (exploration) action to find a better policy (solution), and the exploitation action is when the agent selects the best available action. The exploitation/exploration parameter is labelled with ε. Figure 2.4 shows typical reinforcement learning system.

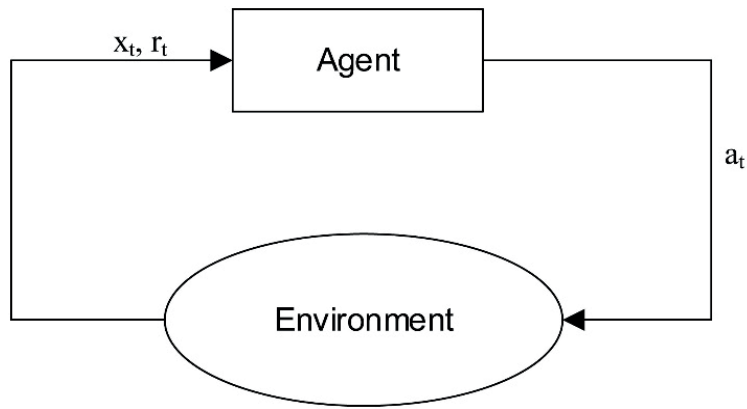

Figure 2.4 Reinforcement learning system

There are two main differences between the RL and SDP in modelling reservoir operation. The first one is that there is no need of describing underlying stochastic processes. The RL agent acquires knowledge of the stochastic environment by learning. The second one is that while SDP makes exhaustive optimization search over all possible state – action space, RL optimization is incremental for the currently visited state shown in Figure 2.5. The nodes in Figure 2.5 represent the states $x_t$ while the arrows represent the actions $a_t$ and consequently the rewards $g\ (x_t, x_{t+1}, a_t)$. Figure 2.5a shows that all possible state-actions are calculated with breadth-first search in stochastic dynamic programming, while Figure 2.5b shows single-step depth-first search in reinforcement learning.

A Markov decision process can formally describe the RL system. One important consideration is to comply with the Markov property that the future state $x_{t+1}$ (next state) is independent of the past states, given the present $x_t$ (current state). Until now all of our equations are based on this assumption, Equation (2.7a). The Markov decision process is a discrete stochastic control process where there are probabilities to select a possible action $a_t$ from the current state $x_t$, that will trigger a transition to the next state $x_{t+1}$ and return reward $g\ (x_t, x_{t+1}, a_t)$ (Bertsekas and Tsitsiklis 1995). This completely complies with the ORO problem described before and the RL algorithm.

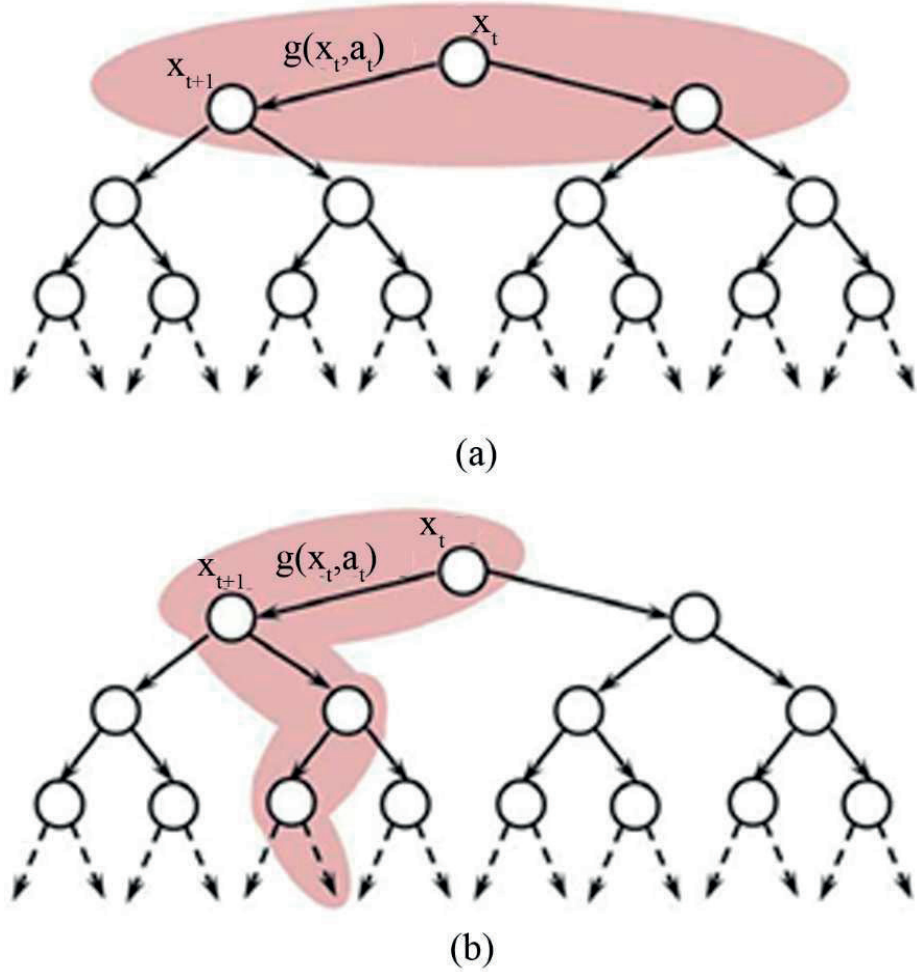

Figure 2.5 Illustration of search methods in Markov decision processes. (a) Breadth-first search in stochastic dynamic programming. (b) Single-step depth-first search in reinforcement learning from (Lee and Labadie 2007)

Although there are several RL methods for solving Markov decision problems, the most popular is the Q-learning method (Sutton and Barto 1998). The Q-learning updates the state-action value function incrementally, rather than performing a complete replacement:

$$Q(x_t,a_t) = Q(x_t,a_t) + \alpha \cdot [g(x_t,x_{t+1},a_t) + \gamma \cdot \max Q(x_{t+1},a_{t+1}) - Q(x_t,a_t)] \qquad (2.15)$$

where $Q(x_t, a_t)$ is the state-action value function; $\alpha$ is the learning rate coefficient; $x_t$, $a_t$, $\gamma$ and $g(x_t, x_{t+1}, a_t)$ are described before. In the context of reservoir operation, the environment in RL can be seen as described by the reservoir inflow and the mass balance Equation (2.1).

Overview of RL algorithms and their development from a programming point of view are presented in (Sutton and Barto 1998). Another important book is Neuro-Dynamic Programming (Bertsekas and Tsitsiklis 1995), which explains in details the RL mathematical foundation and its combination with neural networks. Other useful books to mention are (Mitchell 1997) and (Russell and Norvig 2009) that describe many machine learning and artificial intelligence methods that are often combined with RL.

(Kaelbling et al. 1996) provide details of the advanced features and capabilities, including the state of the art RL applications. One of the conclusions from this work states that "to make a real system work it proved necessary to supplement the fundamental algorithm with extra pre-programmed knowledge." This is a fundamental hypothesis that is somehow forbidden by the RL idea that the agent should learn the system by itself (unsupervised learning), but as the authors have demonstrated in the article, the more pre-programmed knowledge is put into the RL system, the better the agent and the overall system will perform.

Tesauro's backgammon implementation (Tesauro 1994) is one of the most impressive RL demonstrations. The RL agent after playing games against itself (training), reached a level of knowledge close to a human player, and competed in a backgammon tournament with the best players in the world. RL applications have further been developed for managing power consumption and performance in computing systems, which is important research for the new data centres and cloud-computing infrastructure (Das et al. 2008, Tesauro et al. 2007, Tesauro et al. 2006).

In the last decade, there is a significant RL research and applications in ORO. Researchers from Polytechnic University of Milan (Italy) have developed SDP and a number of RL implementations in ORO (Castelletti et al. 2001, Castelletti et al. 2007). The article by (Castelletti et al. 2002) proposes a variant of Q-learning named Qlp (Q-learning planning) to overcome the limitations of SDP and standard Q-learning by integrating the off-line approach, typical for SDP and model-free characteristic of Q-learning. The vast state - actions space in most cases is extremely difficult to express with a lookup table so often a generalization through a function approximation (for example by a neural network) is required (see e.g. (Bhattacharya et al. 2003)). Similar approach, proposed by (Ernst et al. 2006), called 'fitted *Q*-iteration', combines RL concepts of off-line learning and functional approximation of the value function. Recent RL methods (Castelletti et al. 2010) are using tree-based regression for mitigating the curse of dimensionality.

One of the resources that influenced the development of this PhD thesis is (Lee and Labadie 2007) where the three optimization methods implicit stochastic optimization, explicit stochastic optimization and RL are developed and tested on two reservoir system in Korea. This PhD thesis uses a similar logical framework, investigating the nested variants of DP, SDP and RL on the Zletovica hydro system case study.

Several research studies relevant to this PhD thesis have been conducted at UNESCO-IHE Institute for Water Education. The MSc thesis of Geoffrey Wilson (Wilson 1995) presents an overview of the development of a new general control strategy selection technique for real time control. The technique is a learning classifier system that makes state, action -> cost prediction mapping. The learning classifier system is an *if-then* rule-based system that responds almost immediately, and it is particularly appropriate for real time, and model based control. An article related to this MSc thesis (Wilson 1996) presents a successful implementation of real time optimal control of a hydraulic network.

(Bhattacharya et al. 2003) developed a Q-learning algorithm combined with Artificial Neural Network (ANN) for controlling pumps in a large polder system in the Netherlands. In this study, Aquarius DSS was chosen as a reference model for building a controller combined with a machine learning techniques such as ANN and RL, where RL is used to decrease the error of the ANN-based component. The model was tested on a complex water system in the Netherlands, and very good results were obtained.

Although there could be various RL implementations in ORO, for the sake of clarity, a brief explanation of the approach followed in this work is provided here. At the beginning, the available reservoir inflow data $q_t$ is divided into $N$ episodes, one episode per year. The years contain historical data and their number needs to be chosen to cover sufficiently long period to cover different hydrological conditions. The common time steps are monthly, weekly, or daily. The RL system is composed of state variables $x_t$, action variables $a_t$, and reward function $g(x_t, x_{t+1}, a_t)$. The reservoir storage volume $s_t$ and reservoir inflow $q_t$ are taken as a state variable $x_t = \{s_t, q_t\}$, while the next reservoir storage volume $s_{t+1}$ as an action $a_t = \{s_{t+1}\}$. The reward $g(x_t, x_{t+1}, a_t)$ measures the overall objectives satisfaction, which includes water demand users, reservoir critical levels, hydropower production, etc. The RL agent starts from predefined reservoir storage volume $s_1$ and gets the reservoir inflow $q_1$ from the environment. Afterwards the agent makes an exploration/exploitation action $x_{t+1}$, releasing $r_t$ water quantity (calculated by the mass balance equation). Considering the release $r_t$ and other variables (water levels, etc.), the environment calculates the reward $g(x_t, x_{t+1}, a_t)$ and returns it to the agent. The agent goes from the starting state $s_1$ until the end state $s_T$ finishing one episode. After that, another episode is executed with the same starting state $s_1$ and another year of reservoir inflow data. The RL agent with trial and error explores many possible transitions (actions), and learns the optimal policy. In our case, the RL agent uses the Q-learning method to learn the optimal policy. It should be noted that the RL agent learns by obtaining different set of reservoir inflow values at each episode. The pseudo code explaining the RL is shown below:

Algorithm 3. RL pseudo code.

1. Divide the available reservoir inflow data into $N$ episodes for each year, $q_{1t, 2t,..it...Nt}$. The yearly data index is represented by $i$; cycles $k=0$.
2. Set starting reservoir storage volume $s_1$ and $k=k+1$.
3. Get reservoir inflow $q_{it}$ from the environment and define the state $x_t$.
4. Make exploration/exploitation action $a_t$ (select the next state $x_{t+1}$).
5. Calculate the reservoir release $r_t$ from the mass balance equation.
6. Calculate reward $g(x_t, x_{t+1}, a_t)$ concerning multiple objectives.
7. Learn the optimal policy with Q-learning.
8. If t<T, $t=t+1$ and go to step 3.
9. If t=T (end of episode) then i= i+1 and go to step 2.
10. If $i=N$, and agent has not learned the optimal policy, then set $i=1$ and go to 2.
11. Else the agent has learned the optimal policy, stop.

Algorithm 3 presents the RL general logic for the ORO problem. Later in the following chapters the details of the RL algorithms developed in this research are presented.

It can be seen that the RL core is similar to DP and SDP with the major difference that single-step depth-first search is performed instead of breadth-first search. This is an RL advantage because not all state-action space is searched, but on the other hand a curse because significantly more episodes (training) are needed for convergence to the optimal solution. Generally speaking, because RL algorithms search only part of the state-action space, RL can support significantly more state-action variables and higher variables discretization compared to DP and SDP. The DP and SDP with only a few (below 10) state-action variables will provoke the curse of dimensionality. The optimal solution is always found in DP and SDP because the breadth-first search is performed over the entire state-action space. On the other hand, RL searches only a part of state-action space, and it is possible to miss the optimal solution and to find only a suboptimal solution.

The DP and SDP algorithms are based on the Bellman Equation, and after defining the state, actions (decision) vectors it is relatively straightforward to implement and execute optimizations. On the contrary, implementation of RL is not so easy and requires developing an appropriate RL agent and environment, defining variables, adjusting parameters α, γ and ε, etc. In the case of RL there are always many trial and error experiments to find the most optimal RL settings.

The overall conclusion is that DP and SDP algorithms are able to find the optimal solution, but are incapable of handling multiple state-action variables. The RL algorithm is capable of handling multiple state-action variables, but it has complicated development and it can miss the optimal solution.

## 2.5    Approaches to multi-objective optimization

### 2.5.1   Multi-objective optimization by a sequence of single-objective optimization searches

The ORO is in general a MO problem, meaning that multiple objectives are to be considered at the same time. For solving MO optimization problems there exist a large number of MO optimization algorithms – which result in a generation of a Pareto set of optimal solutions (typically containing a large number of them). At the same time, due to the complexity and computational costs of solving full-fledge MO optimization problems some authors use a simplified approach which is generically called "scalarization" (Eichfelder 2008). Scalarization transforms the MO optimization problem to a SO optimization problem (or several of them. The practical implementation of scalarization is presented in (Barreto Cordero 2012) where it is applied for optimization of drainage systems, and it is named MOSS (multi-objective optimization by a sequence of single-objective optimization searches).

MOSS can be applied when the SO optimization algorithm is preferred, e.g. in case of dynamic optimization with a single objective function (a typical case in reservoir optimization), and when there is no need to generate a large Pareto set which is computationally expensive. A user can decide how many MOSS solutions will generate, depending on the practical problem at hand and by choosing a particular number of the weight vectors that are used to weigh the objectives. Although part of the solutions belonging to the Pareto set may not be found with this approach, the generated ones are Pareto optimal, and can be treated as a reasonably good and practically useful approximation of the Pareto set, albeit small. Having this in mind, in this thesis it will be referred as the Pareto set as well. It has to be mentioned that the weighted-sum approach has its known shortcomings because the linear scalar weights will fail to find Pareto-optimal policies that lie in the concave region of the Pareto front (Vamplew et al. 2008).

In the thesis context the considered MOSS approach is implemented as follows: there are $m$ sets of weights $\{w_1^i, \ldots w_n^i\}$ ($i$ starts from 1 to $m$), and $n$ objectives applied to SOAWS functions of DP, SDP and RL. In MOSS problems, there is not a single optimal solution, instead there is always a set of possible solutions. The MOSS optimization pseudo code is:

Algorithm 4. MOSS pseudo code.

1. $i=1$
2. Select $w^i$ from $\{w_1^i, \ldots w_n^i\}$.
3. Execute SO optimization using the SOAWS function (employing DP, SDP or RL) and find the optimal solution.
4. $i=i+1$.
5. If $i \leq m$ go to step 2, else go to step 6
6. Present the $m$ found solutions as the approximation of the Pareto front.

The MOSS scheme allows for parallelization because different sets of objective weights can be assigned to different algorithms that afterwards independently can be calculated on different machines. The implementation is quite straightforward:

Algorithm 5. Parallel MOSS pseudo code.

1. Create $m$ independent SOAWS optimization problems assigning each with a set of weights $\{w_1^i \ldots w_n^i\}$ in DP, $SD$P or RL.
2. Execute of all $m_i$ and find the m optimal solutions.
3. Present the $m$ found solutions as the approximation of the Pareto front.

Another possible solution for MO problems is multi-level programming. Multi-level programming is a MO optimization method that orders the $n$ objectives according to the hierarchy (Caramia and Dell'Olmo 2008). First, the minimizers of the first OF are found; second, the minimizers of the second OF are found; until all of the objectives are optimized. This method has meaning if there is a hierarchical distinction between different objectives. The concept of the nested optimization method is in a way similar to multi-level programming, with one fundamental difference – the nested optimization is built into the optimization method, and all objectives are evaluated at each time step e.g. there is no assumed hierarchy.

### 2.5.2   Multi-objective and multi-agent reinforcement learning

Here we briefly discuss the MO RL and multi-agent RL that is relatively new machine learning optimization algorithm and an attractive scientific topic.

There are MO RL algorithms that try to learn multiple policies at the same time and create a Pareto front. Learning multiple policies at the same time at each state $x_t$ means that there is a multiple optimal actions vector $\{a_t^1, a_t^2, \ldots a_t^n\}$ that create several non-dominant policies. The underlying idea of (Barrett and Narayanan 2008) is to discard the actions $a_t$ that are never optimal no matter the objective weights or importance. One of the problems is to store the optimal action vector at each state that belongs to the Pareto optimal solution. The optimal action vector can have a substantial size, and if considering that often there are many states, the computational power and storage become major limitations.

In the ORO domain, an application for improving water quality by changing the reservoir operation employing an RL agent that works in parallel and in a distributed computer environment is presented in (Rieker 2010, Rieker and Labadie 2012). Interesting research is the MO RL method for ORO that can generate the Pareto front in one single run (Castelletti et al. 2011). This algorithm is an extension of the fitted Q-iteration (FQI) (Castelletti et al. 2010) that enables learning of the operating policies for all the linear combinations of preferences (weights) assigned to the objectives in a single training process. The key idea of MO FQI (MOFQI) is to enlarge the continuous approximation of the value function that is performed by a single objective (SO) FQI over the state-decision space also to the weight space. MOFQI is compared with the reiterated use of FQI and a MO parameterization-simulation-optimization (MOPSO) approach (Castelletti et al. 2013). Results show that MOFQI provides a continuous approximation of the Pareto front with comparable accuracy as the reiterated use of FQI. MOFQI outperforms MOPSO when no a priori knowledge of the operating policy shape is available, while produces slightly less accurate solutions when MOPSO can exploit such knowledge.

There are many advances in independent or cooperating agents (Kok and Vlassis 2006), applications in different fields like optimizing servers' performance (Das et al. 2008, Sridharan and Tesauro 2002), robotics (Yang and Gu 2004), and critics of multi-agent RL implementations (Shoham et al. 2003, 2007).

The MO approach can be extended to multi-agent settings similar like the parallelization before, when each set of weights $w^i$ is assigned to a different RL agent and all of the $m$ agents' work in parallel. The MO and multi-agent RL are attractive new research fields for ORO.

## 2.6   **Conclusions**

This chapter presents the mathematical formulation of the optimal reservoir problem – in general, and with respect to the specifics of the case study considered in this thesis. The main approaches SO (DP, SDP, RL) and MO optimization are presented, with their general overview and applications, especially in ORO. The MOSS algorithmic scheme (a sequence of the single-objective optimization runs), allowing for generating a set of solutions in a MO ORO context is presented as well. The next chapter (3) describes the nested optimization algorithm that has been developed within this research.

# Chapter 3
# Nested optimization algorithms

*"Any intelligent fool can make things bigger, more complex, and more violent. It takes a touch of genius – and a lot of courage – to move in the opposite direction."*

*Albert Einstein*

In this thesis, the problem of ORO is posed in the MO context, and assumes the presence of several water users. If a SO optimization scheme is to be employed as the main optimizer (e.g. DP), this would require the so-called nested approach to optimization. This chapter presents the idea of nesting an optimization algorithm inside each transition of the multi-stage decision problem of reservoir operation that reduces the starting problem dimension and alleviates the curse of dimensionality. This idea is developed and incorporated in the three algorithms: nDP, nSDP, and nRL. These algorithms can solve ORO problem without significant increase in the algorithm complexity. Computationally, the algorithms are efficient and can handle dense and irregular variable discretization. The implementation of the MOSS approach with the nDP, nSDP, and nRL creates MOnDP, MOnSDP, and MOnRL algorithms. At the end of the chapter the general methodology and experiments that follow in the thesis are explained.

______________________________________________________________________

## 3.1    Nested dynamic programming (nDP) algorithm

Typically, in a single-reservoir optimization problem there is only one decision variable at each time step to be identified - the reservoir release. The problem considered in this thesis assumes that this release needs to be allocated to *n* competing users, and this multiplies the total number of decision variables. This problem, if posed in the dynamic programming setup, uses the Bellman Equation (2.8) (Bellman 1957). The mass balance Equation (2.1) governs the reservoir dynamics. The reward $g\ (x_t,\ x_{t+1},\ a_t)$ is a single function, so in order to incorporate the multiple objective functions, a weighted sum of several OFs can be used. The SOAWS function of multiple objectives which includes the terms related to users' releases, deviations from reservoir critical levels, or both releases and levels (e.g. hydropower) is presented in Equation (2.3). The nested DP (nDP) is in essence a DP algorithm, but with the incorporated nested optimization algorithm that at each time step optimally allocates the total reservoir release $r_t$ to different users corresponding to their demands $d_{nt}$, as shown in Figure 3.1.

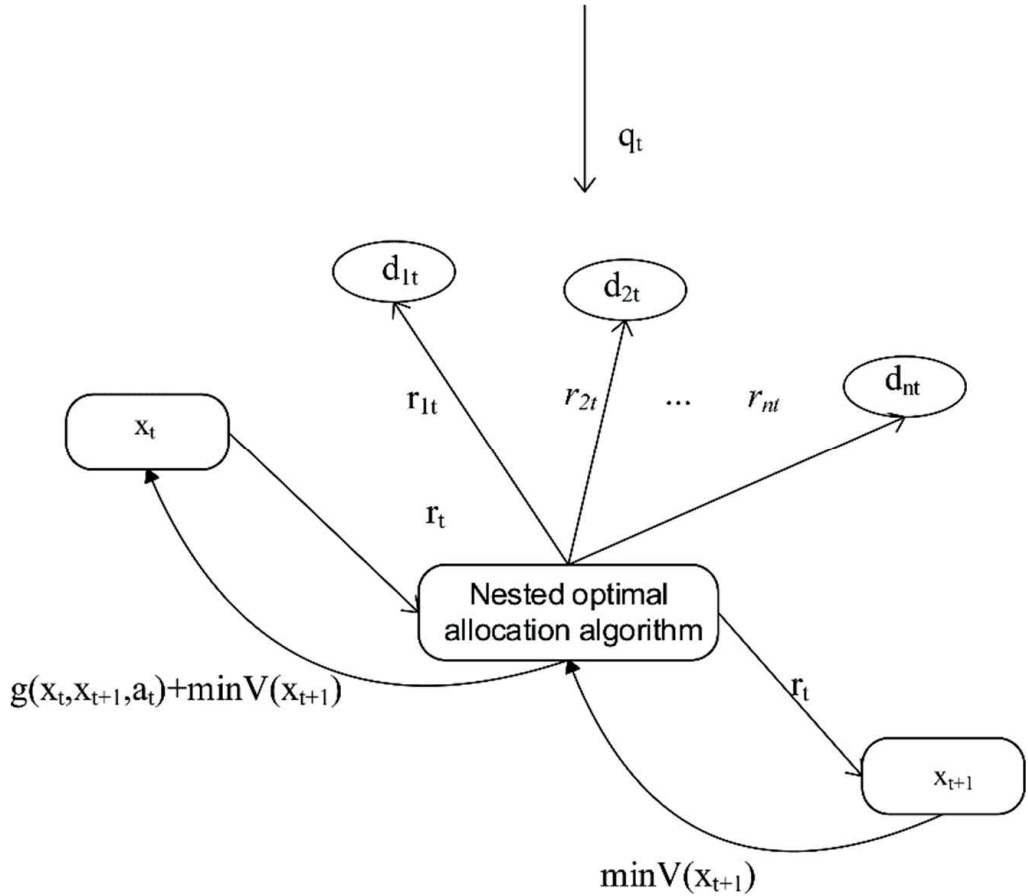

Figure 3.1 Transition at time step t of the nDP algorithm

Figure 3.1 presents a state transition from $x_t$ to $x_{t+1}$. From the mass balance Equation (2.1), the release $r_t$ can be calculated and then nested optimization algorithm is run to identify allocation of $r_t$ between $n$ users. The inputs of nDP are: the reservoir release, the users' demands, their relative importance, and the decision variables are the water volumes allocated to water users (called subsequently the "users' releases") for satisfying their demands. The nDP pseudo code is shown as follows:

Algorithm 6. nDP pseudo code.

1. Discretize storage $s_t$ and $s_{t+1}$ in $m$ intervals, i.e., $s_{it}$ ($i = 1, 2, …, m$), $s_{j, t+1}$ ($j = 1, 2, …, m$) and set $k=0$.
2. Set time at $t=T-1$ and $k=k+1$.
3. Set reservoir level $i=1$ (for time step $t$)
4. Set reservoir level $j = 1$ (for time step $t+1$)
5. Calculate the total release $r_t$ using Equation (2.1).
6. <u>Execute the nested optimization algorithm to allocate the total release $r_t$ to all users $\{r_{1t}, r_{2t}, ... r_{nt}\}$ in an attempt to meet their individual demands $\{d_{1t}, d_{2t}, ... d_{nt}\}$.</u>
7. Calculate $g(x_t, x_{t+1}, a_t)$ and update $V(x_t)$.
8. $j=j+1$.
9. If $j \leq m$, go to step 5.
10. Select the optimal actions (decision variables) $\{a_{1t}, a_{2t}...a_{nt}\}_{opt}$, which consist of the optimal transition $\{x_{t+1}\}_{opt}$ and the users releases $\{r_{1t}, r_{2t}...r_{nt}\}_{opt}$ that give minimal value of $V(x_t)$.
11. $i = i + 1$.
12. If $i \leq m$, go to step 4.
13. $t = t -1$.
14. If $t > 0$, go to step 3.
15. If $t = 0$, compare the optimal actions (decision variables) $\{a_{1t}, a_{2t}...a_{nt}\}_{opt}$, which consist of the optimal transition $\{x_{t+1}\}_{opt}$ and the users releases $\{r_{1t}, r_{2t}...r_{nt}\}_{opt}$ of all states, and check whether they have been changed from the previous episode. If they are changed, go to step 2, otherwise stop.

Algorithm 6 is almost identical to Algorithm 1. The important difference is in step 6 of Algorithm 6 that executes the nested optimization algorithm. This provides the possibility for including several OFs, as will be demonstrated in the next chapters.

The nDP pseudo code presented above is intentionally made generalized as much as possible without specifically defining the OFs. The three types of OFs are envisaged, but not limited to 1) deviations from the reservoir critical levels 2) OF related to users' releases 3) OF related to users' releases and reservoir levels. In the next thesis chapters, a practical example of nDP is demonstrated with the case study.

The optimal allocation algorithm is incorporated (nested) in the DP method and directly updates the state value function $V(s_t)$ at each time step consequently changing the optimal reservoir policy and solving the optimization problem.

The action vector $\{a_{1t}, a_{2t}...a_{nt}\}$ consists of the transition state $x_{t+1}$, and the users' releases $\{r_{1t}, r_{2t}...r_{nt}\}$. The corresponding total release $r_t$, at each transition is calculated from the mass balance. The $e_t$ evaporation losses are calculated at each transition based on the reservoir area that is a function of the reservoir storage volume $s_t$, and other factors. Depending on the formulation, different optimization methods can be used to optimally allocate the total reservoir release $r_t$ between $n$ water users. Two methods have been applied here: Simplex method in case of linear formulation, and a non-linear method in case of the quadratic formulation.

At time step $t$ each water user $i$ is characterized by its demand $d_{it}$ and corresponding weight $w_{it}$. For the nested optimal allocation, the following variables are relevant: $d_{1t}$, $d_{2t}…d_{nt}$ are the users' demands; $w_{1t}$, $w_{2t}…w_{nt}$ are the corresponding demands weights; $r_t$ is the reservoir release; $r_{1t}$, $r_{2t}… r_{nt}$ are the users' releases and $v$ is the release discretization value. Note that at the beginning of each nested optimization, the nDP algorithm checks if the release $r_t$ can fully satisfy the aggregated demand of all users.

$$\text{if } \sum_{i=1}^{n} d_{it} < r_t \text{ then } r_{1t}=d_{1t},\ r_{2t}=d_{2t},…\ r_{nt}=d_{nt,} \qquad (3.1)$$

If the release $r_t$ can satisfy the aggregated demand of all users, the solution is trivial and there is no need to solve the optimal allocation problem.

## 3.2 Nested optimization algorithms

### 3.2.1 Linear formulation

In case the OF is based on the linear combination of deficits, the problem is a linear programming problem, which can be solved by using for example the Simplex method:

$$\min \sum_{i=1}^{n} w_{it} \cdot (d_{it} - r_{it}) \qquad (3.2)$$

subject to:

$$r_{1t} + r_{2t} … + r_{nt} \leq r_t \qquad (3.2a)$$

$$r_{1t} \leq d_{1t}, r_{2t} \leq d_{2t}, …, r_{nt} \leq d_{nt} \qquad (3.2b)$$

$$r_t, d_{t1}, d_{t2}, .. d_{nt} \geq 0 \qquad (3.2c)$$

### 3.2.2 Non-linear formulation

When using sum of weighted quadratic deficits, the OF is expressed as follows:

$$\min \sum_{i=1}^{n} w_{it} \cdot (d_{it} - r_{it})^2 \qquad (3.3)$$

Constraints are same as previously described in Equation (3.3a-c). The weighted quadratic deviation example is presented in (Loucks and Van Beek 2005) pp. 103-113). The reservoir release $r_t$ is assumed to be discretized in $v$ levels. The $v$ value is set at the beginning and stays the same during the execution.

## 3.3 Nested stochastic dynamic programming (nSDP) algorithm

The only difference between the nSDP and the classical SDP is that in the former there is the nested optimization algorithm that executes at each state transition.

The nSDP pseudo code is:

Algorithm 7. nSDP pseudo code.

---

1. Discretize the reservoir inflow $q_t$ into $L$ intervals i.e., $q_{lt}$ ($k=1, 2..., L$)
2. Create the transition matrices *TM* that describe the transition probabilities $p_{q_{t+1}|q_t}$
3. Discretize storage $s_t$ and $s_{t+1}$ in $m$ intervals, i.e., $s_{it}$ ($i = 1, 2, ..., m$), $s_{jt+1}$ ($j = 1, 2, ..., m$) (in this case $x_t = s_t$) and set $k=0$.
4. Set time $t=T-1$ and $k=k+1$.
5. Set reservoir level $i=1$ (for time step $t$)
6. Set reservoir level $j = 1$ (for time step $t+1$)
7. Set inflow cluster $l=1$ (for time step $t$)
8. Calculate the total release $r_t$ using Equation (2.1).
9. <u>Execute the nested optimization algorithm to allocate the total release to all users $\{r_{1t}, r_{2t}, ...r_{nt}\}$ in order to meet their individual demands.</u>
10. Calculate the $g(x_t, x_{t+1}, a_t)$ and update $V(x_t)$.
11. $l=l+1$.
12. If $l \leq L$, go to step 8.
13. $j=j+1$.
14. If $j \leq m$, go to step 7.
15. Select the optimal actions (decision variables) $\{a_{1t}, a_{2t}...a_{nt}\}_{opt}$, which consist of the optimal transition $\{x_{t+1}\}_{opt}$ and the users releases $\{r_{1t}, r_{2t}, ...r_{nt}\}_{opt}$ that give minimal value of $V(x_t)$.
16. $i = i + 1$.
17. If $i \leq m$, go to step 6.
18. If t>0
19. $t=t-1$.
20. Go to step 4.
21. If t = 0, Check if the optimal actions (decision variables) $\{a_{1t}, a_{2t}...a_{nt}\}_{opt}$, are changed from the previous episode (or in the last three consecutive episodes)? If they are changed, go to step 4, otherwise stop.

---

Algorithm 2 and Algorithm 7 are almost identical. The difference is in step 9 (which is underlined) that executes the nested optimization algorithm. The nSDP nested optimization algorithms are either linear or non-linear as explained in the previous section. The state, action, and reward variables are the same as in nDP.

## 3.4   Nested reinforcement learning (nRL) algorithm

The nRL design can support several state $x_t$ and action $a_t$ variables. One of the possible nRL design is to define the state $x_t = \{t, s_t, q_t\}$, action $a_t = \{x_{t+1}\}$ and reward $g(x_t, x_{t+1}, a_t)$.

If we assume there are $N$ years of available historical time series data of reservoir inflow, this data is divided appropriately into $N$ episodes. The *RL* agent includes several parameters settings like previously described: α – the learning rate; γ – the discount factor; *M* – the maximum number of episodes that defines the maximum number of episodes the agent will perform (this is the stopping criterion preventing the *RL* infinite loop); *LT* – learning threshold and *LR* – learning rate. *LR* is the sum of all the learning updates | $Q(x_{t+1}, a_{t+1}) - Q(x_t, a_t)$ | in one episode as shown in Equation (3.4). If *LR* is below some predefined threshold named *LT*, then the *RL* should stop learning.

$$LR = \sum_{t=0}^{T} |Q(x_{t+1}, a_{t+1}) - Q(x_t, a_t)| \qquad (3.4)$$

The nRL pseudo code is:

Algorithm 8. nRL algorithm pseudo code.

1. Divide the inflow into $N$ episodes for each year.
2. Discretize the reservoir inflow $q_t$ into $L$ intervals, making $L$ intervals centers $q_{lt}$ ($k=1, 2..., L$)
3. Discretize storage $s_t$ in $m$ intervals, making $m$ discretization levels $s_{it}$ ($i = 1, 2, ..., m$)
4. Set initial variables: α, γ, maximum number of episodes – *M*, learning threshold - *LT*.
5. Set *T* as period that defines the number of time steps $t$ in episode (in our case 52 for weekly and 12 for monthly).
6. Set *LR*=0;
7. Set *n*=1 (number of an episode)
8. Set *t*=1 (time step of a period)
9. Define initial state $x_t$ with selecting a starting reservoir volume $s_{it}$,
10. Get the reservoir inflow $q_{lt}$ and $t$ from the current episode.
11. Select action $a_t$, (exploration, or exploitation) and make transition $x_{t+1}$.
12. Calculate the reservoir release $r_t$ based on $x_t$, $x_{t+1}$, $q_{kt}$, and the Equation (2.1).

13. Execute the nested optimization with distributing the reservoir release $r_t$ between water demand users using linear or quadratic formulation, calculate the deficits and other objectives, and calculate the reward $g(x_t, x_{t+1}, a_t)$.
14. Calculate the state action value $Q(x_t, a_t)$.
15. Calculate learning update $|Q(x_{t+1}, a_{t+1}) - Q(x_t, a_t)|$ and add it to $LR$.
16. $t=t+1$ and move agent to state $x_{t+1}$.
17. If $t<T$ then go to step 10.
18. If $t=T$ then $n=n+1$.
19. If $n<N$ then set new episode data and go to step 8.
20. If $n=N$ and $LR>LT$ then go to 6.
21. If $n=N$ and $LR<LT$ then Stop.
22. If $n=M$ then Stop.

The essence of Algorithm 8 is the same as that of the Algorithm 3, but it includes also the RL-related settings. The main difference, as explained previously is in step 13 that executes the nested optimization algorithm. The nRL design can support the additional state variables, as will be demonstrated in the case study implementation presented in the following chapters. Additional state variables directly influence *RL* agent learning abilities, because the number of possible states increases exponentially and more training is needed to learn the optimal policy. On the other hand, inclusion of additional state variables often describes the environment better, and consequently derives better policies.

## 3.5    Multi-objective nested algorithms

So far the nDP, nSDP, and nRL presented above are by design the SO optimization algorithms. To be able to use them in the MO setting, it has been chosen to employ the MOSS (Barreto Cordero 2012). As was presented in Chapter 2, the main idea is to use the SO optimization algorithms (nDP, nSDP or nRL) with different vector weights $w_{i,t}$ and merge results into a set of the Pareto front, as described in Chapter 2.5. Assuming $n$ OFs, first, we generate $m$ vectors of $n$ weights $w = \{w^1, w^2 \ldots w^m\}$ where each $w^i = w_1^i \ldots w_n^i$. The nested single-objective algorithm is executed with each set of weights $w^i$ from the vector. This will generate $m$ solutions from which the Pareto-optimal front can be identified. The MO nested algorithm pseudo code is:

  Algorithm 9. MO nested pseudo code.

1. $i=1$
2. Select $w^i$ from $w_1^i \ldots w_n^i$.
3. Execute SO optimization using the SOAWS function (employing nDP, nSDP or nRL) and find the optimal solution.
4. $i=i+1$.
5. If $i \leq m$ go to step 2, else go to step 6
6. Present the $m$ found solutions as the approximation of the Pareto front

The Algorithm 9 is the same as Algorithm 4, however, in step 4 the nested algorithms are executed. These algorithms can be parallelized if they are executed on *m* systems, like in Algorithm 5. The *m* value will define the number of Pareto set points. Depending on the underlying optimization algorithm, the MO algorithms created by MOSS are named MOnDP, MOnSDP and MOnRL.

It should be noted that the m identified solutions present only limited part of the Pareto front. With reasonably large m better approximation of the Pareto front can be obtained, although some parts of it may still not be identified. More rigorous tests of these aspects are left for future research.

## 3.6 Synthesis: methodology and experimental workflow

This section presents the methodology and the experimental workflow. The three SO algorithms (nDP, nSDP and nRL) and the three MO algorithms (MOnDP, MOnSDP and MOnRL) are tested in the PhD thesis. All of these algorithms are applied in a case study of the Zletovica hydro system presented in the next chapter. The Zletovica hydro system is not a classical single reservoir, and has a significant tributary inflow that is not controlled by the reservoir operation. The presented case study has eight objectives and six decision variables. There are two objectives related to minimum and maximum critical reservoir levels, five water users deficit objectives (two municipal water supply, two agricultural irrigation and ecological flow) and one complex hydropower. Chapter 4 describes the decision variables, constrains and defines the aggregated objective function.

Chapter 5 deals with implementation of the nDP, nSDP and nRL on the Zletovica river basin. The algorithms explained in this chapter, can be directly employed into a single ORO problem, but because the Zletovica hydro system is more complicated the nDP, nSDP and nRL need to be somewhat modified. The nDP and nRL are implemented as described in the optimization problem formulation in Section 4.4. The nSDP cannot include the hydropower objective because of the hydrosystem complexity. All implementation issues and problems are addressed in Chapter 5.

Chapter 6 shows the experiments, results and discussion of the nDP, nSDP and nRL and their corresponding MO versions MOnDP, MOnSDP and MOnRL. The experiments start with nDP performed on the Zletovica hydro system with monthly and weekly data 1951-2005, and in particular, include studying the OFs weights influence on the optimization results. The nDP is compared with classical DP and 'aggregated water demand DP' on the Zletovica river basin. The comparison shows that nDP is a different algorithm than classical DP and 'aggregated water demand DP' (Delipetrev et al. 2015).

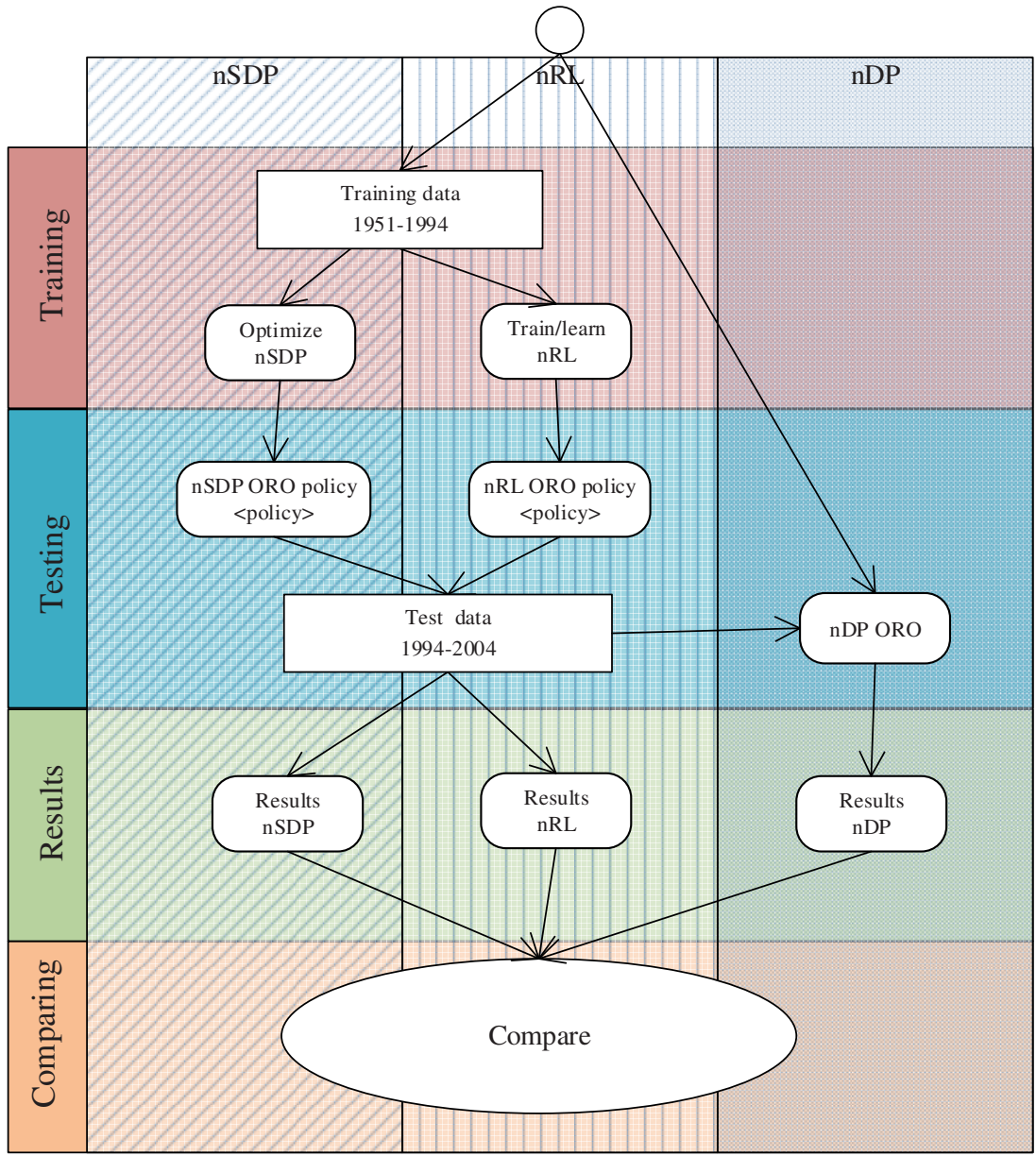

Figure 3.2 Workflow of deriving optimal reservoir policies using the nSDP, nRL and nDP

The nSDP and nRL algorithms are optimizing/learning the ORO policies with weekly training data from the period 1951-1994. After successful completion, the nSDP and nRL have derived one year ORO policy. These two policies are tested on the testing data from the period 1994-2004. At the same time nDP performs ORO on the testing data. Because the nDP is a deterministic optimization algorithm, the solution from nDP is considered the ORO in the testing period. Consequently, the nSDP and nRL results are benchmarked against the nDP ORO results in the testing period. The workflow of deriving ORO policies and their comparison is shown in Figure 3.2.

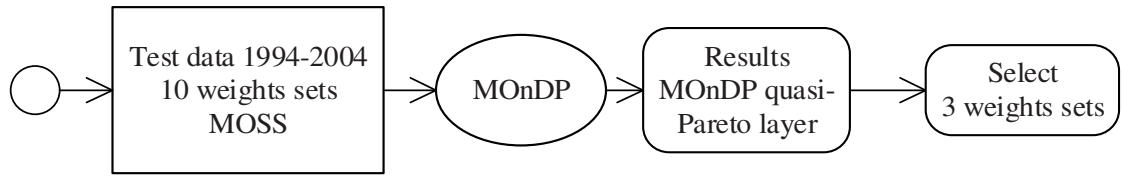

Figure 3.3 MOnDP scanning experiment workflow

The MO experiments are starting with the MOnDP algorithm that is executed with 10 sets of weights on 1994-2004 weekly data as shown in Figure 3.3. The MOnDP scan and explore the possible weights space, because it is significantly faster than nSDP and nRL. The results of this MOnDP scan of the weights space is 10 different ORO in the 1994-2004 period. After analysis of the MOnDP results three weight sets are selected and used in further experiments.

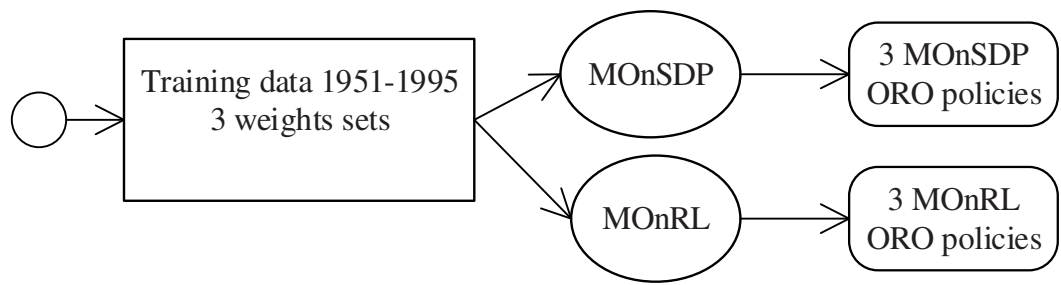

Figure 3.4 MOnSDP and MOnRL calculation of the three ORO policies

These three weights sets are assigned to the MOnSDP and MOnRL to derive the ORO policy on training data, as shown in Figure 3.4.

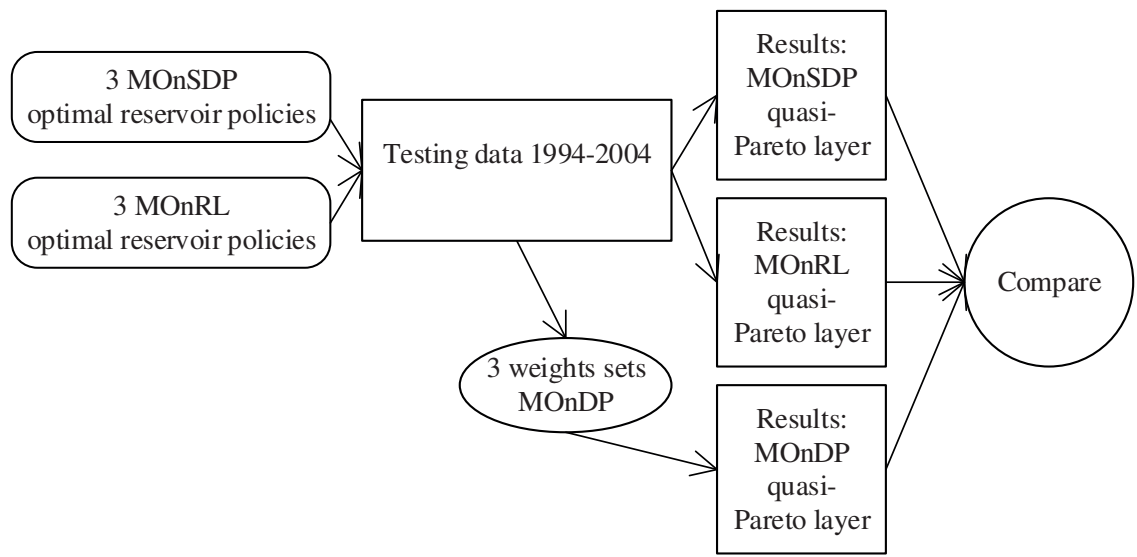

Figure 3.5 Comparing MOnDP, MOnSDP and MOnRL Pareto solutions

The three MOnSDP and MOnRL ORO policies are tested on the testing data, creating the Pareto MO solutions. The three weights sets are applied to MOnDP optimization and executed on the testing data producing three MOnDP ORO solutions. The MOnDP ORO is a benchmark of both MOnSDP and MOnRL ORO policies. All of them were compared and discussed as shown in Figure 3.5.

## 3.7 Conclusions

This chapter presented the three nested SO algorithms – nDP, nSDP, and nRL and their corresponding MO algorithms based on MOSS – MOnDP, MOnSDP, and MOnRL. The general methodology and experimental workflow used in this study is presented as well.

# Chapter 4     Case Study: Zletovica hydro system optimization problem

*"The less you know the more you believe"*

*Bono*

This chapter presents the Zletovica river basin case study located in the northeaster part of Macedonia and being a part of a larger Bregalnica river basin. The Zletovica hydro system and its reservoir Knezevo are explained here in detail, including general climate information, water infrastructure, and design requirements. Knezevo is a multipurpose reservoir that needs to satisfy ecological minimum flows, water supply, irrigation water demands, hydropower, and minimum and maximum reservoir critical levels. The optimization problem is posed by specifying constraints, decision variables, and the OFs.

______________________________________________________________________

## 4.1     General description

Republic of Macedonia covers an area of 25.713 km$^2$ with the population of about 2 million. Considering climate and water resources, Republic of Macedonia has two distinct parts: western and eastern part, divided by the main river Vardar which flows roughly from north to south. The more mountainous western part is richer with water resources compared to the eastern part. In the eastern part, the main river is Bregalnica, which is also left tributary to river Vardar. The total area of the Bregalnica basin is 4,289 km$^2$, located between coordinate boundaries 21° 48' and 23° 3' longitude, and 41° 27' and 42° 51' latitude. Bregalnica has several tributaries: Zletovica, Kamenicka, Osojnica and other smaller rivers. The two main reservoirs in Bregalnica river basin are Kalimanci on river Bregalnica, and Knezevo on river Zletovica shown in Figure 4.1.

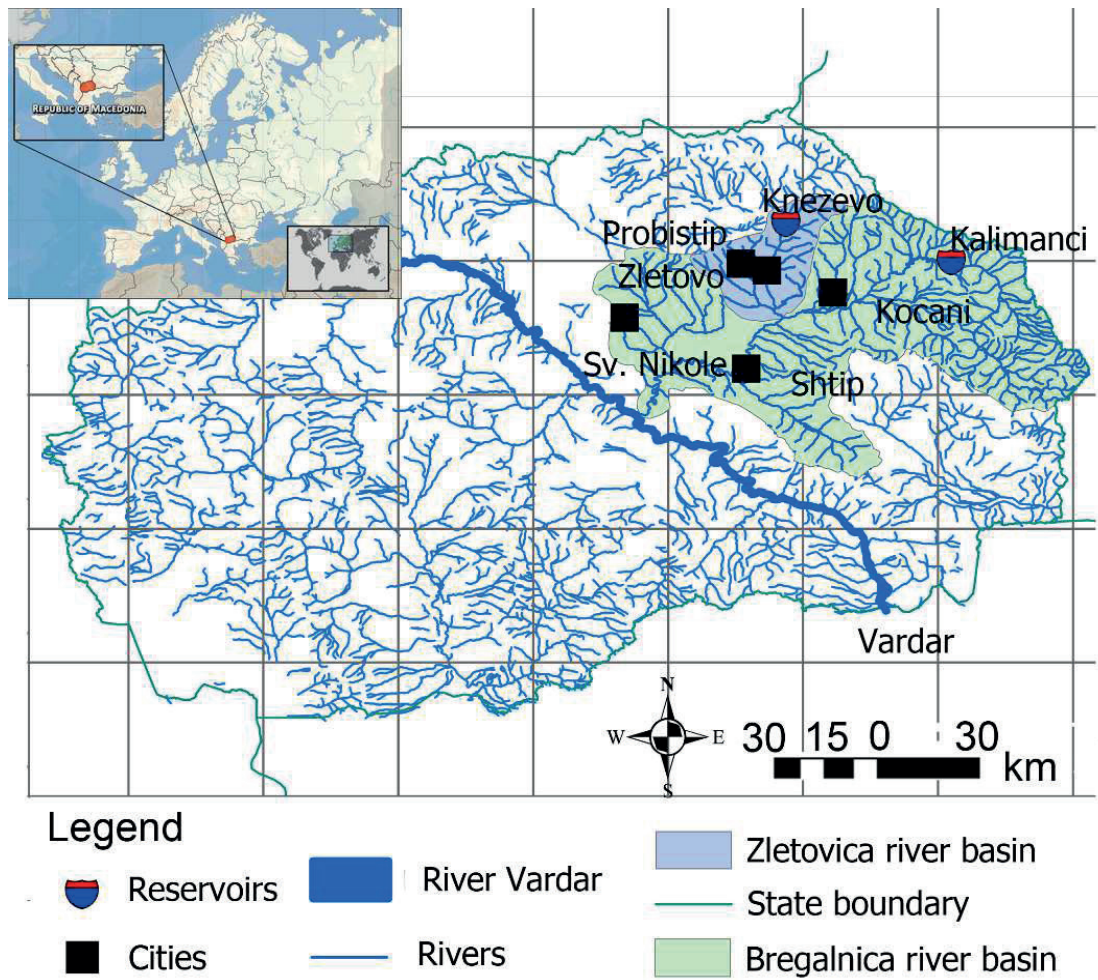

Figure 4.1 Map of Macedonia including Bregalnica and Zletovica river basin

The Bregalnica river basin has developed water infrastructure, which consists of several reservoirs, water supply, irrigation networks, and hydropower stations. Most of the water infrastructure objects were built in the 1960-1970 and in general are not well maintained. During the last 10 years, some new investments in water infrastructure have been provided, including the construction of a new reservoir Knezevo with supporting distribution networks for water supply and irrigation.

The main reservoir in Bregalnica river basin is Kalimanci with $142 \cdot 10^6$ m$^3$ effective storage capacity that is used for irrigation, power generation, and industrial water. Its most important function is the irrigation water demand. Other functions, like power generation and industrial water are of minor importance, but in a period of a rainy year, power generation can have an important value.

The Knezevo reservoir is on river Zletovica, which is the right tributary of Bregalnica. The reservoir has been recently constructed and it is the most important component of the Zletovica hydro system. Knezevo is a multipurpose reservoir with $22 \cdot 10^6$ m$^3$ effective storage capacity that is used primarily for water supply of the towns in the region. Other functions are satisfying the demands of irrigation and hydropower. The Zletovica river basin has been selected as the case study for this research because of the following reasons:

1) Knezevo reservoir is comparatively "more" MO than Kalimanci. The Kalimanci reservoir is almost completely used for irrigation, while Knezevo has water supply, irrigation, and hydropower.

2) Knezevo reservoir and the associated water infrastructure of the Zletovica hydro system are more complex than Kalimanci, and their complexity can be used to demonstrate the possibilities of the nested optimization algorithms.

3) Lastly, since Knezevo is a recent project, the availability of data from recent studies and related projects was significantly larger compared to Kalimanci.

## 4.2   Zletovica river basin

The Zletovica river basin covers an area of around 476 km$^2$, and it is located in the north-eastern part of the Republic of Macedonia between 22°3' и 22°27' geographic longitude and 41°51' and 42°11' geographic latitude as shown in Figure 4.2.

The Zletovica river basin has a typical continental European climate with a little influence of the Mediterranean climate. The high mountains in the basin have colder climate (sometimes even-sub-alpine, or alpine climate), whereas the low-lying valleys in the southern part have continental climate. There are high temperature variations over the year covering range from -20°C in winter until +40 °C in summer. Temperatures in winter are usually between 0-10°C (December - February) with the January temperatures frequently below 0°C. The winter period is characterized with precipitation of rain and snow, especially on higher altitudes. In spring (March - May) the temperatures are increasing from 10°C to 20°C. Precipitation is generally decreasing, but melting snow is the main water contributor in the basin, mainly in the spring period. Summer period (June - August) is dry and hot with temperatures from 25 - 40 °C. This period is typically with very little precipitation, usually coming as short, and localized intensive convective storms. In autumn (September - November) the temperatures are decreasing from 20-10 °C and precipitation is slightly increasing. The average annual precipitation for the whole country is around 730 mm, but the variations in space and with the seasons are very significant. In the Zletovica river basin, the average annual precipitation is around 550-650 mm, but again unevenly distributed within the basin and significantly varying seasonally. The plateau named Ovce Pole that is in-between Shtip and Sv. Nikole and covers the east and south part of the Zletovica river basin is the driest region in Macedonia with 300 mm average annual precipitation measured at Shtip meteorological station. Droughts of over a hundred days in the mounts (July - September) are common in this region. The same area however has fertile soils and developed agricultural activities, which critically depend on irrigation. On the other hand, north-eastern part and the mountains receive higher volumes of precipitation of about 700-900 mm annual average.

Table 4-1. Average climatic data for Shtip region (1971-2010)

| Month | Min temp | Max temp | Humidity | Wind | Sun duration | *Open water $E_p$* |
|---|---|---|---|---|---|---|
| **January** | -2.4 | 5.1 | 80 | 2.2 | 3 | 7 |
| **February** | -0.6 | 8.4 | 75 | 2.4 | 4.1 | 10 |
| **March** | 2.4 | 13 | 69 | 2.7 | 5.5 | 21 |
| **April** | 6.7 | 18 | 64 | 2.7 | 6.5 | 32 |
| **May** | 11.3 | 23.4 | 63 | 2.1 | 7.6 | 43 |
| **June** | 14.5 | 28 | 58 | 2.1 | 9.4 | 54 |
| **July** | 16.4 | 30.5 | 54 | 2.1 | 10.3 | 62 |
| **August** | 16.3 | 30.4 | 55 | 1.9 | 9.9 | 56 |
| **September** | 12.6 | 26.1 | 60 | 1.7 | 8 | 38 |
| **October** | 7.9 | 19.4 | 67 | 1.7 | 5.7 | 23 |
| **November** | 3 | 11.5 | 76 | 2 | 3.6 | 12 |
| **December** | -0.8 | 6.3 | 81 | 1.9 | 2.7 | 7 |
| **Average** | **7.3** | **18.3** | **67** | **2.1** | **6.4** | **30.91** |

*taken from (SWECO 2013a)

In conclusion, continental climatic conditions and high altitude differences are the reasons for vast hydro-meteorological variation in the region, especially concerning rainfall, evapotranspiration, and temperature, which have significant impact on the availability and temporal variability of water resources. This led to the development of the Zletovica hydro system, which includes the Knezevo reservoir, which aims to increase water security in the region. The annual average of the water resources potential of Zletovica river is about 67 $10^6$ m³. Based on previous report (SWECO 2013a), it has been estimated that around 17 $10^6$ m³ will suffice to support the ecological minimum flow regimes and the aquatic ecosystem of river Zletovska. The remaining water resources capacity of around 50 $10^6$ m³ is planned to be used by the multi-purpose hydro system. Knezevo reservoir is controlling about half of the total available water resources and influencing natural hydrological regimes of the watershed of river Zletovska and its tributaries.

## 4.3 Zletovica hydro system

Most of the Zletovica hydro system data concerning reservoir characteristics, tributary inflow, water supply and irrigation demands, ecological minimum flow, hydropower targets, etc. were obtained from (GIM 2008) report. Time series of weekly data for 55 years (1951-2005) are available, which were used in this research. Knezevo reservoir is controlling about half of the total available water resources and influencing natural hydrological regimes of the watershed of river Zletovica and its tributaries. The reservoir Knezevo is a multipurpose reservoir concerning several objectives (listed in order of decreasing priority):

1. Environmental flows in river Zletovica.
2. Water supply to populated areas (towns) in the region: Kratovo, Probishtip, Zletovo, Shtip and Sv. Nikole with a total population of about 100.000 people.
3. Irrigation of 4.500 ha agricultural land.
4. Electric power production with total installed power of about 9.50 MW.

Realization of these objectives is to be achieved with the Zletovica hydro system, consisting of the Knezevo reservoir, together with a network of distribution canals for delivering water for irrigation and urban water supply. A number of small hydropower plants realizes the hydropower generation, one of which is located close to the Knezevo reservoir and the others are distributed along the hydrosystem, operating as derivational (run of river) plants that use the natural head differences created by the topography. The elements of the Zletovica hydro system are presented below in Figure 4.2. (The towns Shtip and Sv. Nikole are slightly shifted from their geographical location to be shown in Figure 4.2.) While the Knezevo reservoir and large parts of the distribution network are already constructed, the number and the locations of the small hydropower plants is still in discussion and the presentation in Figure 4.2 in this regard represents the current design situation.

The town Kratovo and its hydropower station $HEC_7$ shown in Figure 4.2 are not included in the model because the water demand and hydropower are relatively minor. However, to account for this intake of water the demand for Kratovo is deducted from the reservoir inflow $q_t$. Hydropower stations $HEC_4$ and $HEC_5$ are also not included in the model because they do not depend on reservoir operation and are using the natural tributary flow, used to produce hydropower.

In the system, there are four main water intakes for:

- Water supply of Probisthip and Zletovo ($r_{3t}$).
- Upper agriculture zone ($r_{4t}$).
- Water supply of Shtip and Sv Nikole. ($r_{5t}$)
- Lower agriculture zone ($r_{6t}$).

Accordingly, there are four water users, and minimum ecological flow ($r_{7t}$), all with their respective demands.

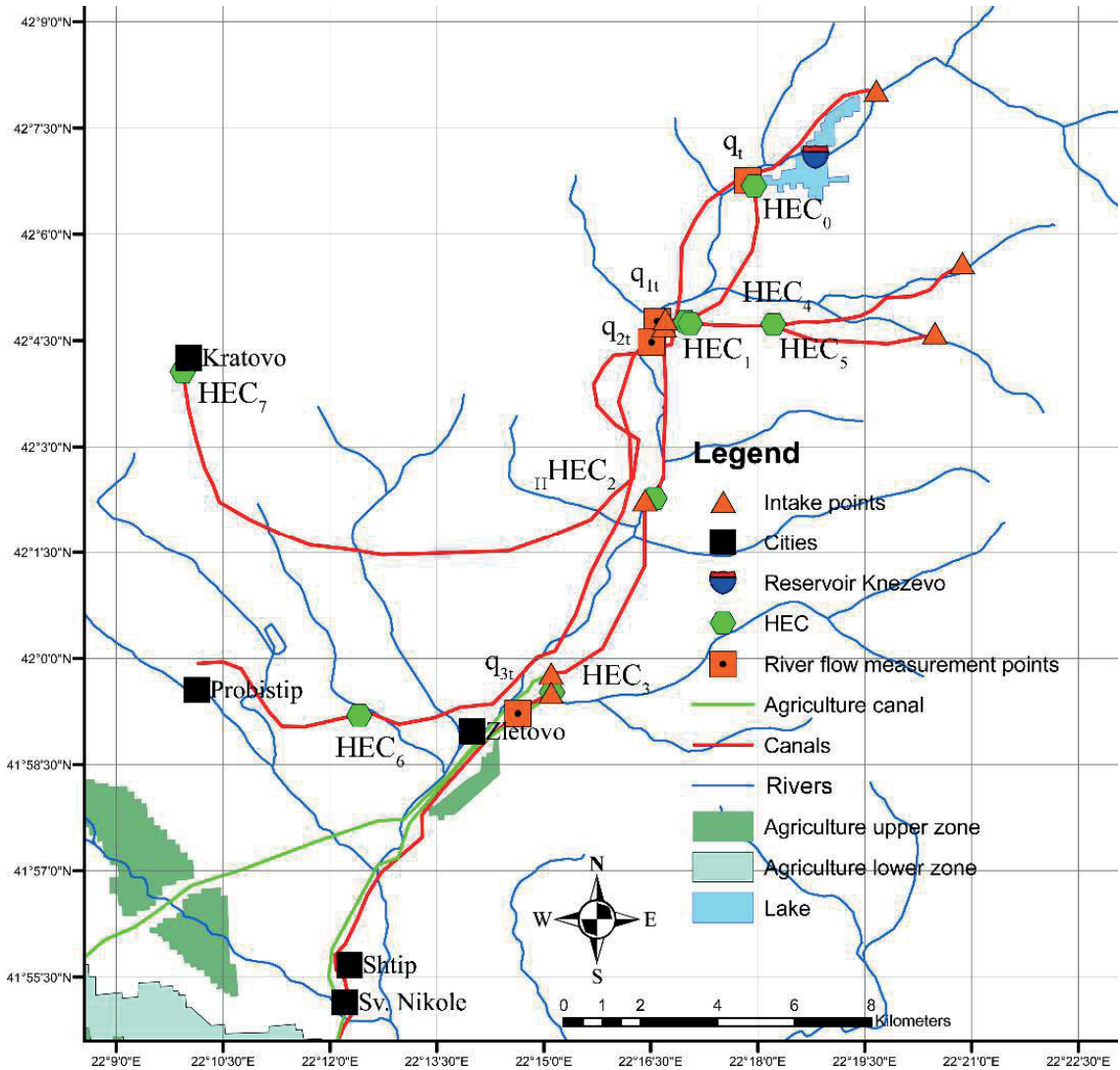

Figure 4.2 Zletovica hydro system

There are four flow measurement points on the river Zletovica and its tributaries, as shown in Figure 4.2 and 4.3. The most upstream river flow measurement point is located upstream of the reservoir location and represents the reservoir inflow $q_t$. The other three flow measurement points are downstream of the reservoir. The $q_{1t}$ is inflow measured at the point before the intake for HEC$_2$. The $q_{2t}$ is the flow measurement point at a point before the intake for HEC$_3$. The last one, $q_{3t}$ is nearby the town of Zletovo. Important notice is that these four flow measurement points $q_t$, $q_{1t}$, $q_{2t}$, $q_{3t}$ register historical flows **before** the construction of the Zletovica hydro system and reservoir Knezevo. The data available from the case study is from 1950 until 2005, and the building started in 2003 with the construction of the access road. So basically everything described in the PhD thesis is based on the design plans and data (GIM 2008, 2010) available for construction of Zletovica hydro system.

As shown in Figure 4.2 there are several tributaries to Zletovica river. The main tributaries that bring large amounts of water in Zletovica river are on the left hand side where HEC$_4$ and $HEC_5$ are located. The river basin has additional complexity because of the inflow coming from these tributaries.

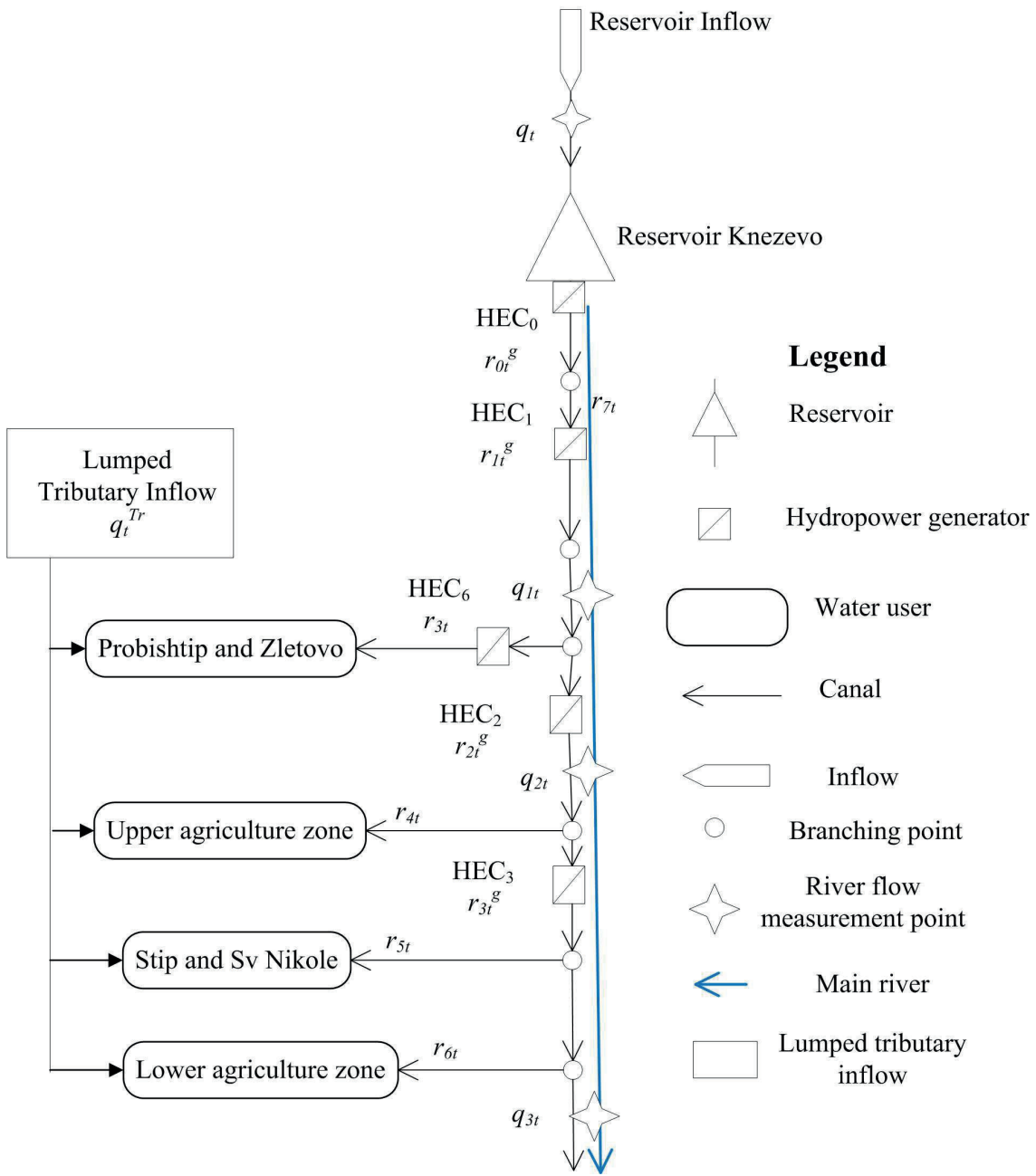

Figure 4.3 Schematic representation of Zletovica hydro system

The simplified, schematic representation of the Zletovica hydro system is presented in Figure 4.3. The $r_{it}$ represent each water user release quantity. The numbering ($r_{3t}$ to $r_{7t}$) is selected to fit the optimization formulation in which objectives related to reservoir water level are numbered with indexes 1 and 2, as will be shown below. The blue line presents the main Zletovica river. What is important to notice is that the water that is not in the canals freely flows in the river. The hydropower stations are also indicated with the water quantity that flows through them. Some of the presented variables are further explained in the following subsection 4.4.3.

The treatment of the inflow coming from the tributaries is dual in nature. When considering the consumptive water users the hydro system is modelled in a lumped way such that water from the tributary inflow $q_t^{Tr}$ is first allocated to all users, and after that the reservoir releases are used to satisfy the remaining user demands. The tributary inflow $q_t^{Tr}$ is calculated as a difference between the last river measurement point $q_{3t}$ and reservoir inflow $q_t$, as in Equation 4.1. Before considering the tributary inflow $q_t^{Tr}$ to comprise the total water quantity that can be distributed among users, analysis was performed to check if this assumption is correct. To satisfy this assumption, $q_{1t}$ always needs to have enough water to satisfy the demands of the towns Probishtip and Zletovo. The analysis proved that this assumption holds and it is possible to consider tributary inflow $q_t^{Tr}$ as the total water quantity available to all users. This approach decreases the number of system variables characterizing the water users. In this system, the main water users are the towns Shtip and Sv Nikole and both agriculture zones.

$$q_t^{Tr} = q_{3t} - q_t \qquad (4.1)$$

$$\sum_{t=1}^{T} q_t^{Tr} \approx \sum_{t=1}^{T} q_t \qquad (4.2)$$

When considering the contributions from the tributaries for the flow through the considered hydropower stations, however, these are accounted separately (not in a lumped way) using the flow differences between two corresponding measuring stations. For example, for $HEC_2$, the contribution from the tributaries is accounted as $q_{t1}-q_t$, which accounts for the additional inflow between the location of the reservoir and the intake for $HEC_2$. Similarly, for $HEC_3$ the tributary contribution is $q_{2t}-q_{1t}$ (accounting for additional inflow between intake for $HEC_2$ and intake for $HEC_3$. This dual representation of tributary inflow does not provide detailed account of all flows through all the canals and rivers of the hydro system, but it is sufficient to accurately represent the different objectives considered in the case study.

Some characteristic data of the Knezevo reservoir are shown in Table 4-2. The maximum water level in the Knezevo reservoir considered for this study is $H_{max}=1061.5$ m amsl, which is in fact the normal operational level, above which overspill occurs. This level corresponds to max storage volume of $V_{max}=23.5 \cdot 10^6$ m³. The minimum storage volume (dead storage) in the Knezevo reservoir is $V_{min}=1.50 \cdot 10^6$ m³ corresponding to $H_{min}=1015.0$ m amsl water level, leaving effectively $22.0 \cdot 10^6$ m³ of storage volume in the Knezevo reservoir for balancing available inflows with downstream water demands.

Table 4-2. Knezevo reservoir data.

| H level [ m amsl] | 990 | 1000 | 1008 | 1020 | 1030 | 1040 | 1050 | 1060 |
|---|---|---|---|---|---|---|---|---|
| $V_{res}$ [$10^6$ m³] | 0.00 | 0.26 | 1.00 | 3.21 | 6.10 | 10.12 | 15.37 | 22.01 |
| $A_{res}$ [km²] | 0.00 | 0.05 | 0.13 | 0.23 | 0.34 | 0.46 | 0.59 | 0.74 |

From the available data the two characteristic curves of the reservoir have been developed: 1) storage volume / head elevation curve and 2) storage volume / area curve. These curves are used for calculation of the hydropower production and the evaporation.

The evaporation from the free water surface is additional loss from the reservoir. There are various methods for calculating the open water evaporation utilizing physically based methods based on mass balance and energy budget, as well as the equilibrium temperature empirical methods and their combinations. The evaporation rates used here are based on the report (SWECO 2013b), where monthly evaporation rates are calculated with the Penman-Monteith formula, using climatic data available for Shtip and Kratovo regions for the period 1971-2010. For the experiments using weekly time steps the monthly evaporation rates are adjusted to weekly values as inputs to the mass balance model. Table 4-3 presents the monthly evaporation rates:

Table 4-3 Monthly evaporation rates

| Month | Jan | Feb | Mar | Apr | May | Jun | Jul | Aug | Sep | Oct | Nov | Dec |
|---|---|---|---|---|---|---|---|---|---|---|---|---|
| $E_t$ [mm/month] | 6.3 | 9.1 | 17.8 | 27.5 | 38.3 | 46.8 | 53.2 | 47.7 | 33.4 | 19.8 | 9.9 | 6.1 |

The evaporated volume is calculated by multiplying the reservoir surface area and the evaporation rates. Since the reservoir area is changing from one-time step to another, the average of the two areas is taken as the representative one:

$$e_t = E_t \cdot \frac{(A(s_t) + A(s_{t+1}))}{2} \tag{4.3}$$

## 4.4 Optimization problem formulation

The Knezevo reservoir is modelled with the mass balance equation introduced before as Equation ((2.1). The difference between a standard single reservoir problem and the one in the Zletovica river basin is in the tributary inflow $q_t^{Tr}$, quite complex water distribution scenario, and the hydropower system. Another important modelling and optimization issue is that first the tributary inflow $q_t^{Tr}$ is used to satisfy water users, and the remaining deficits are requested from the reservoir.

### 4.4.1 Decision variables

The decision variables in the optimization problem are reservoir release *r* and consequently releases $r_{3t}$, $r_{4t}$, $r_{5t}$, $r_{6t}$ and $r_{7t}$ for each water user (two water supply demands, two irrigation demands and the ecological flow). The $s_{t+1}$ is defined by the state transition function univocally in the deterministic case (DP), stochastically in SDP. The future storage $s_{t+1}$ is a decision variable further in the thesis, but only as a substitute of the release $r_t$.

### 4.4.2 Constraints

The optimization problem constraints relate to the dead storage level $H_{dead}$= 1015 m amsl and the corresponding dead storage volume $s_{dead}$= 1.5 $10^6$ m$^6$. The reservoir level cannot go lower than the dead storage level and volume. Another constraint is the reservoir maximum level that is $H_{max}$= 1061.5 m amsl and the corresponding reservoir maximum volume $s_{max}$=23.5 $10^6$ m$^3$. These constraints are expressed as follows:

$$s_{dead} \leq s_t \leq s_{max} \qquad (4.4)$$

If the reservoir level goes above the reservoir maximum level, then overspill occurs, which is uncontrolled water release going directly into Zletovica river, without flowing through $HEC_0$. The actual flows over the spillway are not modelled.

Knezevo reservoir has also a maximum release that is defined by the maximum flow capacity of $HEC_0$. All hydropower plants have also maximum flow constrains described in the following chapter.

### 4.4.3 Aggregated objective function

There are eight objectives considered in the case study: 1) minimum and 2) maximum reservoir critical level deviations, 3) Probishtip and Zletovo water supply deficits, 4) upper zone irrigation deficit 5) Shtip and Sv. Nikole water supply deficits, and 6) lower zone irrigation water deficit, 7) ecological minimum flow deficit, and 8) hydropower deficit. Based on the documentation presented in the report (GIM 2008) the operational targets are a) over 95% satisfaction of the ecological minimum flow, b) 95-98% satisfaction of the water supply and c) 75-80% satisfaction of the irrigation demand. The most important objective is the ecological minimum flow, followed by the water supply, irrigation demand, and hydropower.

The optimization problem posed is the same in all algorithms with slight variations. The Knezevo ORO problem has eight objectives and six decision variables.

The SOAWS function that combines all objectives is considered, being the weighted sum of quadratic deviations over the entire time horizon. Referring to the Bellman Equation the reward function has the following form:

$$g_t(x_t, x_{t+1}, a_t) = \sum_{i=1}^{8} w_{it} \cdot D_{it}^2 \tag{4.5}$$

where $s_t$ is the reservoir storage volume at time step $t$, $w_{it}$ is the objective weight for a given objective $i$ and time step $t$ and $D_{it}$ is the difference between the target value and decision variable for a given objective $i$ and time step $t$.

### 4.4.3.1 Minimum and maximum reservoir critical levels

The main idea of minimum and maximum reservoir critical levels is to control the reservoir level and consequently the volume between the two predefined levels. It is decided to introduce them as objectives and not as hard constraints, in order to treat them as soft targets. With this formulation, small violations of these levels may be acceptable, compared to large violations. The minimum level is set at 1020.5 m amsl and it is 6.5 m above the reservoir dead storage. The minimum level prevents the reservoir to reach a dead storage level and be almost completely empty. The maximum level is set at 1060 m amsl and it is just 1.5 m below the reservoir normal operational level (1061.5 m amsl) above which overspill occurs. The maximum level function is introduced to prevent the reservoir from the overspill level. Overspills are uncontrolled reservoir releases that can produce downstream floods and other problems.

The first two objectives relate to deviations from the minimum and maximum level. These objectives need to be minimized:

$$D_{1t} = \begin{cases} d_{1t} - h_t & \text{if } h_t \leq d_{1t} \\ 0, & \text{if } h_t > d_{1t} \end{cases} \tag{4.6}$$

$$D_{2t} = \begin{cases} h_t - d_{2t}, & \text{if } h_t \geq d_{2t} \\ 0, & \text{if } h_t < d_{2t} \end{cases} \tag{4.7}$$

where $d_{1t}$ and $d_{2t}$ are the minimum and maximum critical levels and $h_t$ is the reservoir level at time step $t$.

### 4.4.3.2  Water demands users

Water demand users are divided into three groups: 1) water supply demands 2) irrigation demands, and 3) ecological flow. The water supply demands of all municipalities (Probishtip, Zletovo, Shtip, Sv. Nikole) are to be supplied by the river Zletovica and the Knezevo reservoir. The water supply demands are obtained from the report (GIM 2008) where they were calculated taking into account the population census data, climatic conditions in the regions of Shtip and Kratovo, seasonality variations and anticipated climate change scenarios. There are two water supply users: 1) Probisthip and Zletovo, and 2) Shtip and Sv. Nikole, denoted as $d_3$ and $d_5$ respectively. The two water supply users' demands in 2005 are shown in Figure 4.4. The two water supply users have almost constant water demand of 75 $10^3$ m$^3$ and 318 $10^3$ m$^3$ per week over the entire 55-year period (1951-2005).

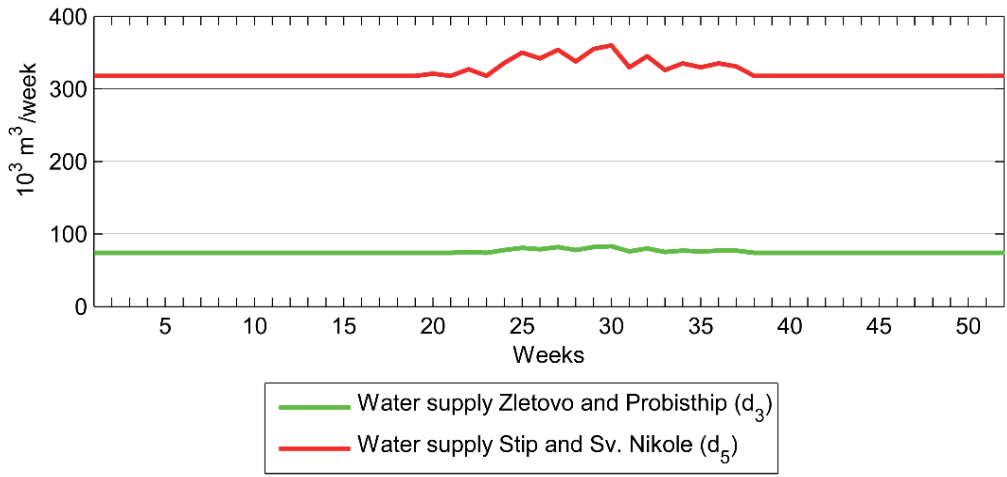

Figure 4.4 Water supply demands in 2005

Similarly, the irrigation demands were calculated by taking into account the different crops and crop rotation schemes, climatic conditions in the regions, soil characteristics and seasonality variations and anticipated climate change scenarios. The two agricultural areas (upper and lower) demands in 2005 are denoted as $d_4$ and $d_6$ and are shown in Figure 4.5. They have relatively different patterns in 55-year period (1951-2005).

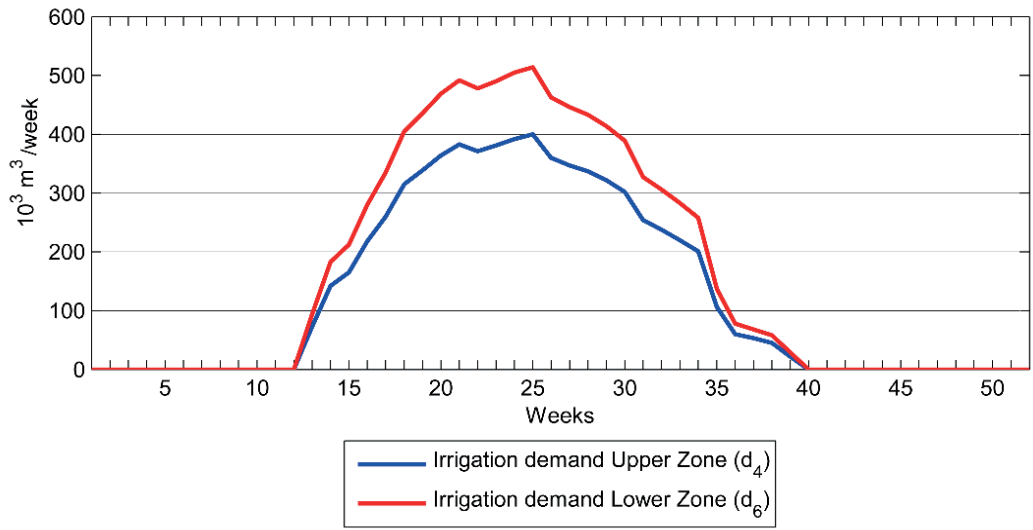

Figure 4.5 Irrigation water demands in 2005

The ecologically guaranteed flows (sometimes termed "biological minimum") enable minimum hydrological flow regimes that sustain the microclimate, the aquatic ecosystem and include the proper groundwater recharge to the Zletovica river basin area. The minimum ecological flow requested from reservoir Knezevo is set at 100 l/s. The ecological water demand is denoted as $d_7$.

Based on the hydro system configuration, our formulation has five water users deficit objectives. These are the following users: 1) the towns of Zletovo and Probishtip (one intake), 2) the upper agricultural zone, 3) the towns of Shtip and Sv. Nikole (one intake), 4) the lower agricultural zone, and 5) the minimum environmental flow, with their respective demands $d_{3t}$, $d_{4t}$, $d_{5t}$, $d_{6t}$, $d_{7t}$. The deficits are calculated using the Equation (4.8):

$$D_{it} = \begin{cases} 0, & \text{if } r_{it} \geq d_{it} \\ d_{it} - r_{it}, & \text{if } r_{it} < d_{it} \end{cases} \quad (4.8)$$

where $r_{it}$ is the release (decision variable) for a given objective $i$ and time step $t$.

### *4.4.3.3 Hydropower OF*

The optimization / simulation results presented in (GIM 2010) contain the hydropower production of the five modelled HECs ($HEC_0 + HEC_1 + HEC_2 + HEC_3 + HEC_6$). In this study, the average of the 55-years simulation hydropower production is used as the hydropower demand. The hydropower demand target is denoted as $d_8$ and is shown in Figure 4.6.

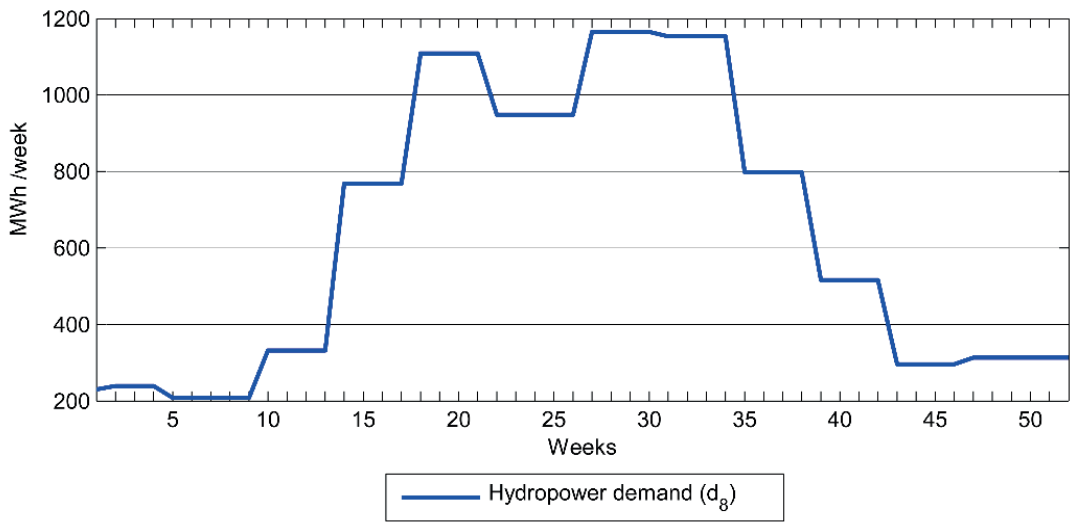

Figure 4.6 Hydropower demand targets

Its corresponding formulation uses $w_{8t}$ as the hydropower weight and the deficits $D_{8t}$ are calculated from:

$$D_{8t} = \begin{cases} 0, & \text{if } p_t \geq d_{8t} \\ d_{8t} - p_t, & \text{if } p_t < d_{8t} \end{cases} \quad (4.9)$$

where $d_{8t}$ is the hydropower demand and $p_t$ is the hydropower production.

HEC$_0$ is positioned at the Knezevo reservoir and the entire reservoir release $r_t$ goes through its turbines. The reservoir release $r_t$ is compared with the generator maximum water capacity HEC$_{0max}$, as in the Equation (4.10).

$$r_{0t}^g = \begin{cases} r_t, & \text{if } r_t \leq HEC_{0max} \\ HEC_{0max}, & \text{if } r_t > HEC_{0max} \end{cases} \quad (4.10)$$

where $r_{0t}^g$ is the reservoir release quantity that goes through the HEC$_0$ turbines. The energy generated by HEC$_0$ (KWh), e.g. in one month is:

$$HEC_{0t} = gen_0 \cdot r_{0t}^g \cdot \frac{h_t + h_{t+1} - 2 \cdot 990}{2} \cdot 24 \cdot (\text{days in month}) \quad (4.11)$$

where $gen_0$ is the coefficient that includes all conversion coefficients and total efficiency and $h_t$ and $h_{t+1}$ are the reservoir levels in time step $t$ and $t+1$. HEC$_1$ uses the same amount of water as HEC$_0$ decreased by the ecological flow. HEC$_1$ head is fixed at 170 m, or water falls from 990 to 820 m amsl.

$$r_{1t}^g = \begin{cases} r_{0t}^g - r_{7t}, & \text{if } r_{0t} - r_{7t} \leq HEC_{1max} \\ HEC_{1max}, & \text{if } r_{0t}^g - r_{7t} > HEC_{1max} \end{cases} \quad (4.12)$$

The amount of hydropower produced by HEC$_1$ is:

$$HEC_{1t} = gen_1 \cdot r_{1t}^g \cdot (990 - 820) \cdot 24 \cdot (\text{days in month}) \quad (4.13)$$

HEC$_2$ is including HEC$_1$ release $r_{1t}^g$, added by the inflows difference $q_{t1}-q_t$, and decreased by the water release for towns Probishtip and Zletovo $r_{3t}$. HEC$_2$ head is 200 m, or from 820 to 620 m amsl.

$$r_{2t}^g = \begin{cases} r_{1t}^g + q_{1t} - q_t - r_{3t}, & \text{if } r_{1t}^g + q_{1t} - q_t - r_{3t} \leq HEC_{2max} \\ HEC_{2max}, & \text{if } r_{1t}^g + q_{1t} - q_t - r_{3t} > HEC_{2max} \end{cases} \quad (4.14)$$

The amount of hydropower produced by HEC$_2$ is:

$$HEC_{2t} = gen_2 \cdot r_{2t}^g \cdot (820 - 620) \cdot 24 \cdot (\text{days in month}) \quad (4.15)$$

HEC$_3$ include HEC$_2$ release $r_{2t}^g$, added by $q_{2t}$-$q_{1t}$ and decreased by the water release for Upper zone agriculture $r_{4t}$. HEC$_3$ head is fixed at 140 m, or from 620 to 480 m amsl.

$$r_{3t}^g = \begin{cases} r_{2t}^g + q_{2t} - q_{1t} - r_{4t}, & \text{if } r_{2t}^g + q_{2t} - q_{1t} - r_{4t} \leq HEC_{3\max} \\ HEC_{3\max}, & \text{if } r_{2t}^g + q_{2t} - q_{1t} - r_{4t} > HEC_{3\max} \end{cases} \quad (4.16)$$

The amount of hydropower produced by HEC$_3$ is:

$$HEC_{3t} = gen_3 \cdot r_{3t}^g \cdot (620 - 480) \cdot 24 \cdot (\text{days in month}) \quad (4.17)$$

HEC$_6$ is on the intake for the towns Probishtip and Zletovo and it uses only release $r_{3t}$. The head is fixed at 200 m, or from 820 to 620 m amsl.

$$r_{6t}^g = \begin{cases} r_{3t}, & \text{if } r_{3t} \leq HEC_{6\max} \\ HEC_{6\max}, & \text{if } r_{3t} > HEC_{6\max} \end{cases} \quad (4.18)$$

The amount of hydropower produced by HEC$_6$ is:

$$HEC_{6t} = gen_6 \cdot r_{6t}^g \cdot (820 - 620) \cdot 24 \cdot (\text{days in month}) \quad (4.19)$$

Table 4-4 Maximum hydropower flows and hydropower generation coefficients

| i | 0 | 1 | 2 | 3 | 6 |
|---|---|---|---|---|---|
| **HEC$_{imax}$ (m³/s)** | 1.5 | 1.5 | 2.1 | 1.8 | 0.14 |
| **gen$_i$** | 8 | 8 | 8.35 | 8.35 | 8.35 |

Both *HEC$_{max}$* and *gen$_i$* are taken from (GIM 2010) and are shown in Table 4-4. The *gen$_i$* coefficients are calculated for each of *i*-th turbine and are function of the efficiency of the *i*-th turbine, ρ - water density (1000 kg/m³) and g - gravitational acceleration (9.81 m/s²)). All hydropower plants HEC$_i$ together produce the following amount of hydropower:

$$p_t = HEC_{0t} + HEC_{1t} + HEC_{2t} + HEC_{3t} + HEC_{6t} \quad (4.20)$$

The produced hydropower $p_t$ is returned to the Equation (4.9) and the hydropower OF is calculated.

### 4.4.4 Objectives weights magnitudes

The main OF combines three distinct objectives types: the minimum and maximum reservoir critical levels that are measured in m, the water user demands that are measured in 10³ m³ /per time step (week or month) and the hydropower energy production that is measured in MWh / per time step (week or month). These different OFs need to be adjusted so they are comparable in their magnitude. The order of magnitude is influenced by the time step (weekly or monthly). Considering that there are five water users deficit objectives, the best approach is to assign higher weights to the objectives related to the minimum and maximum levels. The hydropower objective does not need higher weights because is in the same magnitude range as the water demand users. A similar approach is taken in other previous research studies, e.g. (Pianosi and Soncini-Sessa 2009, Rieker 2010).

## 4.5  Conclusions

In this Chapter 4, the case study of the Zletovica river basin with the Zletovica hydro-system that includes the Knezevo multipurpose reservoir has been described. The optimization problem has been posed with its decision variables, constraints, and SOAWS function. The Zletovica river basin optimization problem requires formulations which are more complex than those for a classical single reservoir. The nDP, nSDP and nRL implementation issues are explained in the next Chapter 5.

# Chapter 5    Algorithms implementation issues

*"Imagination is more important than knowledge"*

*Albert Einstein*

---

This chapter presents the nDP, nSDP, and nRL implementation issues relevant to the case study. The main implementation issues arise from significant tributary inflow $q_t^{Tr}$ and the other two variables $q_{1t}$ and $q_{2t}$ that are used to calculate hydropower. The nDP implements the optimization problem as described in the previous chapter. The nSDP cannot include all stochastic variables and reproduce the optimization problem described before, so some adjustments that are needed are explained further in this chapter. The nRL implements the optimization problem, but has certain drawbacks - complicated implementation, parameters adjusting and convergence criteria.

___________________________________________________________________

## 5.1    nDP implementation

The presented nDP algorithm design in Section 3.1 is for a single reservoir operation. For the case study implementation it needs to include the tributary inflow $q_t^{Tr}$ and other two variables $q_{1t}$ and $q_{2t}$ defined in optimization problem formulation, as described in Section 4.4. The nDP is a deterministic optimization procedure and it is relatively easy to include additional variables. At the beginning of the nDP optimization, the nested optimization algorithm is selected (linear or non-linear). Foremost, the tributary inflow $q_t^{Tr}$ is distributed between the water users' demands ($d_{3t}$, $d_{4t}$, $d_{5t}$, $d_{6t}$, $d_{7t}$), employing the selected nested optimization algorithm. Subsequently, the unsatisfied users' demands are requested from the reservoir. The variables $q_{1t}$ and $q_{2t}$ are included in the computation of the hydropower objective. The nDP completely implements the optimization problem given in Chapter 4. To include handling of the tributary inflow, one significant step has to be added to the nDP pseudo code after step 6, presented in Section 3.1. The nDP implementation pseudo code is presented below, and the added steps are underlined.

Algorithm 10. nDP implementation pseudo code.

1. Discretize storage $s_t$ and $s_{t+1}$ in $m$ intervals, i.e., $s_{it}$ ($i = 1, 2, …, m$), $s_{j, t+1}$ ($j = 1, 2, …, m$) and set $k=0$.
2. Set time at $t=T-1$ and $k=k+1$.
3. Set reservoir level $i=1$ (for time step $t$)
4. Set reservoir level $j = 1$ (for time step $t+1$)
5. Calculate the first group of the OFs, $D_1$ and $D_2$ (related to deviations from reservoir critical levels).
6. Calculate the total release $r_t$ using Equation (2.1).
7. Distribute the tributary inflow $q_t^{Tr}$ using nested optimization between water demand users, calculate their remaining deficits, and set them as the water users' demands.
8. Execute the nested optimization algorithm to allocate the total release to all users $\{r_{1t}, r_{2t}, r_{3t}, r_{4t}, r_{5t}\}$ in order to meet their individual demands.
9. Calculate the second group of the OFs, $D_3, D_4, D_5, D_6$ and $D_7$ (related to users' releases).
10. Using the reservoir levels and the user releases, calculate the third group of the OFs, $D_8$ (related to hydropower production).
11. OFs from step 5, 9 and 10 are combined into the main SOAWS OF $V(s_t)$.
12. $j=j+1$.
13. If $j \leq m$, go to step 5.
14. Select the optimal actions (decision variables) $\{x_{t+1}, r_{1t}, r_{2t}, r_{3t}, r_{4t}, r_{5t}\}_{opt}$ that give minimal value of $V(s_t)$.
15. $i = i +1$.
16. If $i \leq m$, go to step 4.
17. $t = t -1$.
18. If $t > 0$, go to step 3.
19. If $t = 0$, compare the optimal actions (decision variables) $\{x_{t+1}, r_{1t}, r_{2t}, r_{3t}, r_{4t}, r_{5t}\}_{opt}$, and check whether they have been changed from the previous episode. If they are changed, go to step 2, otherwise stop.

Algorithm 10 is a specific implementation of the generalized nDP presented in Algorithm 6. Step 5 of Algorithm 10 calculates the OFs related to deviations from critical levels ($D_1$ and $D_2$) that depend on $x_t$ and $x_{t+1}$ ($s_t$ and $s_{t+1}$). Step 7 distributes the tributary inflow between water users, and decreases their reservoir demands. Step 9 calculates the second group of OFs, users' deficits $D_3, D_4, D_5, D_6$ and $D_7$. With individual users' releases $r_{1t}, r_{2t}, r_{3t}, r_{4t}, r_{5t}$ and reservoir levels $s_t$ and $s_{t+1}$ the hydropower productions and deficits $D_8$ can be determined, described in step 10. All deficits $D_1…D_8$ are included in SOAWS OFs in step 11. The rest of the Algorithm 10 is same as Algorithm 6.

Algorithm 10 calculates the nDP ORO policy. The nDP ORO policy has structure <$x_t, q_t, x_{t+1}$> meaning that at each time step $t$, with the reservoir storage volume $s_t$ and the reservoir inflow $q_t$, there is a rule / policy (denoted by $p$ or $a_t$ actions) for reaching the next state $x_{t+1}$ or next reservoir storage volume $s_{t+1}$. To make a simulation of the nDP ORO policy a starting state $x_1$ (reservoir storage volume $s_1$) needs to be selected.

If the reservoir starting state $x_0$ is unknown, it can be estimated in several ways. One approach to estimate the starting state $x_0$ is to analyse the reservoir historical records and for example take the average reservoir storage in that period. Another approach to estimate the reservoir starting state $x_0$ is to search for a reservoir volume storage $x_1$ that after DP ORO policy simulation will end up in the same reservoir storage volume, or $x_T=x_0$. Afterwards, it is advisable to compare the starting reservoir storage volume $x_0$ with the historical records. If both values are relatively close, then that $x_0$ can be selected as the right starting reservoir storage volume.

In this case because there are no reservoir historical records, as the Zletovica system is still not fully operational, the second approach was used. Based on experience and knowledge from other Macedonian reservoirs, the month of January, when our optimization / simulation period starts, is with lowest reservoir storages. The same applies to Knezevo, and it was confirmed by applying the second approach.

## 5.2 nSDP implementation

The nSDP algorithm presented in Section 3.3 is for a single reservoir problem with the reservoir inflow as a stochastic variable. For the considered case study, the nSDP was re-designed to accommodate the case study optimization problem formulation presented in Section 4.4.

### 5.2.1 Implementation issues

The primary implementation issue in applying nSDP is how to include the four stochastic variables $q$, $q_{1t}$, $q_{2t}$, and $q_t^{Tr}$ shown in Figure 4.3. There is not an example of numerical solution of SDP with four stochastic variables. Perhaps it is possible to design it mathematically, but the practical implementation will probably be very difficult and impractical.

The approach taken here is to investigate the correlation between the reservoir $q_t$ and the tributary inflow $q_t^{Tr}$ shown in Figure 5.1. The correlation coefficient between these two variables is about 0.9 on weekly data. The high correlation gives the opportunity to include the tributary inflow $q_t^{Tr}$ as another stochastic variable in the nSDP algorithm.

If this were not the case (low correlation coefficient) then another approach would be needed. The problem can then be solved by considering the joint probability of the two stochastic variables. There is no theoretically no difficulty in this approach, but an increase of computational complexity would increase.

The nSDP with the two stochastic variables can be implemented only if the reservoir $q_t$ and tributary inflow $q_t^{Tr}$ are clustered into the same number of clusters. It is worth noting that the high correlation coefficient typically suggests that the values of both variables belong to the same cluster interval at each time step over the entire modelling period.

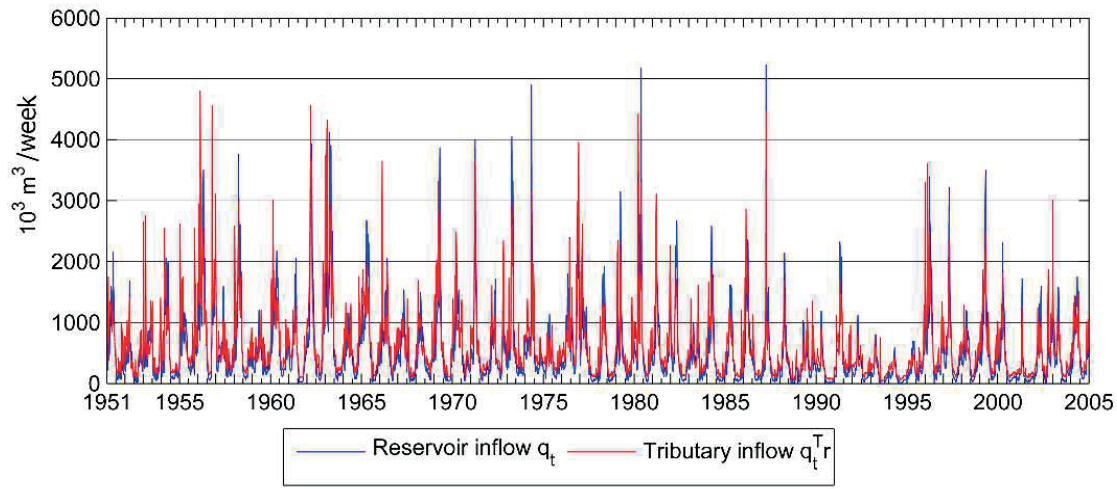

Figure 5.1 Comparison between the reservoir and tributary inflow

The correlation analysis between reservoir inflow $q_t$ and tributary inflow $q_t^{Tr}$ bring us to a possible solution to discard other stochastic variables $q_{1t}$, $q_{2t}$, $q_{3t}$ and simplify the optimization problem formulation. The stochastic variables $q_{1t}$, $q_{2t}$, $q_{3t}$ are only used in calculation of the hydropower OF, as shown in Equations (4.14 – 4.18) and do not affect other OF. The consequence of optimization problem simplification and adjustment is the impossibility to calculate $HEC_2$ and $HEC_3$ power production (and the total hydropower production as well) using nSDP. Therefore, the hydropower aspect is not included in nSDP.

The pseudo-code from Section 3.3 is modified to include the tributary inflow $q_t^{Tr}$ in the nSDP algorithm. The nSDP implementation algorithm pseudo code is presented below, where the modified steps are underlined.

Algorithm 11. nSDP implementation pseudo code, adjusted for the Zletovica case study.

1. Discretize the reservoir inflow $q_t$ into $L$ intervals i.e., $q_{lt}$ ($l=1, 2…, L$)
2. <u>Discretize the tributary inflow $q_t^{Tr}$ into $L$ intervals i.e., $q^{Tr}_{lt}$ ($l=1, 2…, L$)</u>
3. Create the transition matrices $TM$ that describe the transition probabilities $p_{q_{t+1}|q_t}$ of reservoir inflow $q_t$.
4. <u>Create the transition matrices $TM$ that describe the transition probabilities $p_{q^{Tr}_{t+1}|q^{Tr}_t}$ of tributary inflow $q^{Tr}$.</u>
5. Discretize storage $s_t$ and $s_{t+1}$ in $m$ intervals, i.e., $s_{i,t}$ ($i = 1, 2, …, m$), $s_{j,t+1}$ ($j = 1, 2, …, m$) (in this case $x_t = s_t$) and set $k=0$.
6. Set time at $t=T$-1 and $k=k+1$.
7. Set reservoir level $i=1$ (for time step $t$)
8. Set reservoir level $j = 1$ (for time step $t$+1)
9. Set reservoir inflow <u>and tributary inflow</u> cluster $l=1$ (for time step $t$) (<u>the reservoir and tributary inflow clusters are the same</u>)

10. Calculate the first group of the OFs, $D_1$ and $D_2$ (related to deviations from reservoir critical levels).
11. Calculate the total release $r_t$ using Equation (2.1).
12. Distribute the tributary inflow $q_{kt}^{Tr}$ using nested optimization between water demand users and calculate their remaining deficits.
13. Execute the nested optimization algorithm to allocate the total release to all users $\{r_{1t}, r_{2t}, r_{3t}, r_{4t}, r_{5t}\}$ in order to meet their individual demands.
14. Calculate the second group of the OFs, $D_3$, $D_4$, $D_5$, $D_6$, $D_7$ (related to users' releases).
15. OFs from step 10 and 14 are combined into the main SOAWS OF $V(x_t)$.
16. $k=k+1$.
17. If $l \leq L$, go to step 10.
18. $j=j+1$.
19. If $j \leq m$, go to step 9.
20. Select the optimal actions (decision variables) $\{x_{t+1}, r_{1t}, r_{2t}, r_{3t}, r_{4t}, r_{5t}\}_{opt}$ that give minimal value of $V(x_t)$.
21. $i = i + 1$.
22. If $i \leq m$, go to step 8.
23. If t>0
24. $t=t-1$.
25. Go to step 7.
26. If t = 0, Check if the optimal actions (decision variables) $\{x_{t+1}, r_{1t}, r_{2t}, r_{3t}, r_{4t}, r_{5t}\}_{opt}$ have changed in the last three consecutive episodes. If they have changed, go to step 6, otherwise stop.

Algorithm 11 is based on nSDP Algorithm 7. There are additional steps in Algorithm 11, for example step 2 that discretize the tributary inflow $q_t^T$. There are also specific OFs described in steps 10 and 14. Same as in Algorithm 10, the OFs are combined in the main SOAWS function in step 15.

### 5.2.2   Transition matrices

The nSDP algorithm works with the transition probabilities (arranged as matrices) described in Section 2.3. The transition matrices require discretization of the reservoir inflow $q_t$ and tributary inflow $q_t^{Tr}$. Two discretization approaches were tested. The first approach was to discretize the data into equal intervals, with pre-defined number of intervals. This approach produced poor transition matrices because often there were intervals without data, while other intervals (especially the lowest) contained most of the inflow data. A safer approach is to use the K-means to cluster the inflow data. The inflow data is an array of positive real values. The K-means OF is:

$$J = \min \sum_{l=1}^{L} \sum_{i=1}^{n} \left| q_i^{(l)} - c_l \right| \qquad (5.1)$$

where $\left|q_i^{(l)} - c_j\right|$ is the distance measure between the inflow data point $q_i^{(l)}$ and the cluster centre $c_l$. $L$ is the number of cluster centres and $i$ and $l$ are indices. The OF is to minimize the sum of all distances, between the $n$ data points and $L$ cluster centres. Each inflow data belongs to the closest centroid value. The middle value between the centroids defines the intervals' boundaries.

$$I[1] = [0, \frac{(c_1 + c_2)}{2}] \tag{5.2}$$

$$I[j] = \left[\frac{(c_l + c_{l+1})}{2}, \frac{(c_{l+1} + c_{l+2})}{2}\right] \tag{5.3}$$

for $l = 1$ to $L-1$

$$I[L] = \left[\frac{(c_{L-1} + c_L)}{2}, c_L \cdot 5\right] \tag{5.4}$$

The last boundary is set high to include inflow peaks that could occur in simulations. After calculating the inflow intervals, $I[L]$ and the corresponding centroid $c_l$ values, the next step is to compute the transition matrices. The calculation of the transition matrices starts by assigning all inflow data to their corresponding interval value, shown in the equation below.

$$q_t \text{ -> } I_t^l \tag{5.5}$$

where $I_t^l$ is the interval integer number. The period $T$ defines the number of transition matrices (12 if time steps are months, or 52 if time steps are weeks, for one year). The transition matrix dimensions are $L \times L$. An equal transition probability is assigned when there are no transitions from one interval into any other interval.

### 5.2.3  Optimal number of clusters

The K-means OF for different numbers of clusters on the reservoir inflow data $q_t$ is shown in Figure 5.2.

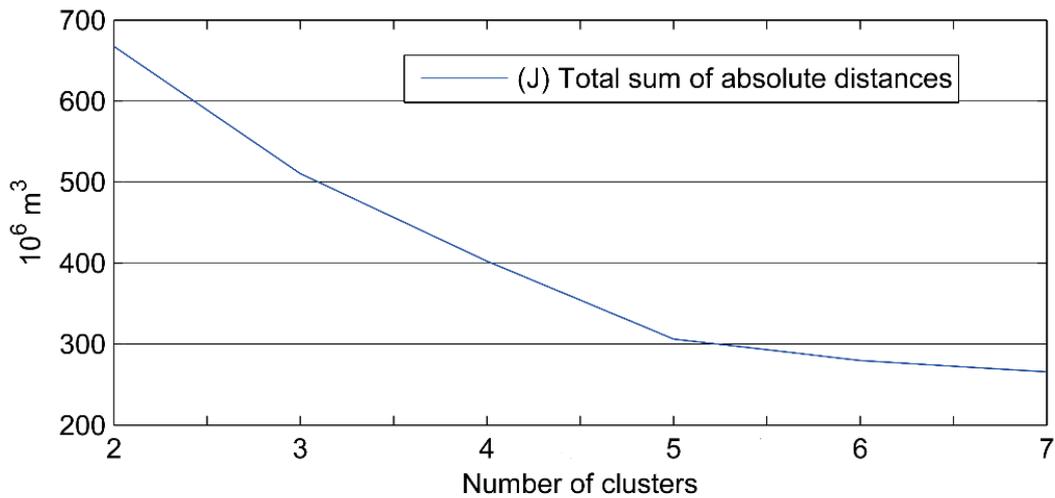

Figure 5.2 K-means OF on reservoir inflow data with different number of clusters

Table 5-1 Inflow distribution with five clusters on reservoir inflow data

| Number of clusters | 1 | 2 | 3 | 4 | 5 |
|---|---|---|---|---|---|
| Cluster centres | 199 | 574 | 1124 | 1933 | 3564 |
| Number of inflows | 1365 | 703 | 374 | 177 | 33 |

Table 5-2 Inflow distribution with seven clusters on reservoir inflow data

| Number of clusters | 1 | 2 | 3 | 4 | 5 | 6 | 7 |
|---|---|---|---|---|---|---|---|
| Cluster centres | 194 | 546 | 1026 | 1589 | 2243 | 3101 | 4107 |
| Number of inflows | 1327 | 694 | 352 | 177 | 75 | 25 | 1 |

Table 5-1 and Table 5-2 show the cluster centres and the number of reservoir inflow distribution with five and seven clusters. From Figure 5.2, Table 5-1 and Table 5-2 it is obvious that the higher number of clusters (seven) does automatically mean better representation. Comparing the two clustering results, the seven cluster results are really similar to those with the five clusters, especially in the first four clusters. The only difference is between the last clusters (5 in five clusters vs 5, 6 and 7 in seven clusters).

The same analysis was applied to tributary inflow $q_t^{Tr}$ and very similar results were obtained. Having in mind that they had a significant correlation coefficient this was expected result. These results confirmed that the optimal number of clusters in tributary inflow is again five. Note that the higher number of clusters increases computational requirements. Five clusters for both the reservoir and tributary inflow have been used in the experiments provided in Chapter 6.

## 5.3 nRL implementation

Similarly to nSDP and nDP, the nRL algorithm described in Section 3.4 is for a single reservoir operation. The nRL design easily included all case study variables ($q_t^{Tr}$, $q_{1t}$ and $q_{2t}$.), and implements the optimization problem formulation as described in Section 4.4. The nRL executes multiple episodes with deterministic variables time series data, where each episode is one year.

The nRL implementation is very specific for this ORO problem described in the case study. That is why often designing and implementing RL (and other machine learning techniques) is an art, because the modellers construct the entire system, define variables, states, actions, rewards, etc. Although most of the subchapters present general settings for RL systems, they are completely dependent on the presented ORO optimization problem. In the following subchapters multiple approaches and possibilities are presented, tested, prototyped and analysed to find the optimal nRL implementation.

### 5.3.1 nRL design and memory implications

The primary design decision in the nRL (and RL in general) is to determine the state, the action, and the reward variables. Three different approaches to define states $x_t$ were tested:

    1) $x_t = \{t, s_t\}$

    2) $x_t = \{t, s_t, q_t\}$

    3) $x_t = \{t, s_t, q_t, q_t^{Tr}\}$.

The nRL action and reward were the same in all three approaches and described as follows:

1) The action $a_t$ with the next state $a_t=\{x_{t+1}\}$ and consequently "nested" releases $a_t= \{x_{t+1}, r_t, r_{3t}, r_{4t}, r_{5t}, r_{6t}, r_{7t}\}$.

2) The reward $g(x_t, a_t, x_{t+1})$ as defined in optimization problem formulation or Equation (4.5). The only difference is that deviation is with a negative sign and the nRL OF is to maximize negative deviation. The maximal gain is 0 when the objective is satisfied.

The state space grows exponentially with the additional state variables, which is shown in Table 5-3. The state space directly influences the computational time and agent's ability to learn. However, the action space stays the same due to the "nested" methodology.

Table 5-3 Number of states considering different designs (concerning weekly data)

| Approach | Time | Reservoir | Reservoir | Tributary | Number |
|---|---|---|---|---|---|
| 1 | 52 | 73 | | | 3,796 |
| 2 | 52 | 73 | 5 | | 18,980 |
| 3 | 52 | 73 | 5 | 5 | 94,900 |

The following example explains the issues with the first and the second definition of state. If the agent is in state $x_t$ and the reservoir inflow $q_t$ is high, then the preferred (optimal) action can be to make the transition to a higher reservoir volume $x_{t+1}$ or to release $r_t$ more water to satisfy all user demands (among other objectives). If the agent is in the same state $x_t$ and the reservoir inflow $q_t$ is low, then the preferred action can be to make the transition to a lower reservoir volume $x_{t+1}$, or to release $r_t$ less water. Using this approach, the agent cannot define the right decision based on the state variables. The reservoir inflow $q_t$ is unknown to the agent and consequently the optimal decision is difficult to be established.

The same discussion applies for the tributary inflow $q_t^{Tr}$. Depending on the tributary inflow $q_t^{Tr}$ the optimal transition (decision) will vary. This simple example demonstrates that it is very important to include the reservoir $q_t$ and tributary inflow $q_t^{Tr}$ into nRL as state variables.

Another important consideration is that the state variable design depends on the modelling time step. If the time step is monthly or weekly, then the values of the reservoir inflow $q_t$ and the tributary inflow $q_t^i$ are relatively comparable with the storage volume $s_t$ in the sense that these inflows bring large changes to the storage volume of the reservoir. Both variables (reservoir inflow $q_t$ and tributary inflow $q_t^i$) influence the agent's ability to observe the environment, learn, and derive the optimal reservoir policy. If the model is based on daily steps, then perhaps the first approach ($x_t = \{t, s_t\}$) is sufficient, because both variables have very small values compared to reservoir volume, and their influence is relatively limited to the agent observation of the environment. In many articles where an RL reservoir model has been developed and the time step is daily, the reservoir inflow is not a state variable (Castelletti et al. 2010). This approach significantly reduces the state – action space.

The three approaches were coded and tested in the case study using monthly and weekly data. The first and second approach had highly fluctuating optimal release policy, especially using monthly data, because the two important stochastic variables: $q_t$ and $q_t^{Tr}$ were not included into nRL, which was reflected in the optimal reservoir policy results. At the end, the obvious solution was to use the third approach.

The nRL implementation pseudo code is shown below. The added steps from the nRL pseudo code presented in Section 3.3, are underlined.

Algorithm 12. nRL implementation pseudo code, adjusted for the Zletovica case study.

1. Divide the available data (usually inflow) into *N* episodes for each year.
2. Discretize the reservoir inflow $q_t$ into *L* intervals, making *L* intervals centres $q_{kt}$ (*k*=1, 2…, *K*)
3. <u>Discretize the tributary inflow $q_t^{Tr}$ into *L* intervals i.e., $q_{lt}^{Tr}$ (*k*=1, 2…, *K*)</u>
4. Discretize storage $s_t$ in *m* intervals, making *m* discretization levels $s_{it}$ (*i* = 1, 2, …, *m*)
5. Set initial variables: α, γ, maximum number of episodes – *M,* learning threshold - *LT*.
6. Set *T* as period that defines the number of time steps *t* in *episode* (in our case 52 for weekly and 12 for monthly).
7. Set *LR=0;*
8. Set *n*=1 (number of episode)
9. Set *t*=1 (time step of a period)
10. <u>Define initial state $x_t$ with an initial reservoir volume $s_t$, read the reservoir $q_{kt}$ and tributary inflow cluster value $q_{kt}^{Tr}$ from the current *episode*, and the time step *t*.</u>
11. Select action $a_t$, (exploration, or exploitation) and make transition $x_{t+1}$.
12. Calculate the reservoir release $r_t$ based on $x_t$, $x_{t+1}$, $q_{ct}$ and the mass balance Equation (2.1).
13. <u>Distribute the tributary inflow $q_{kt}^{Tr}$ using nested optimization between water demand users and calculate their remaining deficits.</u>
14. Distribute the reservoir release $r_{t+1}$ between water demand users using linear or quadratic formulation, calculate the deficits and other objectives targets, and calculate the reward $g(x_t, a_t, x_{t+1})$.
15. Calculate the state action value $Q(x_t, a_t)$.
16. Calculate learning step $|Q(x_{t+1}, a_{t+1}) – Q(x_t, a_t)|$ and add it to *LR*.
17. *t=t+1* and move agent to state $x_{t+1}$.
18. If *t<T* then go to step 10.
19. If *t=T* then *n=n+1*.
20. If *n <N* then go to step 9.
21. If *n=N* and *LR>LT* then go to 7.
22. If *n=N* and *LR<LT* then Stop.
23. If *n= M* then Stop.

Algorithm 12 is based on Algorithm 8. Almost all the steps are the same except the steps for including the tributary inflow $q_t^{Tr}$ (steps 3 and 13). Step 3 discretizes the tributary inflow, and step 13 distribute the tributary inflow using nested optimization between the water users and calculate their reaming deficits. All other parameters and settings of the Algorithm 12 are the same as those of the Algorithm 8.

The $Q(x_t, a_t)$ representation in the third approach requires four dimensional matrixes and needs about 94,900 cells (52 weeks' x 73 reservoir discretization levels x 5 reservoir inflow discretization x 5 tributary inflow discretization). Because the agent explores/exploits the possible actions over the modelling period, it is very likely that some of the $Q(x_t, a_t)$ in the matrix will be unused. The solution selected for dealing with this issue was to use the HashMap function supported in Java.

The HashMap is a data structure used to implement an associative array: it can map keys to values. A hash table uses a hash function to compute an index into an array of buckets or slots. In our case the key is the $<x_t, a_t>$ and the value is $Q(x_t, a_t)$. At each agent action the $Q(x_t, a_t)$ is updated using the HashMap. The HashMap dynamically allocates the computer memory depending on the number of keys/values. This data structure is the best possible solution for our nRL model representation and storage, otherwise, using classical tables and matrices, implementation of the third approach would be computationally time-consuming on a standard PC.

### 5.3.2 nRL parameters

The RL has several parameters that need to be specified: 1) the learning rate parameter $\alpha$, 2) the discount parameter $\gamma$, 3) the exploration/exploitation parameter $\varepsilon$, and 4) the maximum number of episodes *M*. All these parameters influence the nRL agent ability to learn the optimal policy.

The learning rate determines to what extent the newly acquired information will override the old information. If $\alpha$ is 0 then the agent does not learn anything, while if $\alpha$ is 1 then the agent considers only the most recent information. Because nRL is an iterative learning process, often the learning rate $\alpha$ is set higher at the beginning and gradually decreases over the course of learning. This strategy proved to produce good results in our optimization too. More importantly, the decreasing of the learning parameter assures convergence. The strategy taken in the experiments was to set $\alpha$ to 1 at the beginning and afterwards with the increase of the number of episodes decrease $\alpha$ to a value close to 0.

The discount parameter $\gamma$ determines the future rewards importance. If $\gamma$ is 0 then the agent only considers current rewards, while if $\gamma$ is 1 then the agent strives for a high reward in the long-term. If the discount factor meets or exceeds 1, the $Q(x, a)$ values will diverge. The experiments showed that discount parameter of around 0.9 leads to good results and that the discount parameter does not influence learning as much as learning rate parameter.

The exploration/exploitation parameter $\varepsilon$ determines the probability of making an exploration action. If $\varepsilon$ is 1 then only exploration actions are executed, while if $\varepsilon$ is 0 only exploitation actions are executed. This parameter is very important for the tuning of the nRL agent. The $\varepsilon$ parameter influences the state – action space mapping, number of episodes, learning, convergence, etc. The general practice is to set $\varepsilon$ high at the begging of learning and decrease it slowly as the number of episodes grows. The argument for this approach is that at the begging it is good for the agent to explore new actions, while at the end, it is better to more frequently choose the optimal ones to assure convergence and stopping conditions. In the experiments, $\varepsilon$ was set 1 at the beginning and with the increasing number of episodes, it is decreased to a value close to 0.

The maximum number of episodes *M* is another parameter that needs to be set before nRL optimization. This number is difficult to estimate and it depends on the model, the state vector, the learning coefficient, convergence, etc. It is good to set a maximum number of episodes, because it is possible that convergence criteria are never satisfied and the agent will learn indefinitely. Additionally, the episode learning rate over the tuning of the nRL can be reported. When the learning rate sum falls over a predefined threshold, it can define the maximum number of episodes. The maximum number of episodes *M* was set very high (example $10^5$ episodes), taking into account that the experiments were not taking long, and the experiment convergence was always checked. Usually the learning rate sum was stopping the experiments.

All these previously explained parameters are very much dependent on the model, state and action variables, and all other nRL components.

### 5.3.3   Agent starting state, action list and convergence criteria

The agent starting state and boundary conditions are very important issues in nRL. The boundary conditions describe the state-action values at the end of the modelling period, and the solution adopted here is described in Section 2.2, where ending states are connected with the starting states (Bras et al. 1983). The problem is how to define the starting state.

There are several possibilities to deal with the starting state. Let us assume that the only state variables are the reservoir storage $s_t$ and time step *t* and discard others. The obvious solution is to select a fixed starting state $s_1$ and/or a range of possible starting states. If a fixed starting state or range is selected then all other starting states value functions $Q(x_t, a_t)$ are not initialized and they are equal to 0. This creates a problem because the agent does not map the out-of-range starting states (or single state), and often these states will be preferred or avoided depending on maximization/minimization of the agent reward function. When the starting state is out of range, states are indistinguishable for the agent, e.g. an empty or half-full reservoir is the same. RL works well if it is ensured that the agent maps all possible states and actions.

The solution adopted for the starting state problem is a random starting reservoir volume. That means that at each episode the nRL agent randomly selects the starting reservoir volume out of all possible reservoir volumes, leading to a random selection of the starting state. In our case the state is described by $x_t = \{t, s_t, q_t, q_t^{Tr}\}$, time step which is 1, randomly selected reservoir storage volume $s_1$, and $q_1$ and $q_1^{Tr}$ are taken from the current episode. With this approach the agent maps the entire state – action space and even if the starting position is very different from the historical ones, the agent will have an optimal policy (solution).

Forming the agent action list is a problem that requires attention as well. The action is the next state $x_{t+1}$ (transition) in our nRL implementation. The first approach is to allow the agent to select any action – this makes the algorithm convergence faster. If the actions list is fixed after many learning episodes, the best (optimal) action can be easily identified. On the other hand, there are constraints – e.g. the agent cannot select all possible actions due to the necessity of satisfying the mass balance equation. The solution is to introduce a penalty reward for the selected action that violates the mass balance equation. This solution looks reasonable, but the issue arises when some impossible actions in one episode are possible in another episode. An example of this problem is shown in Figure 5.3.

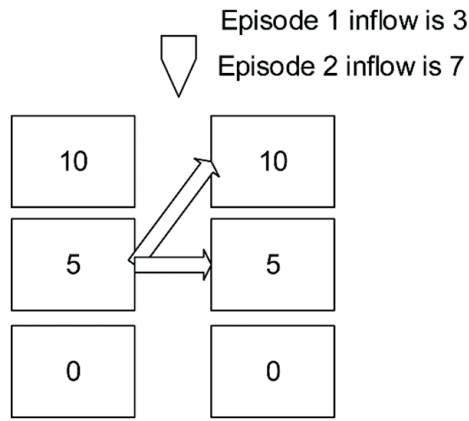

Figure 5.3 Actions list issue

Let us assume that reservoir volume is discretized as shown in Figure 5.3 and the agent is in state 5. The action list is composed of three possible states 0, 5 and 10. If the reservoir inflow in episode 1 is 3 then there are two possible transitions, to state 5 and to state 0. If the reservoir inflow is 7 (episode 2), then there are three possible transitions to 0, 5 and 10. If the agent is in episode 1, makes an exploration action, and chooses the transition to state 10 then the mass balance equation is breached, and the agent is penalized. In the next episode 2 the agent will probably avoid state 10 because in the previous step it was penalized for this transition, but maybe in episode 2 the transition to state 10 is the optimal one. Another important point is the choice of the penalty value. If the penalty value is too high, the agent would never select such actions. On the other hand, if it is too low, this could trigger unwanted agent behaviour with the violation of the mass balance equation. The conclusion is that the approach allowing the agent to take all possible actions is not the best option.

An alternative approach is to generate a list of possible actions at each time step that satisfy the mass balance equation. The dynamic list of possible actions means that at certain state some actions are possible in some episodes and impossible in others episodes, depending on the reservoir inflow and evaporation. The dynamic action list changes at each time step and there is no clear convergence, because the optimal action changes at each time step. The dynamic action list actually incorporates the stochastic process in the nRL optimization, meaning that the agent at each time step gets the information from the environment around its current state (reservoir volume, reservoir inflow, tributary inflow) and produces a list of possible actions with the view of constraints (mass balance equation).

The convergence criteria and stopping conditions topics are an issue in RL like in many other algorithms (artificial neural networks, genetic algorithms, etc.). The DP and SDP method make an exhaustive search over the whole possible state – action space and have clear convergence and stopping criteria, while the nRL explores only a fraction of that space. It is possible that some regions of the state – action space, not explored by nRL, contain the optimal solutions.

The RL convergence criteria vastly depend on the model, the length of the modelling period, state – action space, optimization variables, etc., and the solution is to tune the convergence criteria according to the specific problem. Many different approaches for tackling the convergence and stopping learning conditions were tested in this case study optimization problem. The first approach was to store the optimal reservoir policy at each episode and compare it to the next one. If the optimal reservoir policy is the same over a considerable learning period, then the agent should stop learning, similar to nDP and nSDP. Regrettably, this approach appeared not to be feasible, because the dynamical action list is changing the optimal policy at each episode. The most widely used approach for convergence criteria is to measure the rate of learning shown in Equation (3.4), and when it falls below some threshold μ, the agent stops learning. It is a very straightforward approach, but there is one very important issue, that the learning rate parameter α directly influences the rate of learning. With the decrease of the learning rate parameter α explained in Section 5.3.2, the learning rate also decreases. There is an interplay between RL parameters like defining α, the rate of decreasing α, γ, ε, and the maximum number of episodes. Again, the choice of all these parameters values depends on the model, variables, modelling horizon, etc., and typically requires a lot of experimentation.

## 5.4   Conclusions

This chapter presented the nDP, nSDP, and nRL implementation issues, some of which are relevant to the case study, while others are more general. The case study has significant tributary inflow $q_t^{Tr}$, and two other variables $q_{1t}$ and $q_{2t}$ used to calculate hydropower. This makes the Zletovica case study more complex than a classical single reservoir case study and consequently some of the algorithms (in this case nSDP) have difficulties in implementation. The nDP included all variables as specified in the optimization problem formulation. The nSDP modified the optimization problem and included the tributary inflow $q_t^{Tr}$ with a condition that it is always in the same cluster group as reservoir inflow. The other two stochastic variables $q_{1t}$ and $q_{2t}$ were not included because of increased complexity, and, consequently hydropower was not calculated with nSDP. The nRL included all variables from the optimization problem formulation. The issues concerning the design, parameters, action list and convergence criteria demonstrate the complexity of setting up the nRL algorithm and the need for extensive testing with different settings and parameter values. The following Chapter 6 presents the experiments, results, and discussion related to the use of the nDP, nSDP, nRL, MOnDP, MOnSDP and MOnRL in the case study.

# Chapter 6     Experiments, results and discussion

*"Prediction is very difficult, especially if it's about the future"*

*Nils Bohr*

This chapter presents the nDP, nSDP, nRL, MOnDP, MOnSDP, and MOnRL experiments, results, and discussion. The nDP experiments have been performed for 55 years' monthly data (1951-2005) demonstrating how weights influence the ORO and specific water user satisfaction. The nDP experiments have also been carried out in 55 years' weekly data (1951-2005) and compared with the case study requirements. The nSDP and nRL experiments derive their ORO policies on 45 years weekly training data (1951-1994). These ORO policies are compared with the nDP ORO on 10 years testing weekly data (1995-2004). The nDP is executed with 10 different sets of weights to provide scan of possible MO solutions in the training data. Out of these 10, three sets of weights are selected for MOnSDP and MOnRL. The MOnDP, MOnSDP, and MOnRL identified optimal solution in MO settings. All mentioned experiments and results are analyses and discussed.

___________________________________________________________________

## 6.1     Experiments with nDP using monthly data

The nDP algorithm was tested using 55-year monthly data (1951-2005), with 660 time-steps. The reservoir operation volume is discretized in 73 equal levels (300 $10^3$ m$^3$ each). The minimum reservoir level was set at 1021.5 m amsl, and the maximum reservoir level at 1060 m amsl. The water supply, irrigation, and hydropower are set to their respective monthly demands, derived from the weekly data. The objective function is described by Equation (4.5).

In order to test and evaluate the nDP algorithm and to compare results, the two nested optimization methods and two different sets of weights were considered as shown in Table 6-1, leading to the following test cases:

a) Linear (hence the Simplex method, Equation (3.2)) named nDP-$L_1$ and nDP-$L_2$, and

b) Quadratic (the quadratic method, Equation (3.3)) named nDP-$Q_1$ and nDP-$Q_2$.

The indexes represent the experiments with different weights. In the quadratic Knapsack the discretization of allocation was set to 50. In the first set of weights the higher weights are set for urban water supply objectives ($w_3$ and $w_5$) followed by ecological flow, irrigation, and hydropower. The second set of weights gives higher priority to one of the urban water supply objectives ($w_5$) at the expense of the two irrigation objectives ($w_4$ and $w_6$).

Table 6-1 Two objectives weights sets used in the four considered experiments

| Experiments | $w_1$ | $w_2$ | $w_3$ | $w_4$ | $w_5$ | $w_6$ | $w_7$ | $w_8$ |
|---|---|---|---|---|---|---|---|---|
| nDP-$L_1$ and nDP-$Q_1$ | 25000 | 25000 | 0.2 | 0.1 | 0.2 | 0.1 | 0.26 | 0.04 |
| nDP-$L_2$ and nDP-$Q_2$ | 25000 | 25000 | 0.15 | 0.1 | 0.25 | 0.1 | 0.26 | 0.04 |

The results of 55 years' optimization are shown in Figure 6.1 and Figure 6.2. From the overall 55 years ORO, it is visible that the periods 1987-1995 and 1999-2002 are relatively dry, having impact on all objectives. The nDP-$L_1$ optimization has only several overspills with small volume.

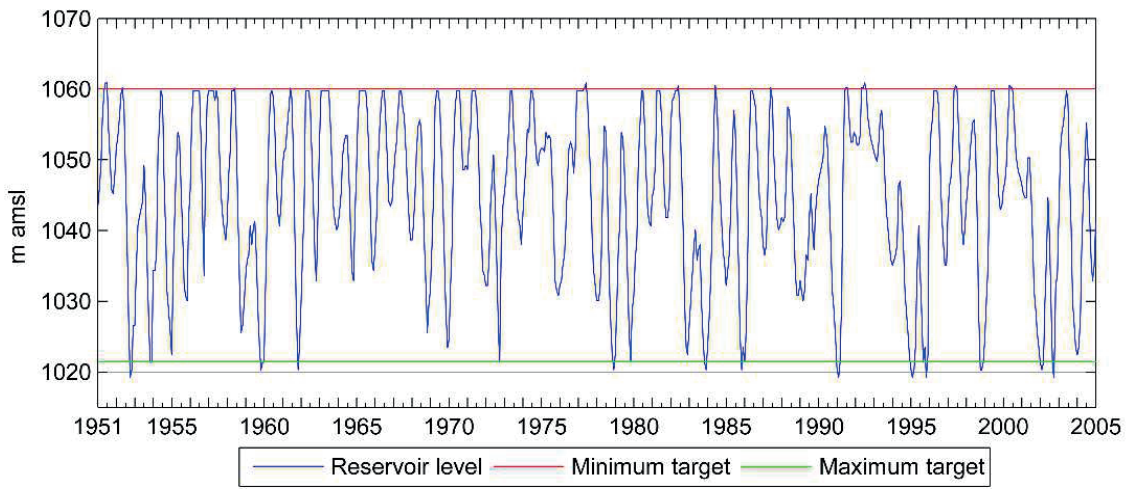

Figure 6.1 nDP-$L_1$ optimal reservoir level in 55 years

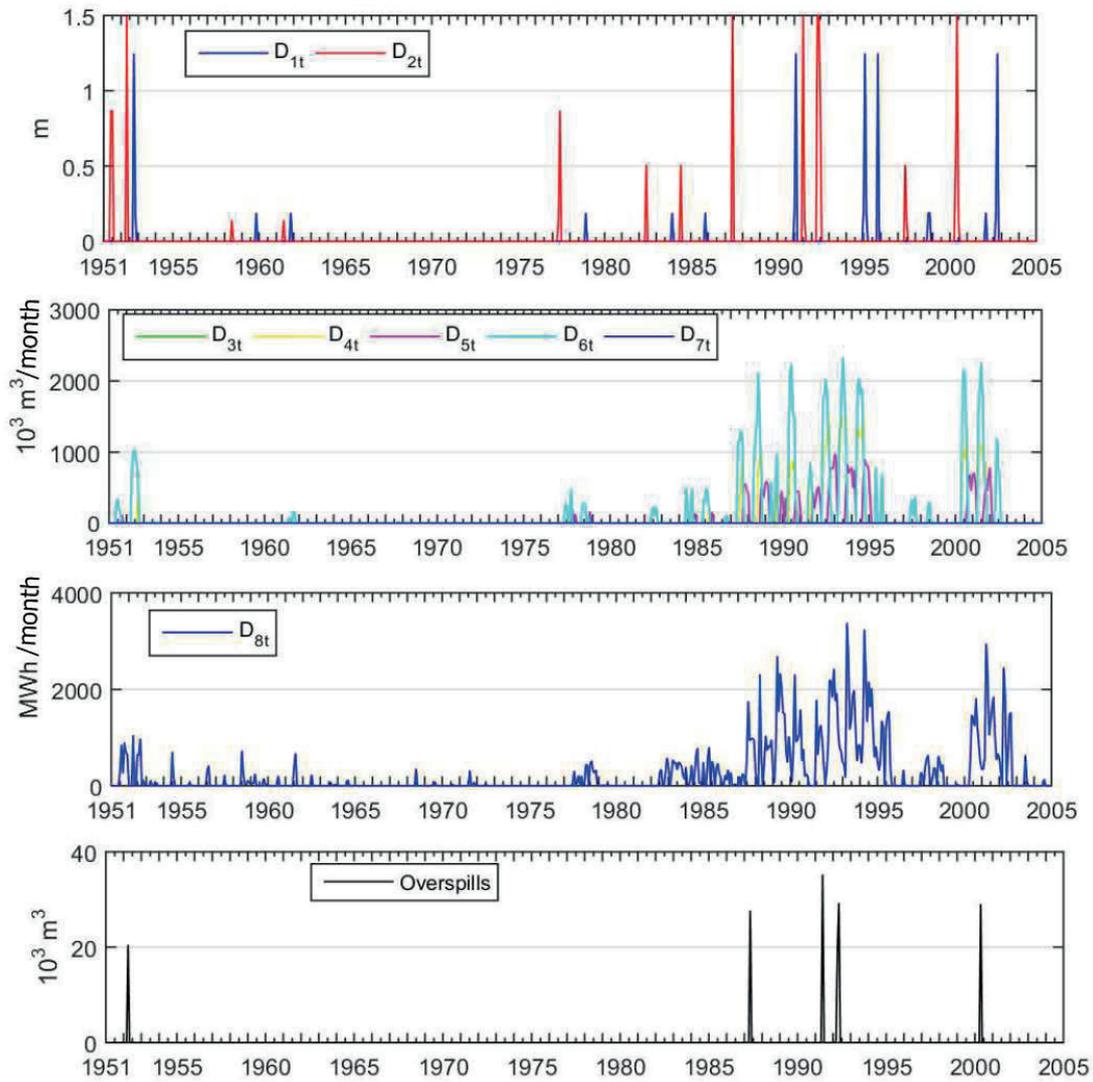

Figure 6.2 nDP-L1 minimum and maximum critical levels deviations, water users' deficits, hydropower deficit, and overspills over entire time horizon of 55 years (monthly)

The nDP-$L_1$, nDP-$L_2$, nDP-$Q_1$, and nDP-$Q_2$ optimization results are also compared with sums of 1) minimum and maximum critical levels deviations, 2) water user's deficits, and 3) hydropower deficit, over the entire time horizon, as shown in Table 6-2 and in Figure 6.3.

Depending on the formulation that is used in the nested optimization, the nDP-$Q_1$ and nDP-$Q_2$ have a better result in $D_1$ and $D_2$ compared with the nDP-$L_1$ and nDP-$L_2$ experiments respectively, as can be seen in Figure 6.3. The $D_3$ and $D_5$ deficits in the nDP-$L_1$ and nDP-$L_2$ experiments are at the expense of $D_4$ and $D_6$ (agriculture), which have higher deficits, as shown in Figure 6.3. When comparing nDP-$L_1$ and nDP-$L_2$ with nDP-$Q_1$ and nDP-$Q_2$ experiments, it can be seen that $D_3$ and $D_5$ have less deficits when a linear formulation is used. The quadratic formulation makes more balanced distribution between the water users, which is due to its objective function and alleviates the $D_6$ peak deficits that occur in nDP-$L_1$ and nDP-$L_2$ experiments. The quadratic formulation gives better results concerning $D_8$ hydropower deficits, as shown in Figure 6.3.

Table 6-2. Total deviation for the nDP-$L_1$, nDP-$L_2$, nDP-$Q_1$ and nDP-$Q_2$ experiments

| Experiment | $D_1$ | $D_2$ | $D_3$ | $D_4$ | $D_5$ | $D_6$ | $D_7$ | $D_8$ |
|---|---|---|---|---|---|---|---|---|
| nDP-$L_1$ | 8.7 | 13.7 | 2 | 41,799 | 31,718 | 86,095 | 0 | 173,481 |
| nDP-$Q_1$ | 4.1 | 10.6 | 30,335 | 41,508 | 36,520 | 41,299 | 23,473 | 146,057 |
| nDP-$L_2$ | 7.9 | 12.7 | 19,356 | 38,354 | 18,968 | 83,725 | 0 | 171,626 |
| nDP-$Q_2$ | 4.1 | 10.4 | 34,066 | 42,791 | 31,133 | 42,240 | 23,535 | 146,219 |

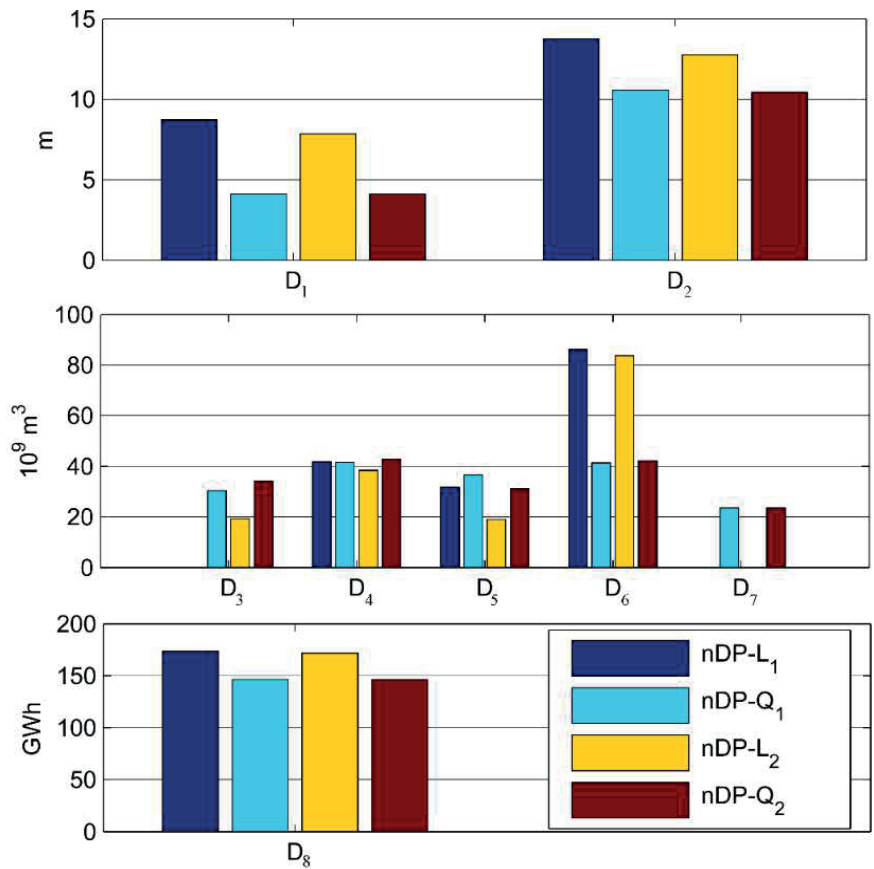

Figure 6.3 Comparison of the sums of minimum level deviations ($D_1$), maximum level deviations ($D_2$), users' water deficits ($D_{3-7}$) and hydropower deficit ($D_8$) over the entire time horizon for the four experiments

The weight $w_5$, which is used for one water supply user (Shtip and Sv. Nikole), is the highest from the water users in the nDP-$L_2$ and the nDP-$Q_2$ experiments and consequently it is the most satisfied water user, as shown in Figure 6.3. The $D_3$ deficit in the nDP-$L_1$ experiment is zero, but it has a significant increase due to the $w_5$ increase in the nDP-$L_2$ experiment, although $w_3$ weight is unchanged, as shown in Figure 6.3. The $D_4$ and $D_6$ deficits are slightly increased in nDP-L$_2$, comparing to nDP-L$_1$ experiment. The $D_3$ and $D_5$ deficits are lower in nDP-Q$_2$ comparing to nDP-Q$_1$ experiment. These deficit reductions are achieved at the expense of $D_4$ and $D_6$ deficits. When comparing nDP-L$_1$, nDP-L$_2$, nDP-Q$_1$ and nDP-Q$_2$ experiments considering $D_1$, $D_2$, $D_3$, $D_7$ and $D_8$ deficits and their weights, it is obvious that the results are different even without changes of the corresponding weights, as shown in Figure 6.3. This happens because the change in one or more weights, in this case $w_4$, $w_5$ and $w_6$, modifies the main objective function and consequently produces different results. The presented experiments demonstrate that by changing the weights in accordance with the user preferences it is possible to create a different optimal reservoir operational schedules.

For more understandable presentation of results, a three-year period (1985-1987) sample was selected from the nDP-Q$_2$ experiment. Figure 6.4 shows the variations of the reservoir and tributary inflow, reservoir volume and reservoir release in this period. Figure 6.5 shows the reservoir level, dead storage level, minimum and maximum levels, and their deviations. Figure 6.6 shows water users' demands and deficits, and Figure 6.7 shows hydropower demands and deficits. This three-year period is a combination of the two dry years with one wet year in between them.

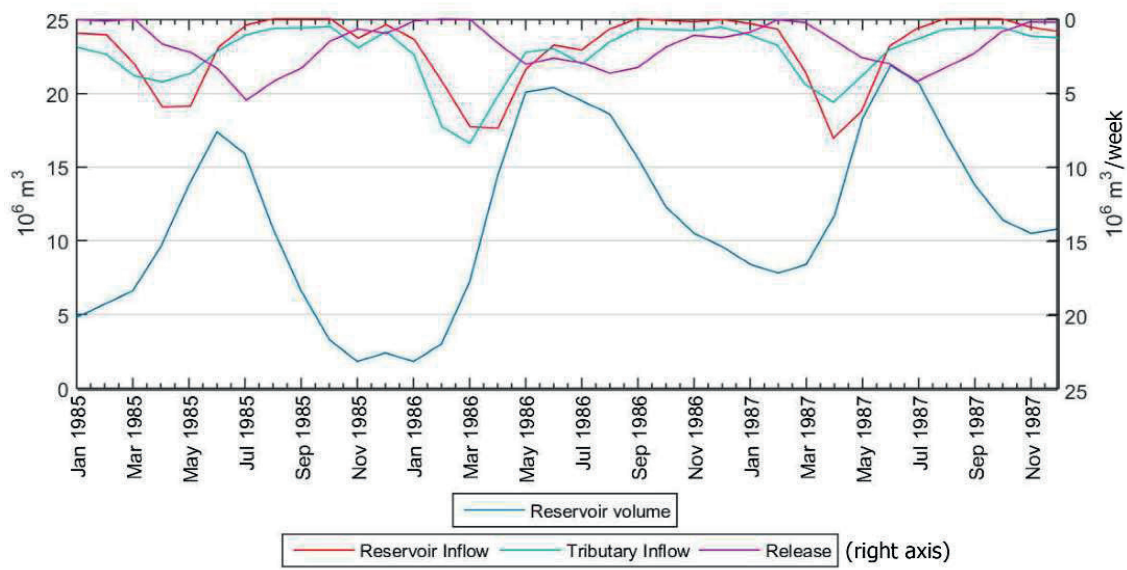

Figure 6.4 nDP-Q$_2$ monthly reservoir volume, release, and reservoir and tributary inflow 1985-1988

This three-year 1985-1988 period is characterized by significant reservoir and tributary inflows in the spring months (from February until May) due to high precipitation and snowmelt, which is relatively small in the other periods, as shown in Figure 6.4. The reservoir releases are rising between June and October because of increased demand of the agricultural users, as shown in Figure 6.4 and Figure 6.6. The urban water supply users ($d_{3t}$ and $d_{5t}$) and the ecology ($d_{7t}$) demands are relatively constant, while the agriculture users ($d_{4t}$ and $d_{6t}$) demands are variable reaching a maximum in the summer months, as shown in Figure 6.6. The reservoir inflow and the tributary inflow in 1986 are larger and more widely distributed between January and June compared to 1985 and 1987 when there are high reservoir inflows in April and May, as shown in Figure 6.4. The difference between 1986 and the other two 1985 and 1987 is shown in: the reservoir release shown in Figure 6.4, the reservoir level shown in Figure 6.5, the water users' deficits shown in Figure 6.6 and hydropower production shown in Figure 6.7.

The majority of users' deficits occur from April until October in 1985 and 1987, as shown in Figure 6.6. The deficits $D_4$ and $D_6$ (agriculture) are larger than $D_3$ and $D_5$ (towns) deficits in 1985 and 1987, because of the smaller weights $w_4$ and $w_6$ and their larger demand, as shown in Figure 6.6. With the linear formulations (experiments $L_1$ or the $L_2$), the difference between the agriculture and the towns' deficits is even larger (not shown in the above figures). The reservoir satisfies (almost) all objectives in 1986, as shown in Figure 6.5, Figure 6.6 and Figure 6.7.

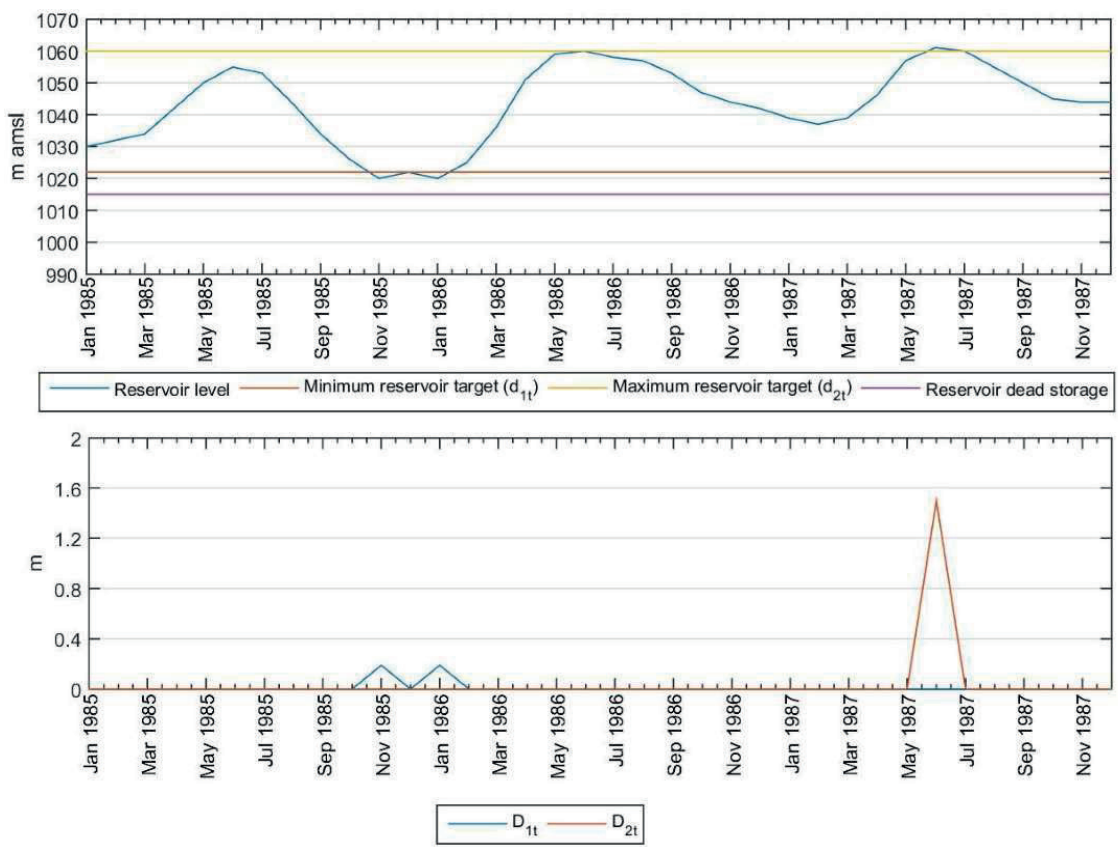

Figure 6.5 nDP-$Q_2$ minimum and maximum deviations 1985-1988

There are reservoir level maximum deviations ($D_2$) in May 1987 because of peak reservoir inflows, as shown in Figure 6.4 and Figure 6.5. On the other side, there are minimum level deviations ($D_1$) in October and December 1985. This coincides with the reservoir level shown in Figure 6.5, which is lowest in periods between November and January. The last months of 1987 suggest that probably the next year will be very dry. The reservoir stores additional volume in winter months (October till December 1987) at the expense of water demand users' and the hydropower, as shown in Figure 6.6 and Figure 6.7.

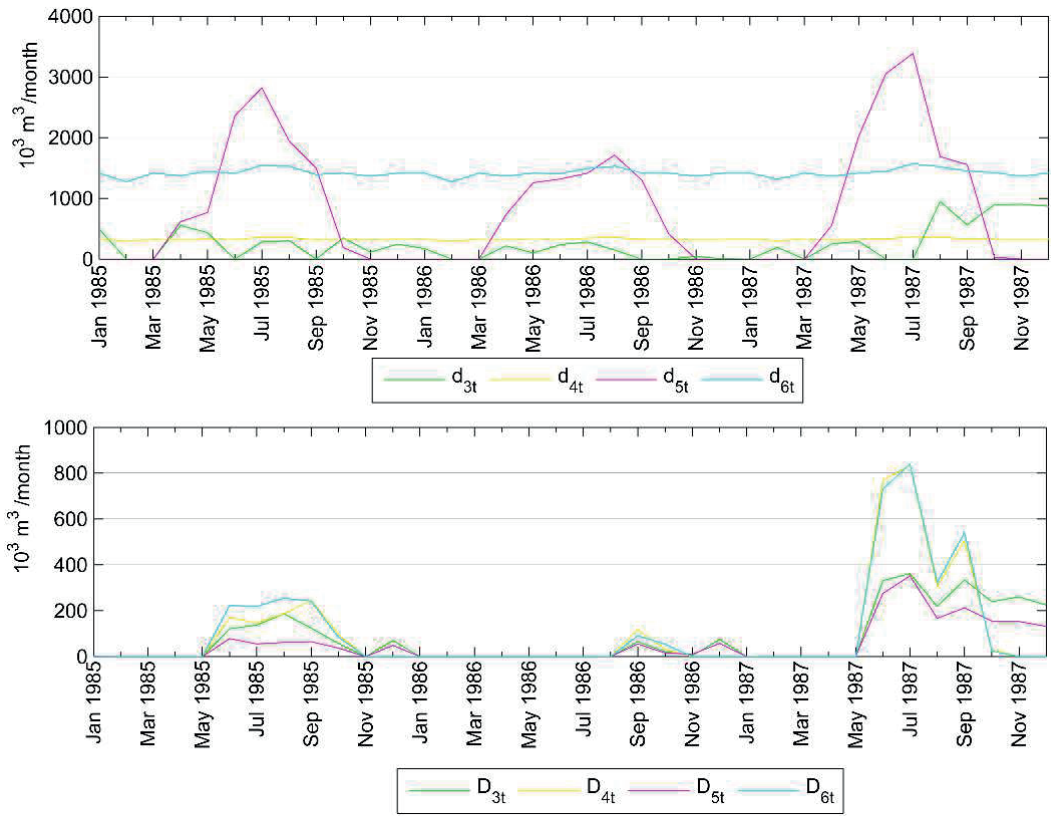

Figure 6.6 nDP-$Q_2$ water users' demands and deficits 1985-1988

The nDP algorithm was also tested with a variable reservoir volume discretization using the nDP-$Q_2$ experiment settings. The reservoir storage volume was discretized as follows: a) from 0 to 5 $10^6$ m$^3$ and from 19 $10^6$ m$^3$ to 22 $10^6$ m$^3$ in intervals of 200 $10^3$ m$^3$, and b) the rest was discretised in intervals of 150 $10^3$ m$^3$ (resulting in the 144 discretization levels). The higher discretization has directly increased the computational resources and the running time, but it has improved the results (1-4%). Additionally, the nDP algorithm was also tested with variable objective weights at different time steps.

In designing the optimal reservoir operation, one needs to be aware of the objectives interdependence, in our case users' releases, reservoir level, and hydropower. The hydropower production is a function of the total users' releases (reservoir release) and the reservoir level. The hydropower weight influences both: the total users' releases and the reservoir level. An additional experiment was executed based on the nDP-$Q_2$ experiment settings, where the highest priority (weight) was set to the hydropower. The experiment result showed that the reservoir was filled to the highest possible degree and subsequently made significant releases, increasing the hydropower production. The increased hydropower production directly influenced the users' releases and the reservoir levels. This indicates that although the main objective function is composed of a summation of weighted objectives, there is extra complexity because of the objectives interdependencies, which means that altering the objective weight influences other dependent objectives and the overall outcomes. The desired optimal reservoir operation can be designed by tuning the weights.

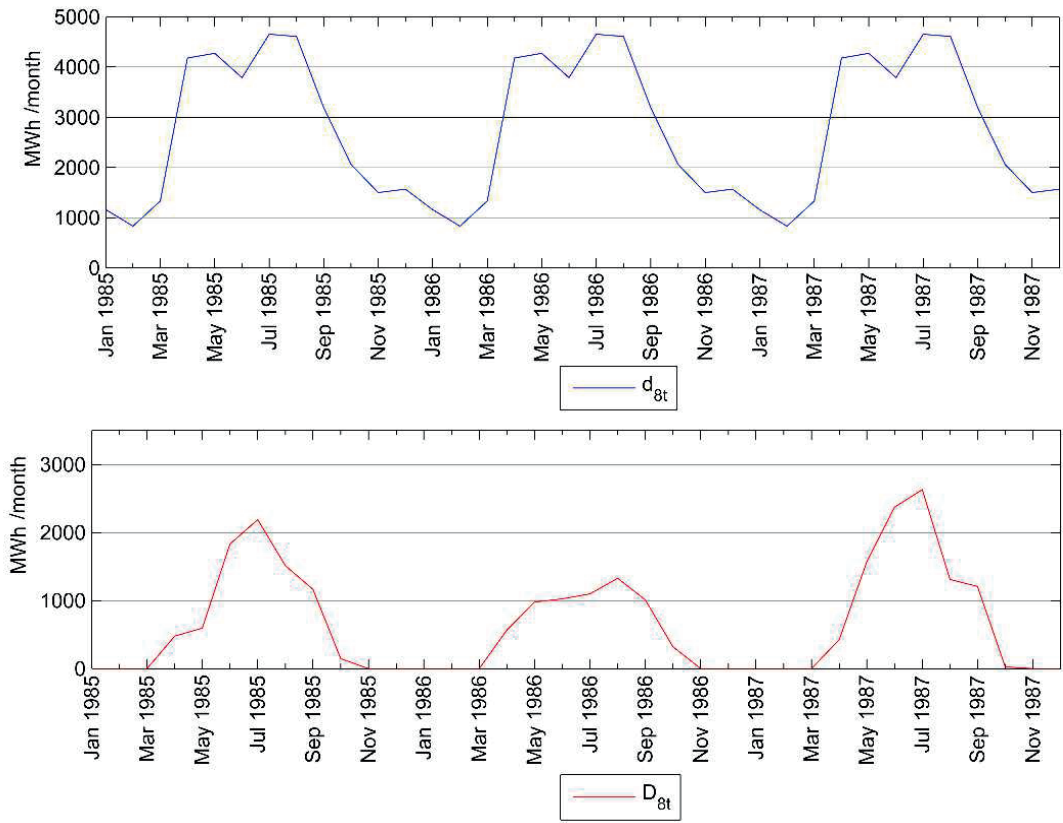

Figure 6.7 nDP-$Q_2$ hydropower demands and deficits 1985-1988

The nDP makes an exhaustive search over all potential (discretized) states and actions, carrying out the nested optimization at each transition, and thus requiring a substantial number of computations. For the presented setup, the experiment execution time was under five minutes on a standard desktop computer. As expected, the execution time of the nDP-$Q_1$ and the nDP-$Q_2$ experiment was longer because of the more complex calculation of the quadratic than the Simplex optimization algorithms. A minor technical issue during the nDP implementation was the memory requirements that in our case were solved by increasing the maximum memory allocation pool for the Java Virtual Machine.

## 6.2  Comparison of nDP with other DP algorithms

### 6.2.1  nDP compared with a classical DP algorithm

The presented case study has five water demand objectives. The classical DP approach would model these objectives as five different releases as shown in Figure 2.3. Since the DP algorithm is an exhaustive search, all possible states and actions need to be evaluated. Let us assume that the reservoir discretization is the same as in the previous experiments of 300 $10^3$ m$^3$, and that the five releases are discretized on 10 $10^3$ m$^3$. The DP algorithm makes a transition between all states. Let us consider only one transition from full to empty reservoir. In this simple calculation, the possible combinations of five releases are below 2300[5]. With 480 time steps and over 3000 transition at each time step, the total number of DP transitions is around $10^{23}$ - making this problem computationally unsolvable. For the applications considered, the usage of classical DP becomes impossible when there are several decision variables.

The Chapter 6 of the book Soncini-Sessa et al. (2007) describes decomposition method on multi reservoir system of Piave project, that decompose the problem into several sub-problems, decreasing the model complexity and computational demand. The difference is that Piave is a multireservoir complex system while Zletovica is a single reservoir system. The nested optimization algorithms are related to the decomposition approach described in this reference, but it is not a particular case of it. In the nested approach proposed in this thesis the 'secondary', decomposed problem is solved at each time step, rather than sequentially.

### 6.2.2   nDP compared with an aggregated water demand DP algorithm

To decrease the optimization problem dimension and to allow for DP implementation a simplified sub-optimal approach can be taken: the water demand objectives can be combined into a single aggregated objective so that at the first stage the standard DP optimization can be employed. The resulting releases are optimal since only the aggregated demand is optimized rather than the individual ones. At the second stage, these optimal reservoir releases are distributed between the different water demand users for each time step according to their importance. Such an approach can be called the aggregated water demand (AWD) DP algorithm. It would good to compare AWD DP with nDP but using the case considered in this paper it appeared not be a straightforward task.

Let us consider a case similar to the Knezevo reservoir case with 8 objectives, minimum level, maximum level, five water demand users (two towns, two agriculture and ecology), and hydropower. The aggregated water demand approach will combine the water demand objectives, into a single objective, and the rest of the problem will remain same, Equations (6.1) and (6.2):

$$w'_{3t} = w_{3t} + w_{4t} + w_{5t} + w_{6t} + w_{7t} \qquad (6.1)$$

$$d'_{3t} = d_{3t} + d_{4t} + d_{5t} + d_{6t} + d_{7t} \qquad (6.2)$$

The new optimization problem now has four objectives: 1) minimum level, 2) maximum level, 3) (aggregated) water demand user, and 4) hydropower, and it can be solved by AWD DP. In the first stage the DP algorithm is executed with the four objectives and the total optimal reservoir releases $R_{3t}'$ identified. In the second stage, these releases are distributed at each time step between the five users, forming allocations $r_{3t}'$, $r_{4t}'$, $r_{5t}'$, $r_{6t}'$ and $r_{7t}'$.

Let us assume that hydropower is a priority, and only the agriculture-related allocations $r_{4t}'$ and $r_{6t}'$ are used for producing hydropower. In this case, the AWD approach cannot be applied because it is difficult to create an objective function in DP that will represent the hydropower weight when all water demand users are aggregated (since hydropower is generated only by the two out of four allocations). The solution for this problem would be to divide the water demands users and their allocations in the two groups: (a) $r_{4t}'$, $r_{6t}'$, and (b) $r_{3t}^{i}$, $r_{5t}'$, $r_{7t}'$, and in this way to represent the hydropower in the objective function. This will introduce the second decision variable in DP that considerably increases the dimensionality of the problem. If we make this case more complicated and assume that the water demand user $d_{3t}'$ is very important compared to the other users, and needs to be represented as a separate decision variable in DP then the application of the classical AWD would become even more computationally demanding.

An alternative is to use the nested approach, as introduced in this research, which makes it easy to introduce a large number of separate objective functions (e.g. hydropower or $d_{3t}$ prioritization as discussed above).

This consideration permits to see the conceptual difficulty in the accurate comparison of nDP and AWD on the case study Zletovica: the main problem is that the objective functions formulations for these two algorithms are different. Indeed, the AWD DP combines all water demands into a single objective in Equations (6.1) and (6.2) while the other objectives are the same, so the water demand objectives (and consequently the hydropower) are represented differently. The first stage of AWD DP uses the following objective function:

$$g_t(s_t, s_{t+1}, a_t) = \sum_{i=1}^{4} c_i \cdot w_{it}' \cdot D_{it}'^{2} \qquad (6.3)$$

The OF used by nDP is shown in Equation (4.4). The difference is obvious in the user demand objectives, and:

$$w_{3t}' \cdot (d_{3t}' - R_{3t}')^2 \neq \sum_{i=3}^{7} w_{it} \cdot (d_{it} - r_{it})^2 \qquad (6.4)$$

where

$$R_{3t}' = \sum_{i=3}^{7} r_{ir} \qquad (6.5)$$

Even if we consider the results of the AWD DP second stage and combine them into one objective function with the eight sub-objectives, the problem setting will not be the same as in nDP. Another difference between the nDP and the AWD DP is that the second stage of the AWD DP does not change the reservoir releases, because they are already calculated in the first stage, while in the nDP at each transition the releases are calculated taking into consideration all objectives. This interplay between individual objective functions brings additional complexity into the problem of comparing these algorithms.

## 6.3    Experiments with nDP using weekly data

The nDP weekly optimization was executed on the entire 55-year period, and it was aimed to satisfy the case study requirements explained in Section 4.4.3. These requirements demand for minimization of ecology flow deficits $D_7$ and two towns' deficits $D_3$ and $D_5$ at the expense of irrigation deficits $D_4$ and $D_6$. Because of these requirements, the ecology $w_7$ and water supply weights $w_3$, $w_5$, need to be significantly higher than the irrigation users' $w_4$ and $w_6$. The question is how to assign the appropriate weights to all objectives, including minimum and maximum critical reservoir levels and hydropower. A possible solution is to perform several experiments with different sets of weights ($w_1^i...w_8^i$) and afterwards choose the most appropriate weight set. More about this MO problem solution is explained in Section 6.5. The selected set of weights in this experiment is shown in Table 6-2.

Table 6-2 nDP objectives weights

| Experiments | $W_1$ | $W_2$ | $W_3$ | $W_4$ | $W_5$ | $W_6$ | $W_7$ | $W_8$ |
|---|---|---|---|---|---|---|---|---|
| **nDP-L$_3$ and nDP-Q$_3$** | 20,000 | 20,000 | 200 | 1 | 200 | 1 | 300 | 0.01 |

The minimum ecology flow $w_7$ has the highest weight out of all water demand users, followed by water supply users' $w_3$, $w_5$ and irrigation demands $w_4$ and $w_6$. Considering that the deviation ranges of minimum and maximum levels $D_1$ and $D_2$ are between 0.3-6.5 m, while the water users range is between 0-350 that is afterwards being squared, the minimum and maximum weights need to balance this vast difference in ranges. On the other hand, the minimum $w_1$ and maximum $w_2$ weights should not shrink the available reservoir volume and significantly influence the optimal reservoir operation. The hydropower weights $w_7$ are set extremely low because of the previously elaborated reasons in Section 4.4.3. The reservoir discretization, minimum, and maximum levels are the same as in nDP monthly settings. The overall optimal reservoir operation results are shown in Table 6-3.

Table 6-3 nDP-L$_3$ and nDP-Q$_3$ weekly optimization results over 55 years (1951-2005)

|  | $D_1$ | $D_2$ | $D_3$ | $D_4$ | $D_5$ | $D_6$ | $D_7$ | $D_8$ |
|---|---|---|---|---|---|---|---|---|
| nDP-L$_3$ | 128.69 | 37.64 | 1 | 75,222 | 5,566 | 104,613 | 0 | 250,628 |
| nDP-Q$_3$ | 50.40 | 35.97 | 2,801 | 94,322 | 2,757 | 86,689 | 1,823 | 240,924 |

The deviations from the minimum and maximum levels, users' and hydropower deficits, and overspills are shown in Figure 6.8. The total overspills are 146.78 $10^3$ m$^3$.

Figure 6.8 shows that overall there is enough water to satisfy most of the objectives. Only the periods between 1987-1995, and 1999-2002 are dry and objectives are suffering. The same dry periods are occurring in monthly data as shown in Figure 6.2. Figure 6.9 shows the average weekly annual inflows, demand, and deficit in the Knezevo reservoir using the optimization results from nDP-$L_3$.

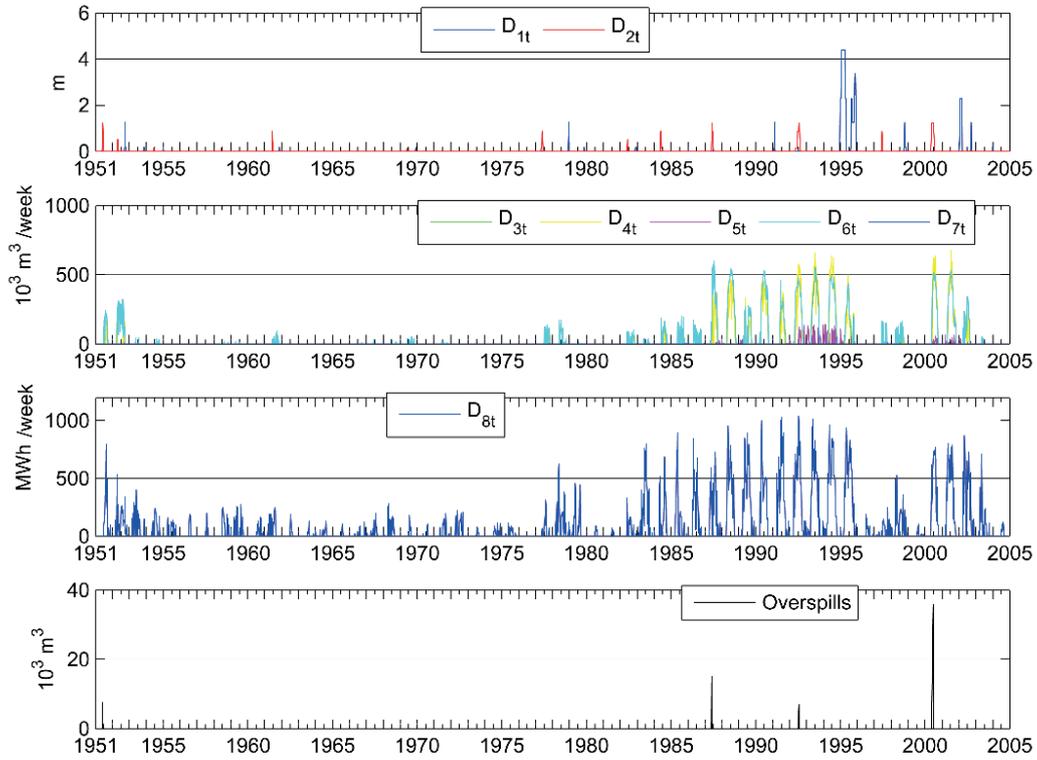

Figure 6.8 nDP-$L_3$ minimum and maximum levels deviations, users' and hydropower deficits, and overspills over 55 years (1951-2005)

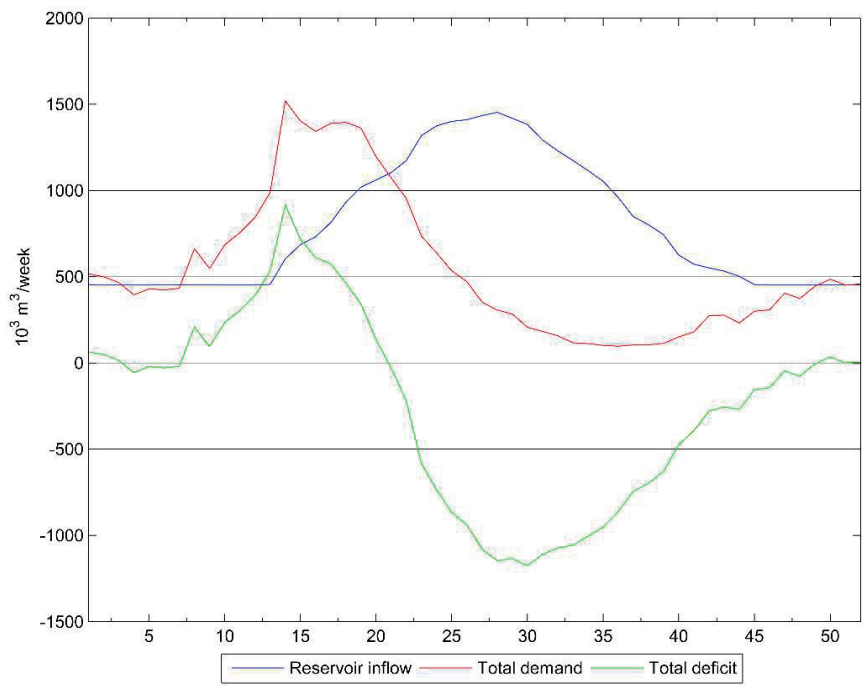

Figure 6.9 Averaged annual inflows, demands and deficits in the Knezevo reservoir, results from nDP-$L_3$

Two additional optimization trials were executed with increasing the weights $w_1$ and $w_2$ to 2.000.000 each (see Table 6-4), and they are named nDP-$L_4$ and nDP-$Q_4$.

Table 6-4 nDP-$L_4$ and nDP-$Q_4$ weights

| Experiments | $w_1$ | $w_2$ | $w_3$ | $w_4$ | $w_5$ | $w_6$ | $w_7$ | $w_8$ |
|---|---|---|---|---|---|---|---|---|
| **nDP-$L_4$ and nDP-$Q_4$** | 2,000,000 | 2,000,000 | 200 | 1 | 200 | 1 | 300 | 0.01 |

Because of the imposed high weights for minimum and maximum levels, the volume of overspills is zero. The nDP-$L_4$ and nDP-$Q_4$ optimization results are shown in Table 6-5 and in Figure 6.10.

Table 6-5 nDP-$L_4$ and nDP-$Q_4$ weekly results with increased minimum and maximum critical levels weights

| | $D_1$ | $D_2$ | $D_3$ | $D_4$ | $D_5$ | $D_6$ | $D_7$ | $D_8$ |
|---|---|---|---|---|---|---|---|---|
| **nDP-$L_4$** | 4.4 | 4.5 | 1 | 77,157 | 6,687 | 108,023 | 0 | 248,826 |
| **nDP-$Q_4$** | 4.4 | 2.3 | 3,150 | 96,825 | 3,084 | 88,745 | 2,056 | 240,776 |

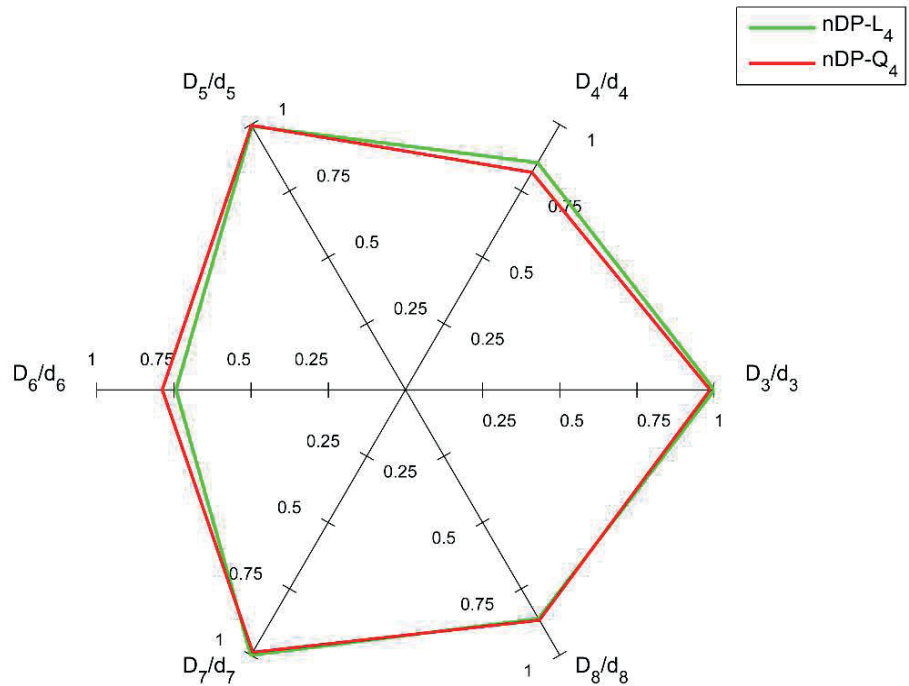

Figure 6.10 nDP-$L_4$ and nDP-$Q_4$ overall objectives satisfaction

The minimum and maximum deviations are very low. The consequences of $w_1$ and $w_2$ increase is the decrease on the available reservoir storage volume that is visible from Table 6-5 where the $D_4$, $D_5$, and $D_6$ are slightly increased.

The nDP optimization results show that ecological flow and the two water supply demands are entirely satisfied in both algorithms (over 99.5%). The irrigation demands are less satisfied with 86% of upper agriculture zone and 75% lower agriculture zone in nDP-$L_3$, and 82% upper agriculture zone and 79% lower agriculture zone in nDP-$Q_4$.

## 6.4 Experiments with nSDP and nRL using weekly data and their comparison to nDP

Optimal reservoir policies are derived using nSDP and nRL algorithms. The available 55 years' weekly data is separated in two parts 1) training and 2) testing. The data from 1951-1994 (2340 time steps) is used for training and 1994-2004 (520 time steps) for testing. The nSDP training data consists of reservoir inflow $q_t$ and tributary inflow $q_t^{Tr}$ in the previously mentioned period. The nRL training data consist of reservoir $q_t$ and tributary inflow $q_t^{Tr}$, and the two other flows $q_{1t}$ and $q_{2t}$ used for hydropower calculation. The nSDP and nRL data for minimum and maximum levels, water supply, irrigation demands, ecological flow, and hydropower are set to the 2005 weekly data presented in Section 4.4, and they are the same in training and testing period. The reservoir operation volume is discretized in 73 equal levels (300 $10^3$ m$^3$). The minimum level was set at 1021.5 m amsl and the maximum level at 1060 m amsl. The weights applied in these experiments are shown in Table 6-6. At the beginning, the nested optimization algorithm (linear or quadratic) and the number of clusters (in our case five) are selected in both nSDP and nRL.

Table 6-6 nDP-L$_5$, nDP-Q$_5$, nSDP-L$_5$, nSDP-Q$_5$, nRL-L$_5$ and nRL-Q$_5$ experiments weights

| Experiments | W$_1$ | W$_2$ | W$_3$ | W$_4$ | W$_5$ | W$_6$ | W$_7$ | W$_8$ |
|---|---|---|---|---|---|---|---|---|
| nDP-L$_5$, nDP-Q$_5$ | 2,000,000 | 2,000,000 | 200 | 1 | 200 | 1 | 300 | 0.01 |

The result of both nSDP and nRL is one-year optimal reservoir policy. The optimal reservoir policy has this structure <time step, storage volume, reservoir inflow, tributary inflow, next reservoir storage> ($<t, s_t, q_t, q_t^{Tr}, s_{t+1}>$) where the output is the next reservoir storage $s_{t+1}$. With the next reservoir storage $s_{t+1}$, it is possible to calculate all other decision variables like a total release $(r_t)$, releases for each user $(r_3, r_4, r_5, r_6, r_7)$, minimum and maximum level $(h_t, h_{t+1})$, and hydropower production $(p_t)$. The optimal reservoir policy internally in nSDP and nRL looks like <72, 2, 4, 2, 68>, and are explained in the Table 6-7 which presents one policy sample.

Table 6-7 Optimal reservoir policy sample

| Optimal reservoir policy (internal) | Optimal reservoir policy | Explanation |
|---|---|---|
| <72,2,4,2,68> | <21,300; 7-14 Jan; 2008.62; 775.98; 20,400> | |
| 72 | 21,300 | The starting reservoir storage discretization level 72 and its corresponding volume 21,300 $10^3$ m$^3$. |
| 2 | 2 | The time-step (second week of January from 7-14). |
| 4 | 2008.62 | The reservoir inflow cluster number 4 and its corresponding value 2008.62 $10^3$ m$^3$ / week. |
| 2 | 775.98 | The tributary inflow cluster number 2 and its corresponding value 775.98 $10^3$ m$^3$ / week. (in case of nSDP both reservoir and tributary cluster numbers are the same) |
| 68 | 20,400 | The next optimal reservoir (transition) level 68 and its corresponding volume 20,400 $10^3$ m$^3$. |

In addition, the nDP produces the optimal reservoir operation using the testing data. This nDP results are "the optimal operation" meaning that because nDP is deterministic optimization and knows the future (all variables in advance) can calculate the optimal reservoir operation. The nSDP and nRL on the other hand are trained on training data and have not seen the testing data (future). The nDP results are used as a benchmark for the nSDP and nRL policies. The nDP also calculates the starting state or in our case starting reservoir volume $s_t$ when t=1. The starting reservoir volume is assigned as a starting state in the nSDP and nRL testing period. Beginning from the starting state, until the end, the nSDP reads the reservoir and tributary inflow, calculates in which cluster they belong, and from the nSDP optimal reservoir policy gets the next optimal state. When next state is known, all other variables are calculated and the nSDP goes to the next state. For clarification, the nSDP in training calculate with cluster centres of $q_t$ and $q_t^{Tr}$, while in testing with the real values of $q_t$ and $q_t^{Tr}$. The nRL works in the same manner with the difference that additional data from $q_{1t}$ and $q_{2t}$ are included for the hydropower calculation. The $q_{1t}$ and $q_{2t}$ are not a part of the nRL state variables or optimal reservoir policy. The nDP optimal reservoir operation is a target for both the nSDP and nRL policies. The closer the policies derived by nSDP and nRL are to nDP, the better they are.

The algorithms are additionally labelled to denote the deficit formulations used in the nested optimization used. For example, nDP-$L_5$ stands for nDP using the linear deficits formulation, and nDP-$Q_5$ stands for nDP using the quadratic deficits formulation. The nRL parameters at the beginning are set at: $\alpha_0$=0.8, $\gamma$=0.5 and $\varepsilon$=0.8. The parameter $\alpha$ is set to decrease linearly with the number of episodes. This is achieved by using $\alpha$ function, which linearly decreases from the starting $\alpha_0$ to $\alpha_{min}$, as described with the following equation:

$$\alpha_n = \alpha_0 - \frac{\alpha_0 - \alpha_{min}}{M} \cdot n \qquad (6.6)$$

where $\alpha_{min}$ is the minimum value set at 0.001, M is the maximum numbers of episodes, and $n$ is the episode number. The value of $\alpha$ is updated at each 500 episodes and the maximum number of episodes is set to $M$=400,000. There were various approached tested for decreasing $\varepsilon$, and the one used in the experiments is shown in Table 6-8.

Table 6-8 $\varepsilon$ values in the experiments.

| n | $\varepsilon$ |
|---|---|
| 0 | 0.8 |
| *M/4* (100.000 episodes) | 0.4 |
| *M/2* (200.000 episodes) | 0.2 |
| *3M/4* (300.000 episodes) | 0.05 |
| *3.5M/4* (350.000 episodes) | 0.0001 |

Table 6-8 shows how $\varepsilon$ is decreased making less exploration and more exploitation actions with the increasing number of episodes, and insuring convergence to the optimal solutions.

The nRL agent learns and after a number of episodes (after 10.000) episodes, the policy is tested. The nRL-$L_5$ agent learns the optimal policy and gets closer to nDP-$L_5$ ORO as the number of episodes' increases, as shown in Figure 6.11. However, after a large number of episodes, the learning deteriorates. The nRL-$L_5$ agent optimal reservoir policy is poor at 30,000 episodes as shown in Figure 6.11. From 50,000 to 80,000 episodes, the nRL-$L_5$ policy improves, and between 80,000 and 160,000, the policy is the closest to the nDP-$L_5$ ORO. After 250,000 episodes the policy slightly deteriorates. The reason for this is overtraining, which is a known issue when using machine learning algorithms.

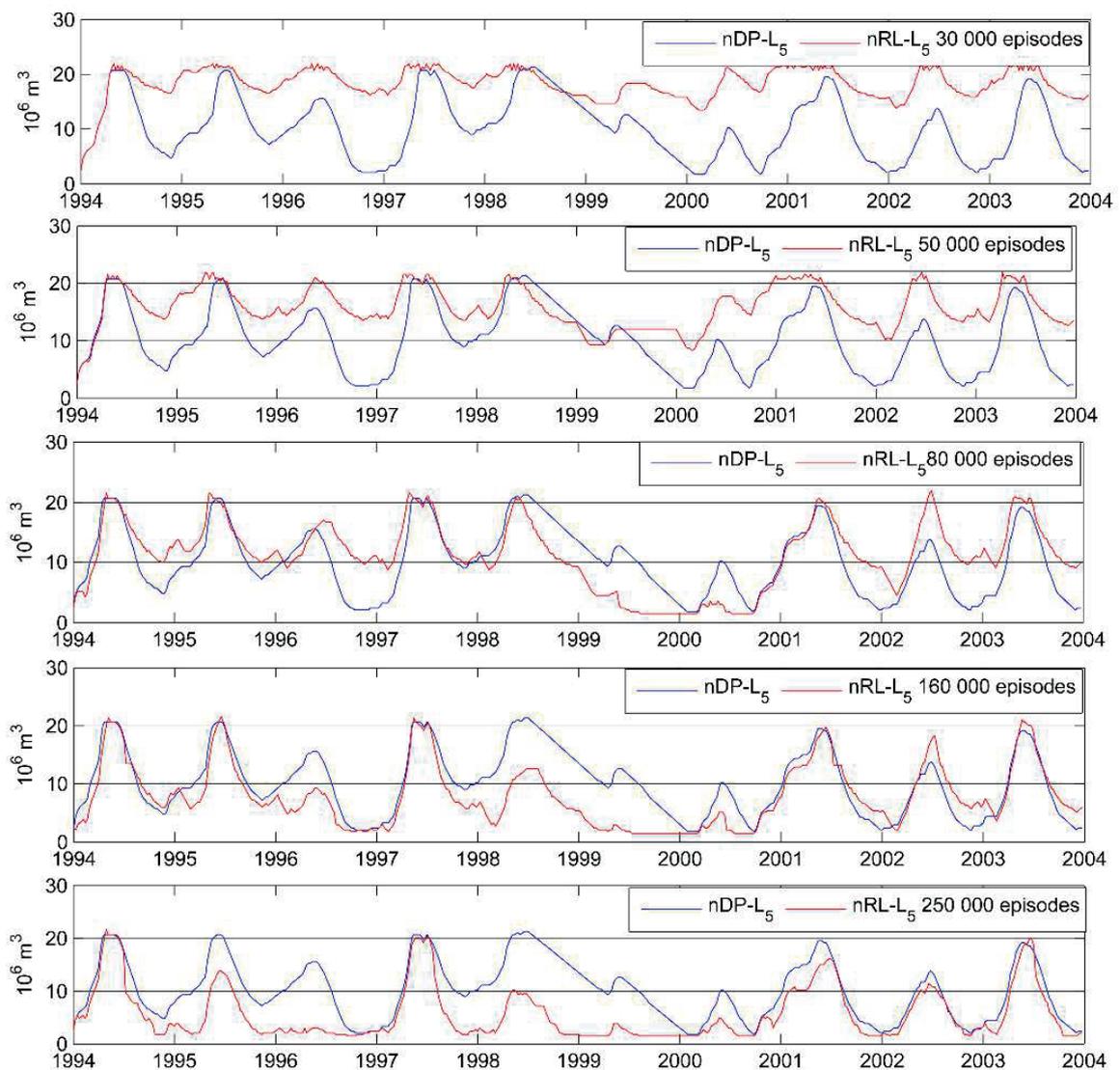

Figure 6.11 nRL-$L_5$ agent learning with increasing the number of episodes: nDP-$L_5$ target reservoir storage (blue) and nRL-$L_5$ obtained reservoir storage (red) (testing period)

Another benchmark for the nRL optimal reservoir policy is to sum up the absolute difference between nDP optimal reservoir volume and nRL optimal reservoir volume at each time-step in testing period. The formula used is presented below:

$$S_n = \sum_{t=1}^{T} \left| s_t^{nDP} - s_t^{nRL} \right| \qquad (6.7)$$

where $s_t^{nDP}$ is the nDP-$L_5$ reservoir volume at time $t$ of and $s_t^{nRL}$ is the reservoir volume of nRL-$L_5$ at time $t$.

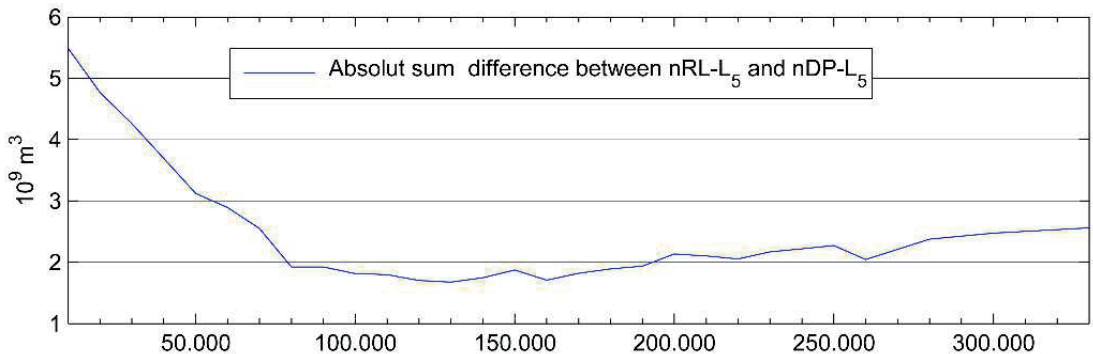

Figure 6.12 Sum of absolute difference between nRL-$L_5$ and nDP-$L_5$ measured as difference in reservoir volumes in the period 1994-2004 as a function of the nRL number of episodes.

Results presented in Figure 6.12 coincide with those from Figure 6.11 regarding nRL-$L_5$ learning with different number of episodes. The absolute difference between the nRL and nDP optimal reservoir volumes can be used as the stopping criterion. Obviously, the nRL-$L_5$ optimal reservoir policy performs best between 80,000 - 160,000 episodes of training. Afterwards, the policy has somewhat deteriorated, although it is still relatively good.

The nDP-$L_5$, nSDP-$L_5$ and nRL-$L_5$ reservoir optimal volume in the testing period is shown in Figure 6.13. The optimal reservoir volume curves of nDP-$Q_5$, nSDP-$Q_5$ and nRL-$Q_5$ are similar to these and are not presented here.

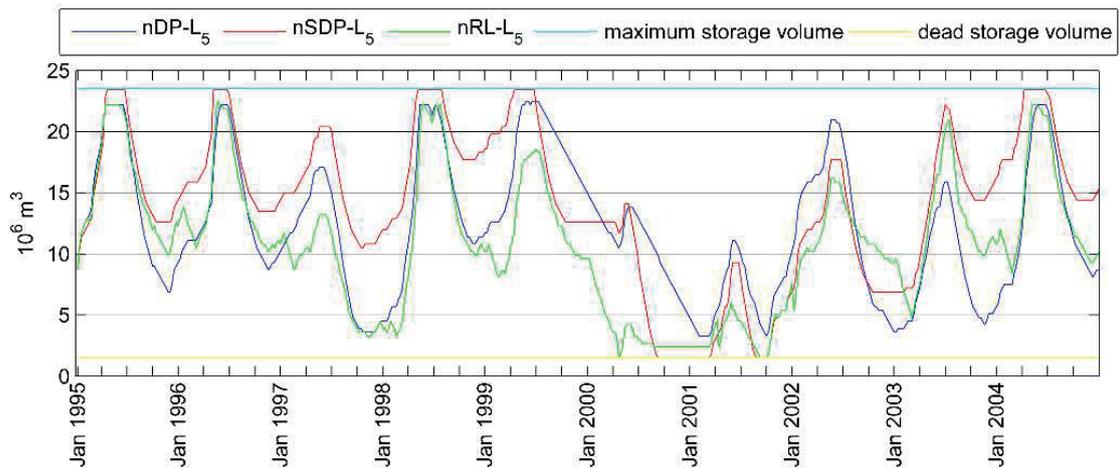

Figure 6.13 The nDP-$L_5$, nSDP-$L_5$ and nRL-$L_5$ optimal reservoir volume in testing period 1994-2004.

Figure 6.14 presents the reservoir and the tributary inflow in the testing period and Figure 6.15 presents the optimal reservoir level curves for nDP-$L_5$, nSDP-$L_5$ and nRL-$L_5$.

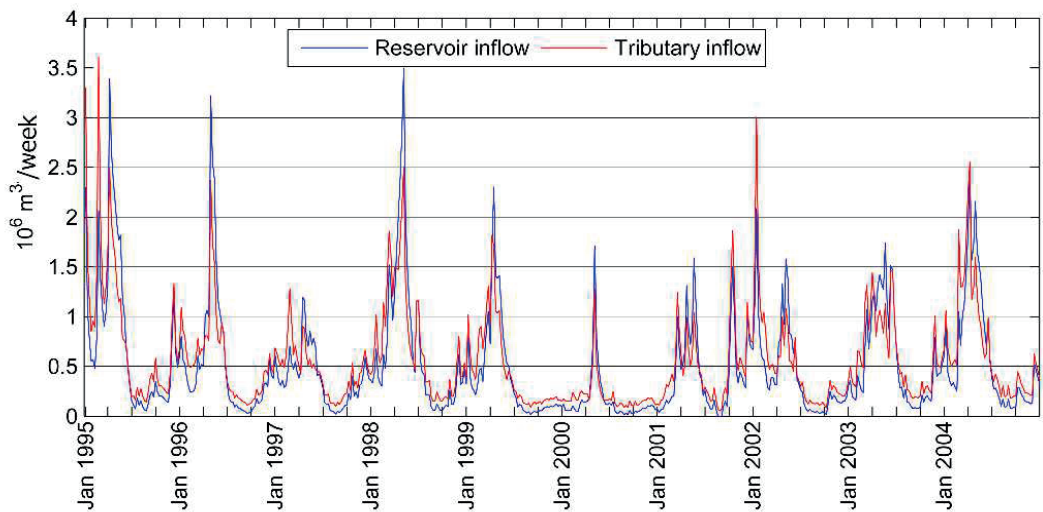

Figure 6.14 Reservoir and tributary inflow in testing period 1994-2004

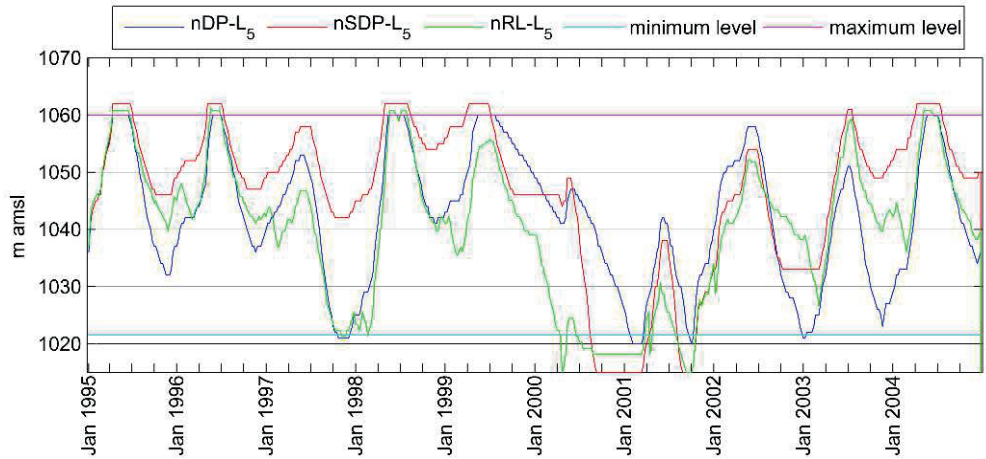

Figure 6.15 nDP-$L_5$, nSDP-$L_5$, and nRL-$L_5$ optimal reservoir level and minimum and maximum levels in testing period 1994-2004

The period 1999-2001 is very dry, because of low reservoir and tributary inflow, as shown in Figure 6.13, Figure 6.14, and Figure 6.15. The reservoir level (and reservoir volume) in all algorithms goes to the lowest possible level. During this period the nDP-$L_5$ has a very limited minimum level violation, while nSDP-$L_5$ and nRL-$L_5$ have significant minimum level violation as shown in Figure 6.16. The reason for this behaviour is that nDP is a deterministic optimal algorithm that has perfect knowledge (forecast) of the future. When nDP is applied to the 1994-2004 testing data, it means that an exhaustive search is performed to find the optimal deterministic solution. On the other side, both nSDP and nRL train/learn on training data. Both nSDP and nRL have not seen the testing data. The training data 1951-1994 is the combination of wet, average, and dry years that is fed to nSDP and nRL algorithms. Based on these data, both algorithms derive the optimal policy, which is a universal one-year policy (for wet, average, and dry years).

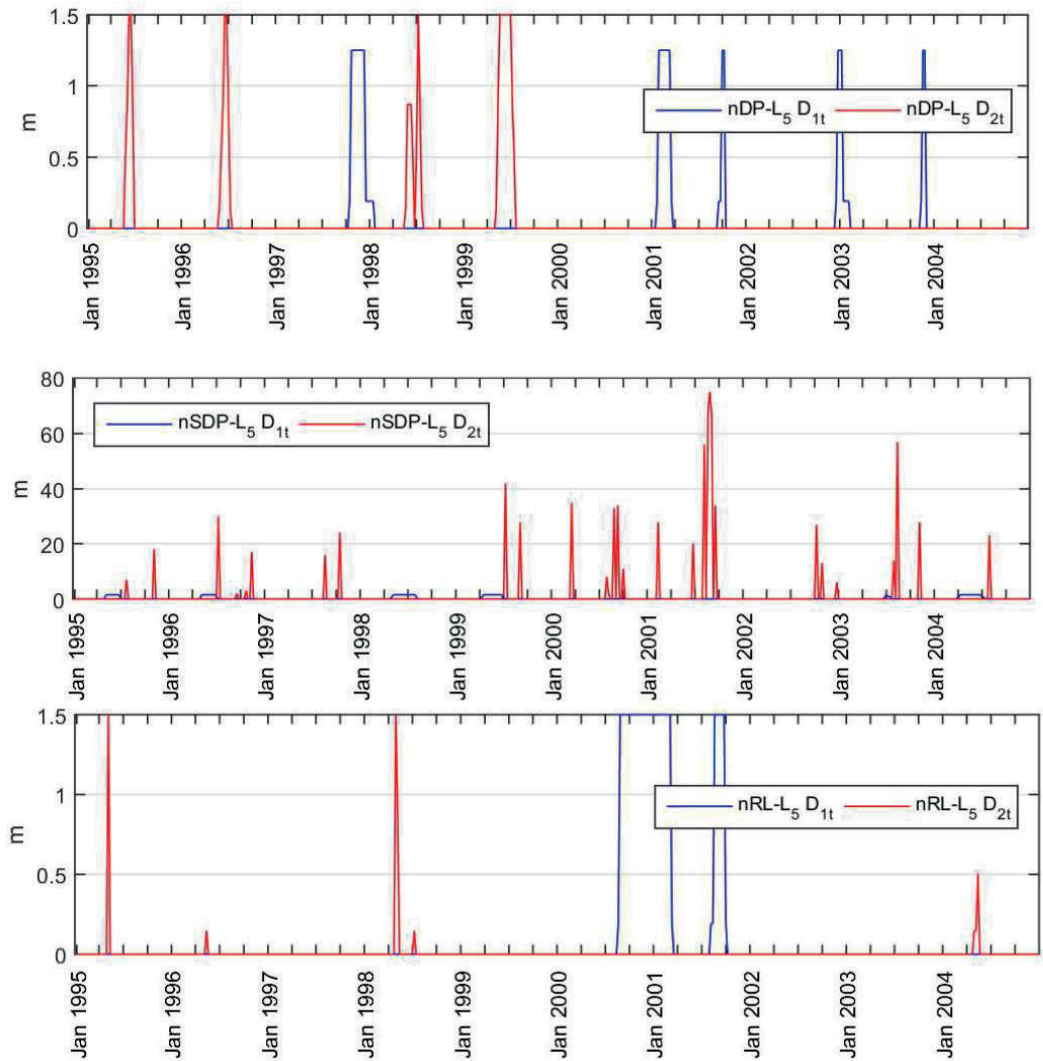

Figure 6.16 nDP-$L_5$, nSDP-$L_5$ and nRL-$L_5$ minimum and maximum level deviations ($D_{1t}$ and $D_{2t}$) in testing period 1994-2004

The nSDP and nRL performs poorly in dry years because their policy is a yearly universal policy, or it is not a reservoir policy for a dry or a wet year. A possible solution for this problem can be to use a committee of nRL agents that will train on, for example wet, average, and dry years and provide the possibility to switch the policy according to a particular year. A similar approach can be applied to nSDP.

The water users' demands in the testing period are shown in Figure 6.17. The deficits are different depending on the algorithm used as shown in Figure 6.18. The nDP-$L_5$ is the target and the deficits are relatively low. The nRL-$L_5$ has smaller deficits than nSDP-$L_5$. The pattern in all three algorithms is the same. The minimum ecological flow $d_7$ and water supply $d_3$ are almost completely satisfied. The water supply $d_5$ suffers in nSDP-$L_5$ and nRL-$L_5$.

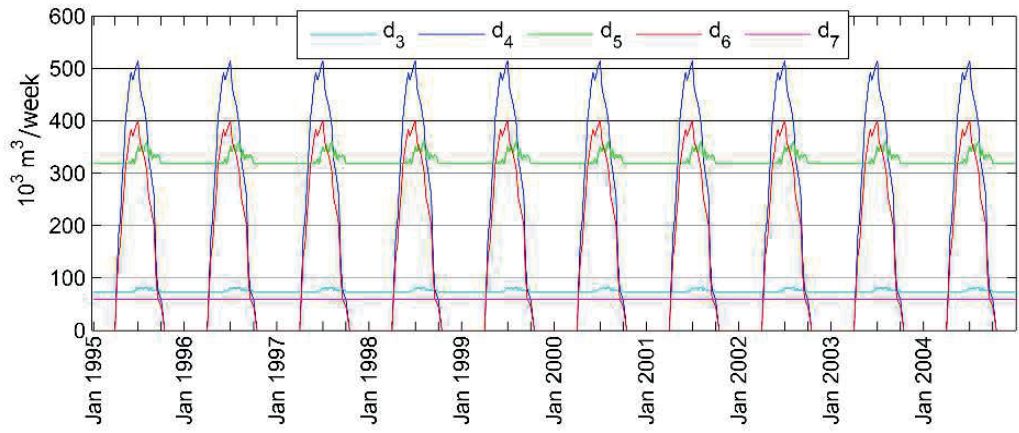

Figure 6.17 Water users' demands in testing period 1994-2004

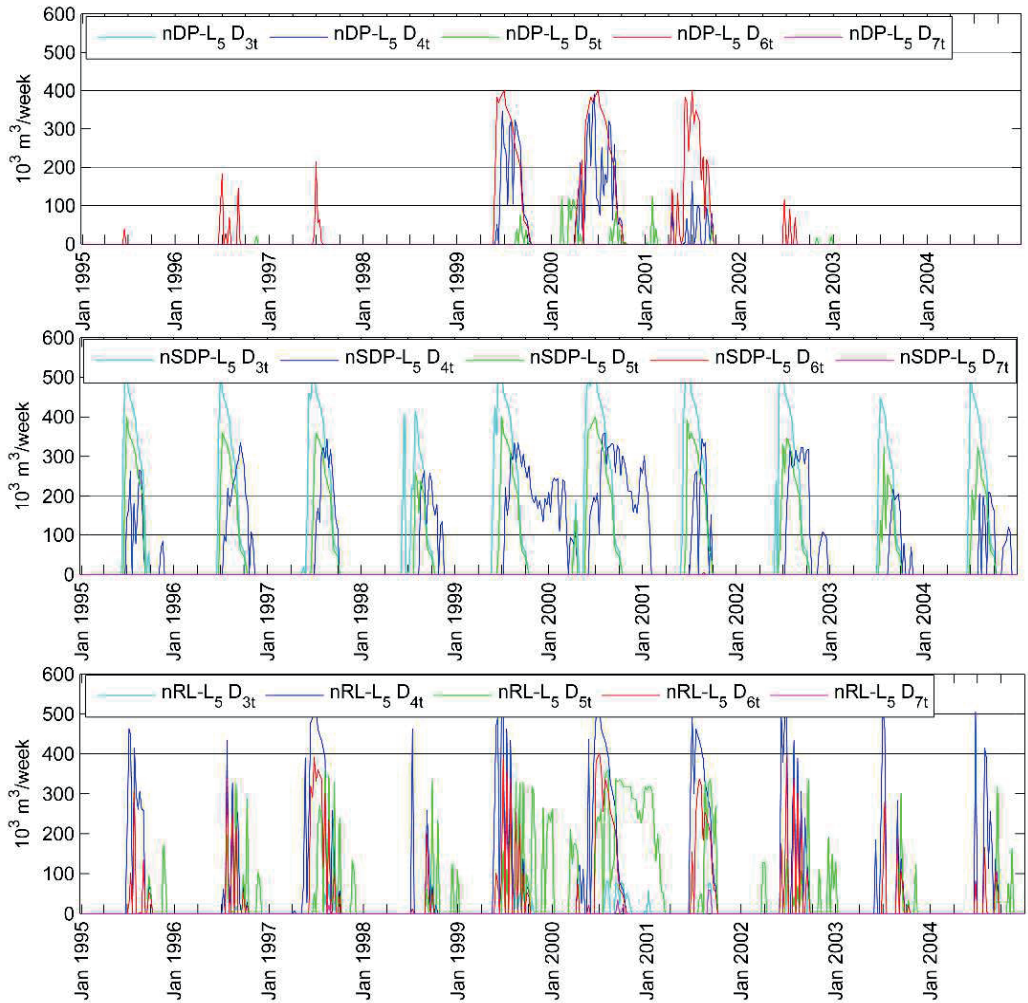

Figure 6.18 nDP-$L_5$, nSDP-$L_5$ and nRL-$L_5$ users' deficits

The hydropower deficits and the water users' deficits are correlated, as shown in Figure 6.18 and Figure 6.19 . During the period 1999-2002, the hydropower deficits, increases due to lower reservoir releases.

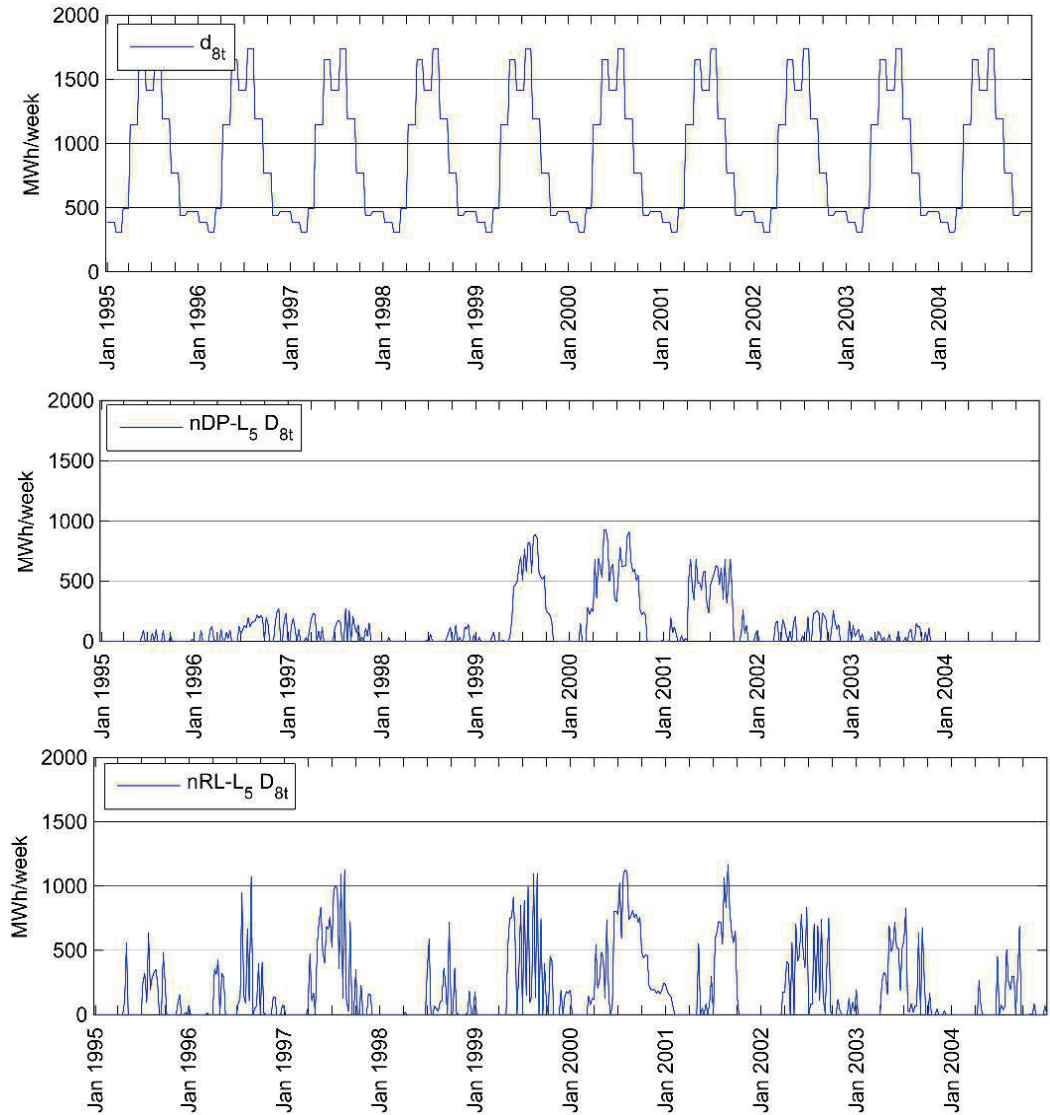

Figure 6.19 Hydropower demand and deficit 1995-2004

The nDP-$L_5$, nSDP-$L_5$, nRL-$L_5$, nDP-$Q_5$, nSDP-$Q_5$, and nRL-$Q_5$ optimization results and comparison of the sum of minimum level ($D_1$) and maximum level ($D_2$) deviations, sum of users' deficit ($D_{3-7}$), and sum of hydropower deficit ($D_8$), in testing period 1994-2004 are shown in Table 6-9 and Figure 6.20.

Table 6-9 nDP-$L_5$, nSDP-$L_5$, nRL-$L_5$, nDP-$Q_5$, nSDP-$Q_5$ and nRL-$Q_5$ optimization results

| Experiments | $D_1$ | $D_2$ | $D_3$ | $D_4$ | $D_5$ | $D_6$ | $D_7$ | $D_8$ |
|---|---|---|---|---|---|---|---|---|
| **nDP-$L_5$** | 1.33 | 0.84 | 0 | 12,143 | 769 | 15,517 | 0 | 43,623 |
| **nSDP-$L_5$** | 250.48 | 70.48 | 827 | 41,557 | 36,609 | 40,569 | 5 | 0 |
| **nRL-$L_5$** | 187.68 | 0.15 | 1,339 | 21,345 | 21,350 | 26,631 | 141 | 83,190 |
| **nDP-$Q_5$** | 1.33 | 0.98 | 338 | 14,655 | 351 | 13,365 | 236 | 43,071 |
| **nSDP-$Q_5$** | 395.52 | 55.58 | 11,569 | 45,174 | 17,687 | 37,302 | 8,361 | 0 |
| **nRL-$Q_5$** | 144.97 | 1.16 | 4,466 | 21,412 | 21,209 | 24,924 | 778 | 79,265 |

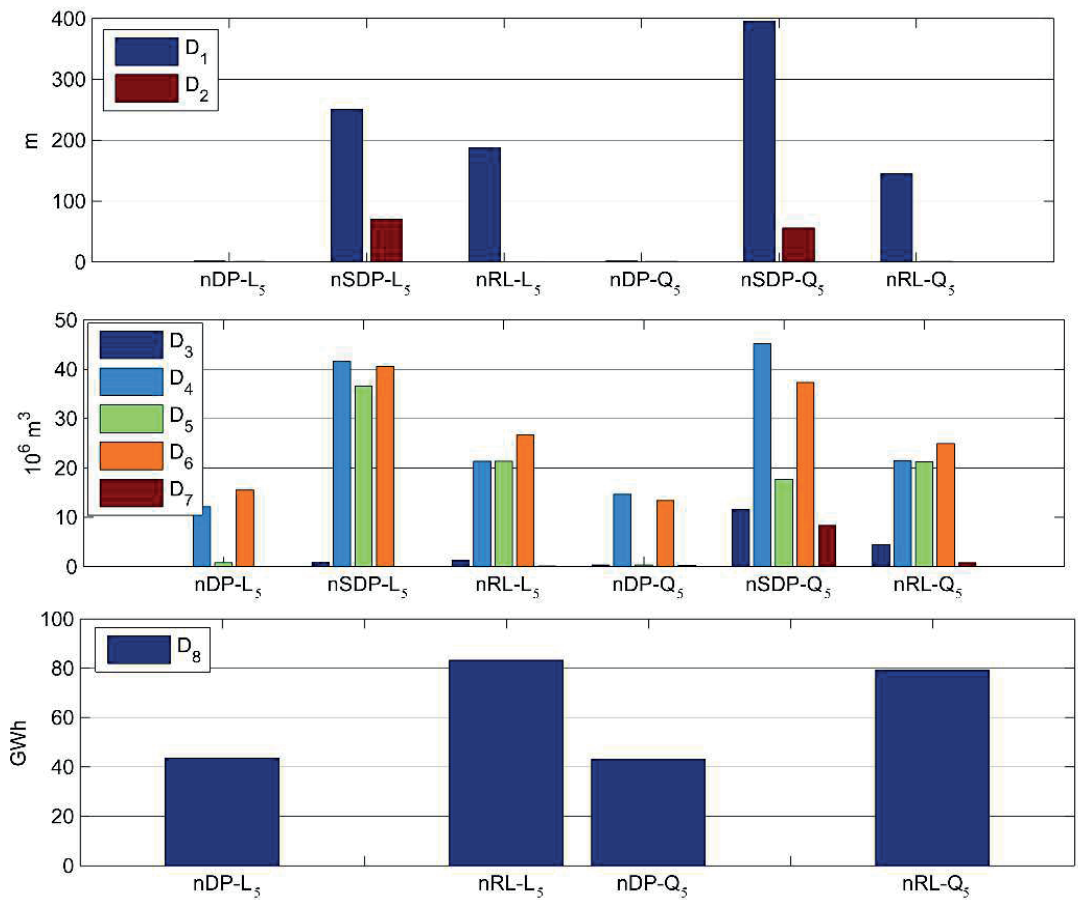

Figure 6.20 nDP-$L_5$, nDP-$Q_5$, nSDP-$L_5$, nSDP-$Q_5$, nRL-$L_5$ and nRL-$Q_5$ comparison of the sum of minimum level ($D_1$) and maximum level ($D_2$) deviations, sum of users' deficit ($D_{3-7}$) and sum of hydropower deficit ($D_8$) in the testing period 1994-2004

The nDP-$L_5$ and nDP-$Q_5$ are the targets, and the nRL-$L_5$ and nRL-$Q_5$ are better than nSDP-$L_5$, nSDP-$Q_5$ optimization results in all objectives, shown in Figure 6.20. The conclusion is that nRL produces better reservoir policies than nSDP. Additionally, nRL is more capable and can include additional variables (like $q_{1t}$ and $q_{2t}$) and model more complex systems than nSDP. The nRL disadvantages are however in its complex settings and implementation.

## 6.5 Identification of optimal solutions in multi-objective setting using MOnDP, MOnSDP and MOnRL

Until now we have been considering the single objective function being the (aggregated) weighted sum of several (eight) functions. As it has been presented earlier, the second approach to deal with the MO problem is the MO optimization by a sequence of single-objective optimization searches (MOSS). MOSS can be used when the use of a SOO algorithm is preferred and when there is no need to generate a large Pareto set which is computationally expensive. Depending on the problem at hand, a user can decide how many solutions are needed and generate a corresponding number of the weight vectors used to weigh the objectives in MOSS. A limited set of solutions is generated, but they can be treated as a useful approximation of the Pareto optimal set, albeit small. Having this in mind, in this manuscript we will be referring to his set as the Pareto set as well.

In the context of the case study considered MOSS approach is implemented as follows: there are several sets of weights $[w_1^j...w_8^j]$ used to form the weighted sum of the aggregated single objective function to identify the desired optimal reservoir policy. In our case, the desired optimal reservoir policy is the one that decreases the ecological, water supply and irrigation deficits, and minimum and maximum deviations, and hydropower deficits. In multi-objective problems, there is not a single optimal solution, instead there is always a set of possible solutions.

There are three MO algorithms that need to be analysed and tested: MOnDP, MOnSDP and MOnRL. The MOnDP is significantly faster than MOnSDP and MOnRL, and it is used to scan the possible weights sets space. The possible weight space is scanned with 10 weights sets, as shown in Table 6-10. Out of these 10 weights sets, three are selected and applied to MOnSDP and MOnRL. The MOnDP was executed on the testing data with the setting from the previous subchapter. One may argue that the MOnDP should be utilizing the training data like MOnSDP and MOnRL. However, for scientific reasons and proper comparison between all algorithms the approach of using only the test data was taken. The optimization results are compared not on the basis of all eight objective functions separately, but using the following criteria:

1) The irrigation deficits *($D_4+D_6$)* compared the water supply deficit *($D_3+D_5$)*.

2) The minimum and maximum deviations *($D_1+D_2$)* compared to a total deficit *($D_3+D_4+D_5+D_6+D_7$)* and

3) The hydropower deficit *($D_8$)* compared with total deficit *($D_3+D_4+D_5+D_6+D_7$)*.

Table 6-10 MOnDP sets of weights

| Weights set | w₁ | w₂ | w₃ | w₄ | w₅ | w₆ | w₇ | w₈ |
|---|---|---|---|---|---|---|---|---|
| 1 | 2,000,000 | 2,000,000 | 500 | 1 | 500 | 1.5 | 700 | 0.01 |
| 2 | 200 | 200 | 500 | 1 | 500 | 1.5 | 700 | 0.01 |
| 3 | 2,000,000 | 2,000,000 | 200 | 2 | 300 | 1 | 400 | 0.1 |
| 4 | 200 | 200 | 200 | 2 | 300 | 1 | 400 | 0.1 |
| 5 | 200 | 200 | 5 | 2 | 5 | 1 | 6 | 0.1 |
| 6 | 200 | 200 | 5 | 2 | 5 | 1 | 6 | 100 |
| 7 | 200 | 200,000 | 5 | 2 | 5 | 1 | 6 | 100 |
| 8 | 200,000 | 200 | 5 | 80 | 5 | 80 | 6 | 0.1 |
| 9 | 200,000 | 200 | 5 | 150 | 5 | 80 | 6 | 0.1 |
| 10 | 2,000,000 | 2,000,000 | 200 | 1 | 200 | 1 | 300 | 0.01 |

Table 6-11 MOnDP results

| Ex. | D₁ | D₂ | D₃ | D₄ | D₅ | D₆ | D₇ | D₈ |
|---|---|---|---|---|---|---|---|---|
| 1_L | 1.34 | 2.1 | 0 | 19,406 | 691 | 8,343 | 0 | 43,129 |
| 1_Q | 1.34 | 1.01 | 306 | 16,566 | 316 | 11,626 | 234 | 43,057 |
| 2_L | 106.78 | 15.7 | 0 | 18,735 | 236 | 6,423 | 0 | 40,366 |
| 2_Q | 106.78 | 14.97 | 117 | 15,286 | 133 | 10,472 | 86 | 40,429 |
| 3_L | 1.34 | 1.16 | 761 | 10,958 | 20 | 16,663 | 0 | 42,266 |
| 3_Q | 1.34 | 1.01 | 459 | 12,896 | 314 | 15,242 | 219 | 42,508 |
| 4_L | 123.56 | 17.58 | 236 | 9,219 | 0 | 16,226 | 0 | 39,133 |
| 4_Q | 123.56 | 18.81 | 233 | 10,903 | 160 | 14,550 | 117 | 38,597 |
| 5_L | 125.66 | 16.21 | 0 | 4,618 | 2,723 | 15,949 | 0 | 36,786 |
| 5_Q | 113.69 | 14.47 | 3,440 | 4,282 | 3,493 | 8,296 | 2,851 | 31,956 |
| 6_L | 205.2 | 24.02 | 0 | 2,819 | 8,018 | 14,137 | 0 | 30,864 |
| 6_Q | 155.13 | 25.25 | 4,272 | 4,465 | 5,017 | 8,846 | 3,548 | 27,692 |
| 7_L | 205.2 | 14.69 | 0 | 2,819 | 8,018 | 14135 | 0 | 30,922 |
| 7_Q | 172.99 | 13.67 | 4,278 | 4,488 | 5,007 | 8916 | 3,555 | 27,749 |
| 8_L | 1.82 | 12.01 | 4,095 | 0 | 18,788 | 0 | 42 | 31,294 |
| 8_Q | 0.57 | 12.15 | 7,031 | 342 | 9,982 | 388 | 5,669 | 33,430 |
| 9_L | 1.82 | 12.01 | 4,095 | 0 | 18,788 | 0 | 42 | 31,294 |
| 9_Q | 0.57 | 11.29 | 7,070 | 216 | 10,073 | 406 | 5,685 | 31,031 |
| 10_L | 1.33 | 0.84 | 0.00 | 12,142 | 769 | 15,516 | 0.00 | 43,622 |
| 10_Q | 1.33 | 0.98 | 338 | 14,655 | 351 | 13,365 | 235 | 43,071 |

The weight vectors were chosen to reflect various scenarios. The 1, 2, 3, 4 and 10 scenarios have a set of weights in which the water supply weights are significantly higher than agriculture. The other way around where agriculture has significantly higher weights than water supply is captured in scenarios 8 and 9. Scenarios 1, 3 and 10 have significant minimum and maximum critical level weights compared to the other scenarios. Scenarios 6 and 7 have high hydropower weights compared to the other scenarios.

The MOnDP optimization results are presented in Table 6-11, where the number denotes the scenario (particular set of weights) r and the indices L or Q denote linear or squared formulation of deficits.

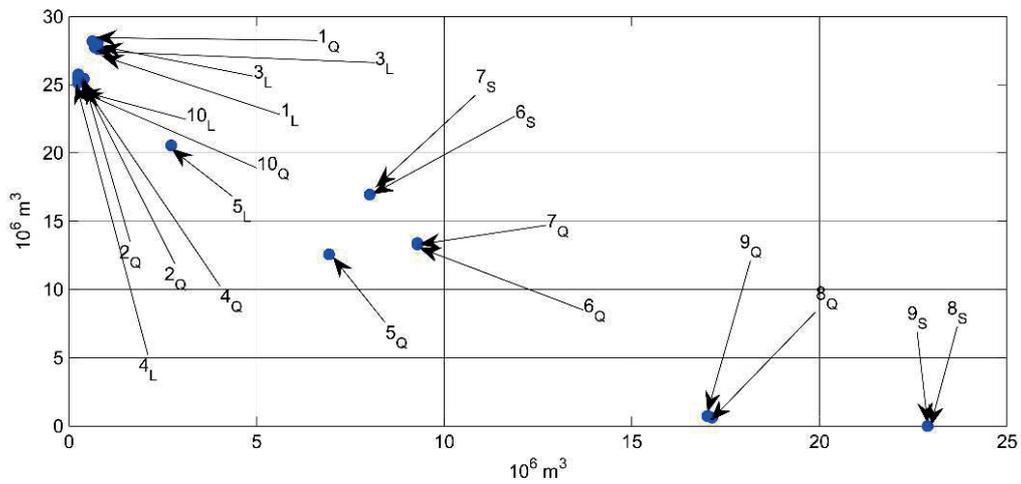

Figure 6.21 MOnDP. Water supply deficits (D3+D5, x-axis) compared with irrigation deficits (D4+D6, y-axis) using nDP with 10 sets of weights in testing period 1994-2004

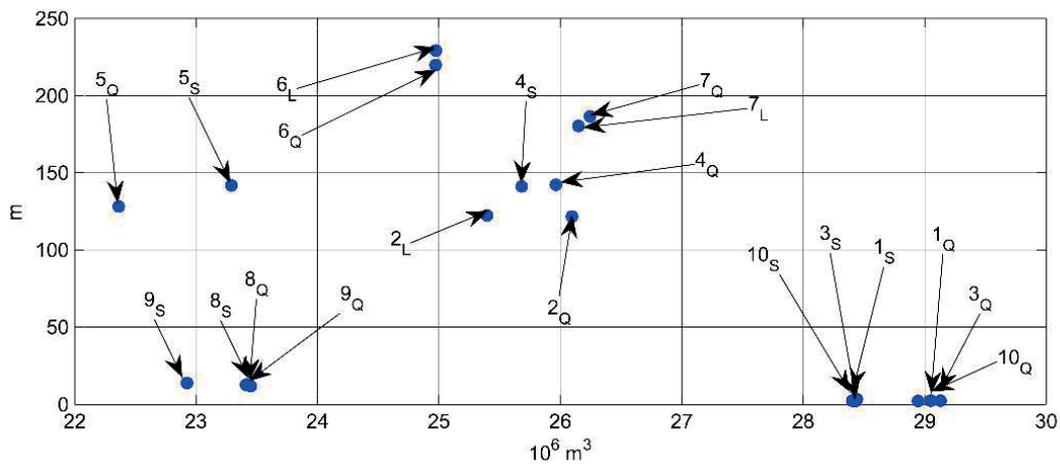

Figure 6.22 MOnDP. Total users' deficits (D3+D4+D5+D6+D7, x-axis) compared with minimum and maximum deviations (D1+D2, y-axis) using nDP with 10 sets of weights in testing period 1994-2004

The MOnDP optimization results comparing water supply and irrigation deficits are presented graphically in Figure 6.21. The blue dots are MOnDP solutions and are marked with the arrows that show the experiment. As expected, depending on the weighting scheme shown in Table 6-11 results are varying from almost completely satisfied water supply in experiments 1, 2, 3 and 4 (L and Q), to a balance between water supply and irrigation in experiments 5, 6 and 7, and almost completely satisfied irrigation in experiments 8 and 9. The results are completely reflecting the change of different weight sets.

Figure 6.22 shows the total users' deficits compared with minimum and maximum deviations combined for the 10 MOnDP scenarios. Again, there are varieties of possible solutions depending on the weight sets. Experiments 1 and 3 with the highest minimum and maximum weights have the smallest deviations. On the other side experiments 2, 4, 5, 6 and 7 have a relatively significant minimum and maximum deviations. The experiments 8 and 9 are in between the previously mentioned two groups.

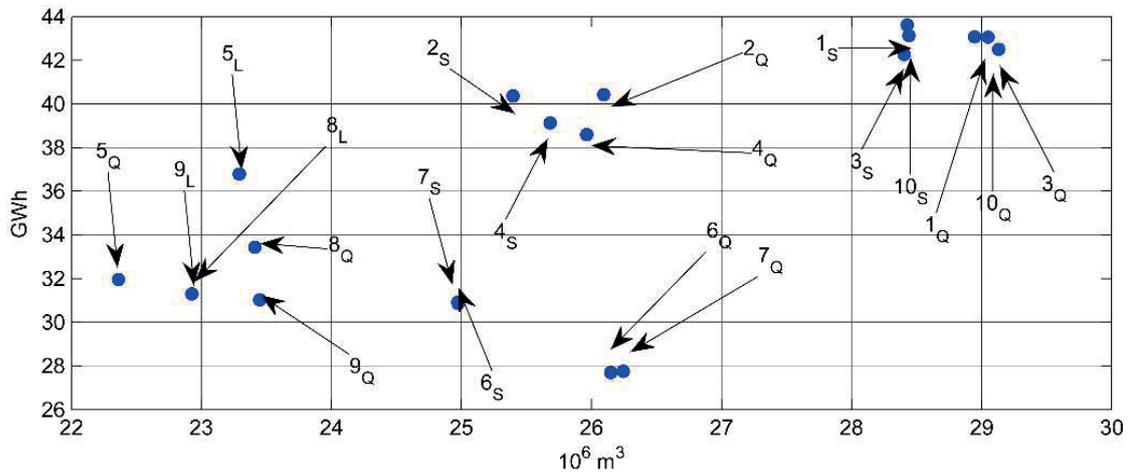

Figure 6.23 MOnDP. Total deficits (D3+D4+D5+D6+D7, x-axis) compared with hydropower deficits D8

Hydropower deficits are lowest in scenarios 6 and 7 and are highest in scenarios 1, 3 and 10, as shown in Figure 6.23, corresponding to the highest weight sets as shown in Table 6-10. Hydropower weights in scenarios 1, 10 are ten times smaller than scenario 3 while hydropower deficits are similar, shown in Table 6-10 and Figure 6.23. A similar argument applies to the scenarios 7 and 8, where the hydropower weights difference is 1000 times, shown in Table 6-10. Scenarios 5, 8 and 9 also have a lower hydropower deficit, although the hydropower weights are low. This is because the hydropower production is a complex function in the presented case study.

Experiments 2 and 4 are slightly better than 1 and 3 with respect to the water supply and irrigation deficits, as shown in Figure 6.21. When compared with the following Figure 6.22, experiments 2 and 4 have significant minimum and maximum deviation, while experiments 1 and 3 have very low minimum and maximum deviations. These experiments clearly demonstrate that increasing of minimum and maximum deviations provides additional storage that is used to satisfy water users. There is an obvious trade-off between minimum and maximum levels and (total) water users' deficits. Another clearly visible trade-off is between water supply and irrigation deficits as shown in Figure 6.21.

Several weight sets from the list shown in Table 6-10 need to be selected and applied to MOnSDP and MOnRL. These weight sets needs to cover the space of possible solutions. Because some of the weights sets in Table 6-10 are similar, further investigation is not needed. The analysis of scenario 4 weights sets shows that these are similar to scenarios 2 and 5, as shown in Table 6-10. Scenario 4 has low water supply deficits as show in Figure 6.21, and it is balanced compared to other scenarios in minimum and maximum deviations, hydropower deficits and total deficits as shown in Figure 6.22 and Figure 6.23. Scenario 7 results are similar (with weights also) with scenario 6. It has the lowest hydropower deficits as shown in Figure 6.23, and it is balanced compared to other scenarios in total deficits, water supply, and irrigation as shown in Figure 6.21 and Figure 6.22. Scenario 10 is similar to 1, 3 (weights also), with low water supply deficit as shown in Figure 6.21, low minimum and maximum deviations as shown in Figure 6.22, and high hydropower and total deficit as show in Figure 6.23. Scenarios 8 and 9 have a significant water supply deficit compared to irrigation, and are not suitable for further investigation. Based on this analysis, scenarios 4, 7 and 10 are selected for further experiments as shown in Table 6-12. Previous experiments have shown that the solutions found by linear or quadratic deficit formulations are relatively similar as shown in Figure 6.21, Figure 6.22, and Figure 6.23. Therefore, in the following experiments only the linear deficit optimization formulations are used.

Table 6-12 MOnSDP-L and MOnRL-L weights sets

| Weights set | $w_1$ | $w_2$ | $w_3$ | $w_4$ | $w_5$ | $w_6$ | $w_7$ | $w_8$ |
|---|---|---|---|---|---|---|---|---|
| 4 | 200 | 200 | 200 | 2 | 300 | 1 | 400 | 0.1 |
| 7 | 200 | 200,000 | 5 | 2 | 5 | 1 | 6 | 100 |
| 10 | 2,000,000 | 2,000,000 | 200 | 1 | 200 | 1 | 300 | 0.01 |

The resulting table with nine experiments is shown in Table 6-13, below. For comparison purposes, the table includes results of two additional experiments (last two rows), in which MOnSDP and MOnRL were executed with scenario 4 weights sets, but with four intervals for inflows classification, instead of five.

Table 6-13 MOnDP, MOnSDP and MOnRL results

| Experiment. | $D_1$ | $D_2$ | $D_3$ | $D_4$ | $D_5$ | $D_6$ | $D_7$ | $D_8$ |
|---|---|---|---|---|---|---|---|---|
| **MOnDP-$L_4$** | 123.56 | 17.58 | 236 | 9,219 | 0 | 16,226 | 0 | 39,133 |
| **MOnSDP-$L_4$** | 250.48 | 70.45 | 13,793 | 41,451 | 23,643 | 40,675 | 5 | 0 |
| **MOnRL-$L_4$** | 266.51 | 5.28 | 7,986 | 22,967 | 14,752 | 27,518 | 284 | 87,743 |
| **MOnDP-$L_7$** | 205.20 | 14.69 | 0 | 2,819 | 8,018 | 14,135 | 0 | 30,922 |
| **MOnSDP-$L_7$** | 450.98 | 42.17 | 827 | 41,451 | 36,609 | 40,675 | 5 | 0 |
| **MOnRL-$L_7$** | 378.34 | 4.56 | 1,766 | 22,147 | 25,067 | 27,773 | 269 | 86,870 |
| **MOnDP-$L_{10}$** | 1.33 | 0.84 | 0 | 12,142 | 769 | 15,516 | 0 | 43,622 |
| **MOnSDP-$L_{10}$** | 250.48 | 70.45 | 827 | 41,557 | 36,609 | 40,569 | 5 | 0 |
| **MOnRL-$L_{10}$** | 187.68 | 0.15 | 1,339 | 21,345 | 21,350 | 26,631 | 141 | 83,190 |
| **MOnSDP-$L_{10}$-4c** | 286.14 | 53.09 | 827 | 41,557 | 36,609 | 40,569 | 5 | 0 |
| **MOnRL-$L_{10}$-4c** | 244.8 | 0.86 | 1,601 | 22,887 | 25,835 | 27,659 | 110 | 85,468 |

The results allowing for selecting the best policy are presented on Figure 6.24, Figure 6.25 and Figure 6.26:

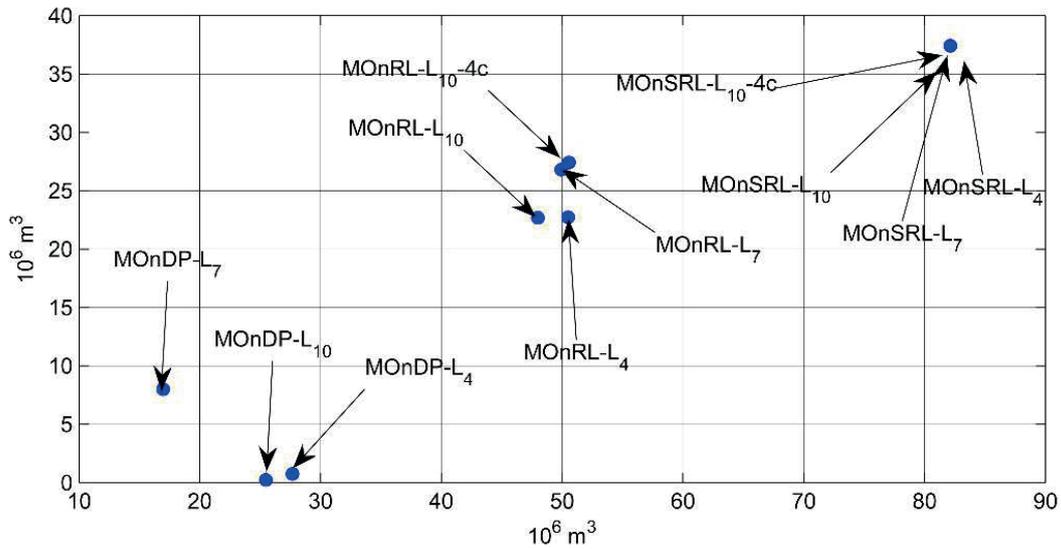

Figure 6.24 MOnDP-L, MOnSDP-L and MOnRL-L. Comparison between irrigation deficit (D4+D6, x-axis) and water supply (D3+D5, y-axis)

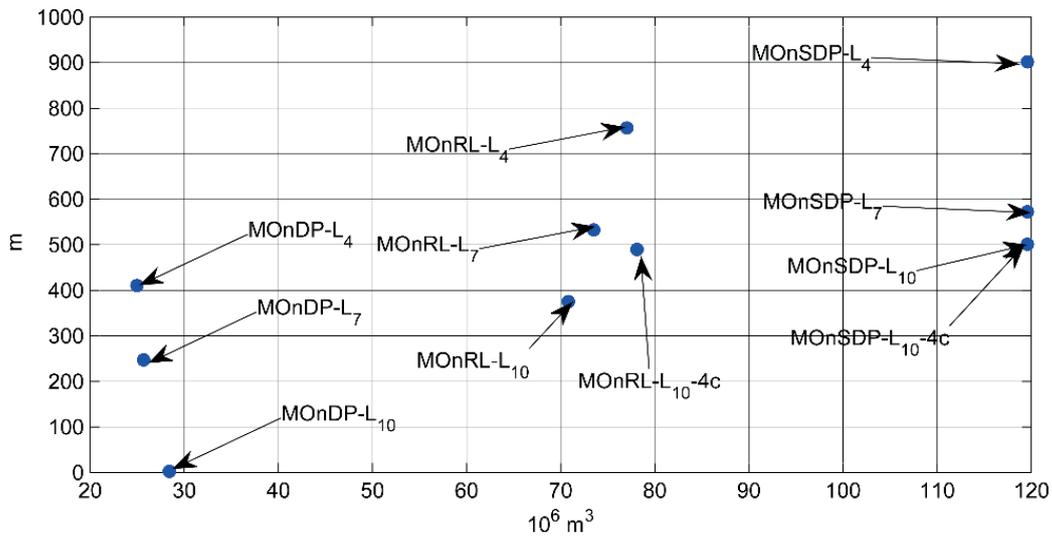

Figure 6.25 nDP-LM, nSDP-LM and nRL-LM. Comparison between total deficit (D3+D4+D5+D6+D7, x-axis) and minimum and maximum deviations (D1+D2, y-axis)

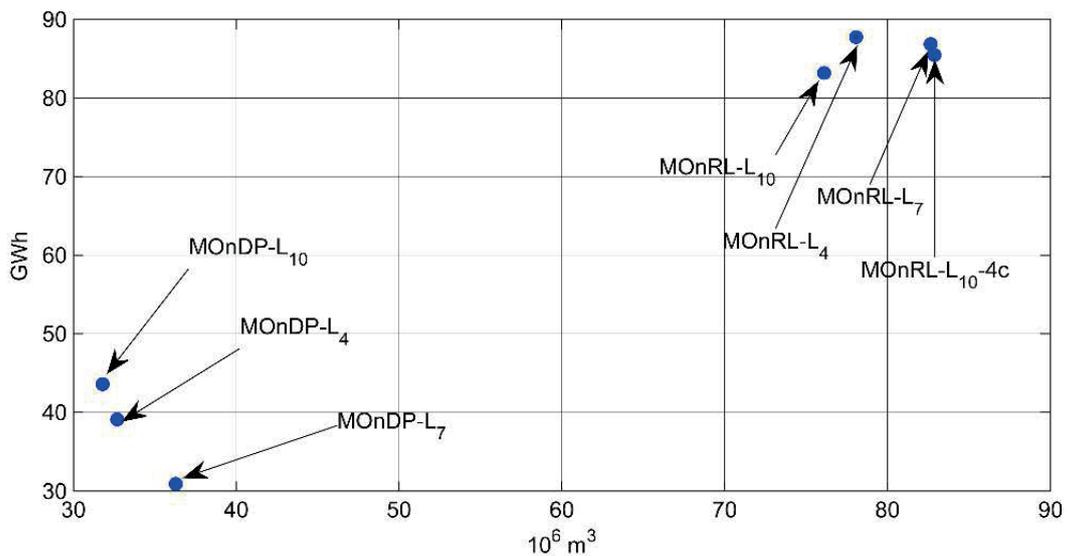

Figure 6.26 MOnDP-L and MOnRL-L. Comparison between the hydropower deficit (D8) and total deficit (D3+D4+D5+D6+D7)

The first conclusion to be drawn is that there are three groups of solutions corresponding to MOnDP, MOnSDP and MOnRL. The MOnDP group of solutions are considerably better than MOnSDP and MOnRL, which is expected because MOnDP calculates the optimal reservoir operation, as shown in Figure 6.24, Figure 6.25 and Figure 6.26. The MOnDP can be considered as an optimal target for MOnSDP and MOnRL as described in the previous subchapters.

The MOnRL group performs considerably better than MOnSDP comparing irrigation and water supply deficits, as shown in Figure 6.24. All four MOnRL results are better than any other MOnSDP. Each of the MOnRL optimization results is better than corresponding MOnSDP considering the total deficit and minimum and maximum deviations, as shown in Figure 6.25. The hydropower deficit is not calculated in MOnSDP (neither in nSDP previously) so the MOnRL and MOnSDP cannot be compared considering hydropower.

These results are expected having in mind all previous results and explanations in this chapter. One of the best solutions considering irrigation deficits, water supply deficits, minimum deviations, maximum deviations, and total deficits is produced by MOnRL-$L_{10}$, as shown in Figure 6.24, Figure 6.25 and Figure 6.26 and Table 6-13.

The results from the experiments with four inflow intervals (last two rows in Table 6-13) presented in Table 6-13, Figure 6.24, Figure 6.25 and Figure 6.26 demonstrate that they are quite similar to those with five intervals.

## 6.6     Conclusions

This chapter started with nDP optimization on 55 years monthly and weekly data (1951-2005). The nDP monthly experiments demonstrated that each objective weight influences the optimal reservoir operation. The nSDP and nRL derived the ORO policies on 45 years training data (1951-1994) and were compared to the nDP ORO policy on 10 years testing data (1995-2004). The results showed that nRL results produced better ORO policy than nSDP. The MOnDP experiments were performed on 10 different weight sets. Because of its speed, the MOnDP was used to scan the space of potential appropriate set of weights. Three sets of weights were selected and applied to MOnSDP, and MOnRL. Three groups of solutions were produced corresponding to MOnDP, MOnSDP and MOnRL. The group of solutions found by MOnRL was better than those found by MOnSDP. This confirmed the previous statement than nRL is better and more capable reservoir optimization algorithm than nSDP.

.

# Chapter 7     Cloud decision support platform

*"He who controls the past controls the future. He who controls the present controls the past."*

*George Orwell*

The previously developed reservoir optimization algorithms have been embedded in a prototype cloud decision support platform. The cloud platform has a web service for water resources optimization that provides access to previously developed algorithms, and web interface for presentation of the results. Additionally, the cloud platform has three other web services: (1) for data infrastructure, (2) for support of water resources modelling, (3) for user management. The presented cloud system has several main advantages: it is available all the time, it is accessible from everywhere, it creates real time multi-user collaboration platform, the programming languages code and components are interoperable and designed to work in a distributed computer environment, it is flexible for adding additional components and services and it is scalable depending on the workload. The cloud platform was deployed and tested in a distributed computer environment running on two virtual machines ($VM_1$ and $VM_2$). The platform was successfully tested in the Zletovica case study with concurrent multi-users access.

___________________________________________________________________________

## 7.1     Background

The presented cloud decision support platform is a continuation of the research for development of a web application for water resources based on open source software, published in the article (Delipetrev et al. 2014). The material in this chapter is based on the material presented in that article, with an upgrade of the existing web application with additional user management service and distribution of the web application to two virtual machines ($VM_1$ and $VM_2$) from which $VM_1$ runs on the Amazon web services platform (AWS) and $VM_2$ runs of server in University Goce Delcev in Shtip, Republic of Macedonia. The presented cloud-platform is based on three pillars: cloud computing, Service Oriented Architecture (SOA) and web Geographic Information Systems (GIS).

The first pillar is cloud computing (Armbrust et al. 2010), which is a technology through which everything from computing power, applications, infrastructure, systems, models are delivered to users wherever and whenever they are needed, without end-user knowledge of the physical location of system components. The idea is similar to the electric grid, where users utilize electrical power without the need for understanding system components (Carr 2008). Cloud computing makes computation and information always available. Of course, the drawback of cloud computing is network dependence i.e. if there is no network connection the systems will not work.

There is still a debate about the unique definition of cloud computing, or the system characteristics that need to be satisfied so the system can be referred to as "cloud." A recent definition of cloud computing by the National Institute of Standards and Technology NIST (Mell and Grance 2011) outlines the essential characteristics, service and deployment models. The essential characteristics of cloud computing are on-demand self-service, broad network access, resource pooling, rapid elasticity and measured services. There are three different cloud service models: software as a service (SaaS), Platform as a Service (PaaS) and Infrastructure as a service (IaaS) and four different deployment models Private, Public, Community, and Hybrid. Cloud computing is currently the main trend in information and communication technologies (ICT) and all major companies are heavily investing in new cloud infrastructure and services.

The second pillar is SOA (Erl 2005), which emerged as a preferred design methodology in internet applications and cloud computing. SOA is a set of design principles for development, integration, and implementation of a software system by using interoperable services provided by heterogeneous components. These principles and concepts can also be applied to integrate heterogeneous applications and platforms into web based environment. Instead of defining an application programming interface (API), SOA defines interface as a group of protocols and functionalities usually using eXtended Markup Language (XML). Interfaces and services designed by SOA often communicate asynchronously or "on demand" between each other with XML messages.

The third pillar is web GIS, which is in fact shifting of GIS to the internet. Opening of the Global Positioning System (GPS) signal to the public and embedding GPS receivers in many devices (mobile phones, tablets, computers, etc.) created an enormous new market and increased interest for geographic information. Rapid development of ICT and GIS provoked the creation of new geospatial web standards and application. The OGC (Open Geospatial Consortium) geospatial standards (Web Mapping Service – WMS; Web Feature Service – Transactional WFS-T) are increasingly being used for realising interoperable web applications that utilise spatial data and they have also been used in this research.

By making use of technologies from the three pillars mentioned above a cloud platform was developed and successively tested with data from the Zletovica river basin and with multi-users collaborating in real time. The previously developed nested optimization algorithms nDP, nSDP, and nRL have been embedded in the platform as components of envisaged cloud-based decision support system. The trial confirmed that the platform is functional and can be utilized as a prototype, demonstrating the possibilities of cloud and distributed computing. The cloud system URL is www.delipetrov.com/his/. The cloud application has a help page that contains web links with video presentations of the system components, guides about how to use the web services, etc.

## 7.2    Architecture and implementation

The cloud decision support platform architecture is presented in Figure 7.1. The arrows represent the data communication links between the web services. The data communication is asynchronous or "on demand." The system has four web services for:

1. Data Infrastructure (DI).

2. Support for Water Resources Modelling (WRM).

3. Water Resources Optimization (WRO).

4. User management.

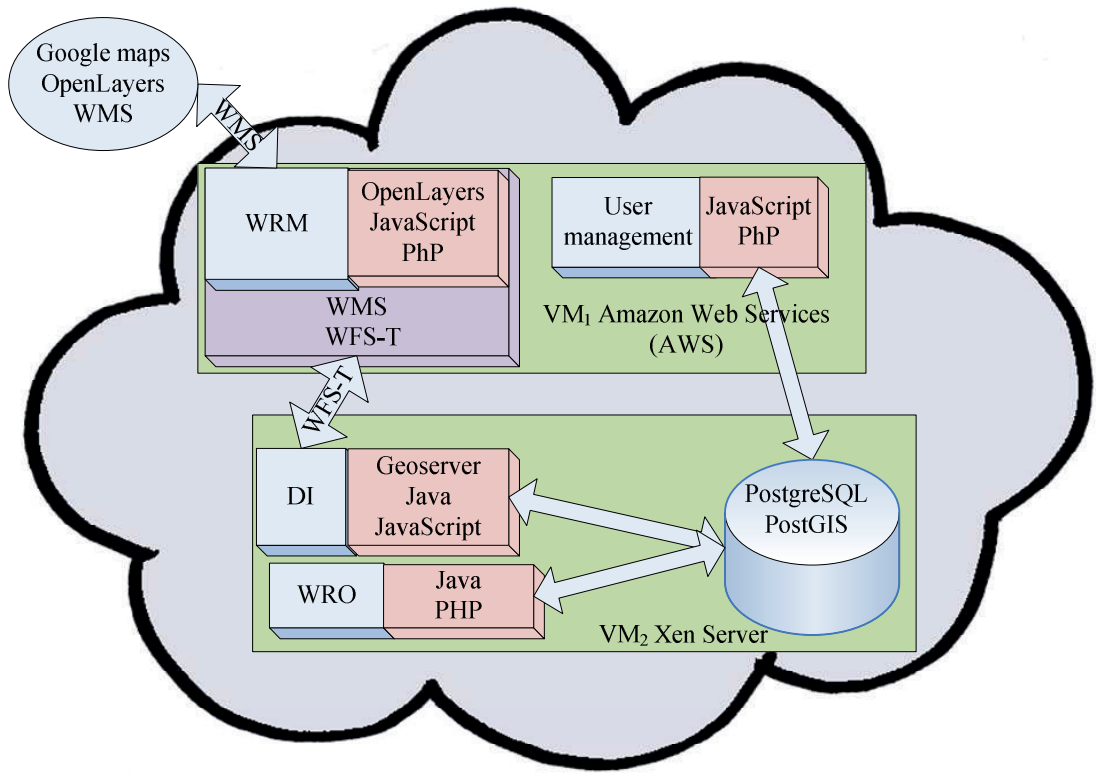

Figure 7.1 Cloud-based decision support platform architecture

### 7.2.1 Data infrastructure web service

The DI web service has two data sets and functions. The first data set is comprised of geospatial data where DI provides spatial data infrastructure (SDI) services. The SDI (Infrastructures 2004) web service consists of two components: (1) the data repository built from the relational database HMak created in PostgreSQL and PostGIS, (2) and the web application GeoServer. The HMak stores the six vector geospatial layers used by the web service for supporting WRM.

GeoServer is a powerful open source web application that manages, stores and presents geospatial data on the internet. The main purpose of the GeoServer as a middle tier application is to connect the relational database HMak on the backend with the developed web services on the frontend of the application. In the platform implementation, the GeoServer provides WFS-T interface connections for the web service for supporting WRM.

The second dataset contains mainly time series data, storage discretization data, etc., which are organised in 73 tables used by the WRO web service.

### 7.2.2 Web service for support of Water Resources Modelling

The web service for support of WRM user interface is shown in Figure 7.2. The web interface has tools to work with the objects from the six-vector geospatial layers that are representing water resources components and infrastructure. These geospatial objects are the basic building blocks to create WRM (thus the name of the service, "for support of WRM"). The existing stand-alone WRM applications have user interfaces for creating and editing such elements (building blocks) with corresponding attributes and associated data (see for example, Mike Basin of DHI (DHI, 2015) water and Environment, RIBASIM of Deltares (Deltares, 2015) and other similar applications). This web service is intended for a similar purpose, with one important difference that is now accessible via a web browser. The web service does not have all model set-up capabilities (most importantly, there is no computational engine), but the developed solution already allows for creating and editing of the needed elements.

All elements provided by the web service for support of WRM are geospatial objects. The geospatial objects are defined by their geographical information and corresponding attributes e.g. rivers are polylines that are defined by geographical coordinates and their corresponding attributes such as name, unique identifier, category, and 'goes in' attribute for defining downstream direction. The corresponding attribute tables are not explained here because they are relatively simple and serve just as a demonstration to assign additional information for each modelling object. The web service allows only specific type of geospatial object to be inserted in the layers corresponding to the intended elements, e.g. points for users, reservoirs, and inflows, polylines for rivers and canals, and regions (polygons) for agricultural areas.

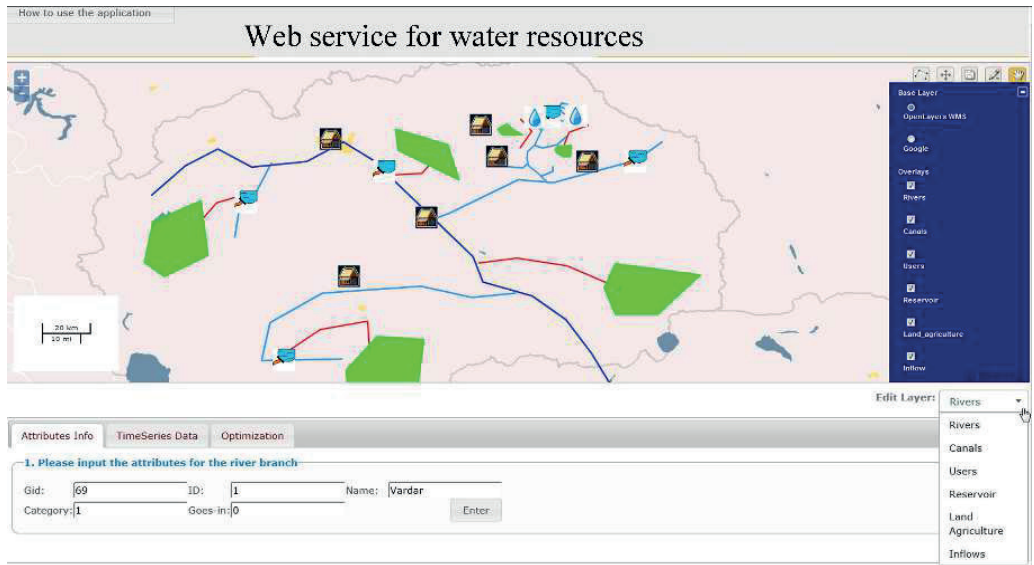

Figure 7.2 Web service for support of water resources modelling (WRM) interface

The web service for supporting WRM is built using JQuery, OpenLayers library, and additionally developed prototype source code written in PHP, Ajax and JavaScript programming languages. OpenLayers is an open source JavaScript library that supports the Open Geospatial Consortium (OGC) standards from which Web Map Services (WMS) and Web Feature Services - Transactional (WFS-T) are used in the application. The web service is using WMS to connect with different base map providers (such as Google Maps or OpenLayers WMS) and uses these layers as background maps. The users can select the background map from the menu e.g. as shown in Figure 7.2, where OpenLayers WMS is used.

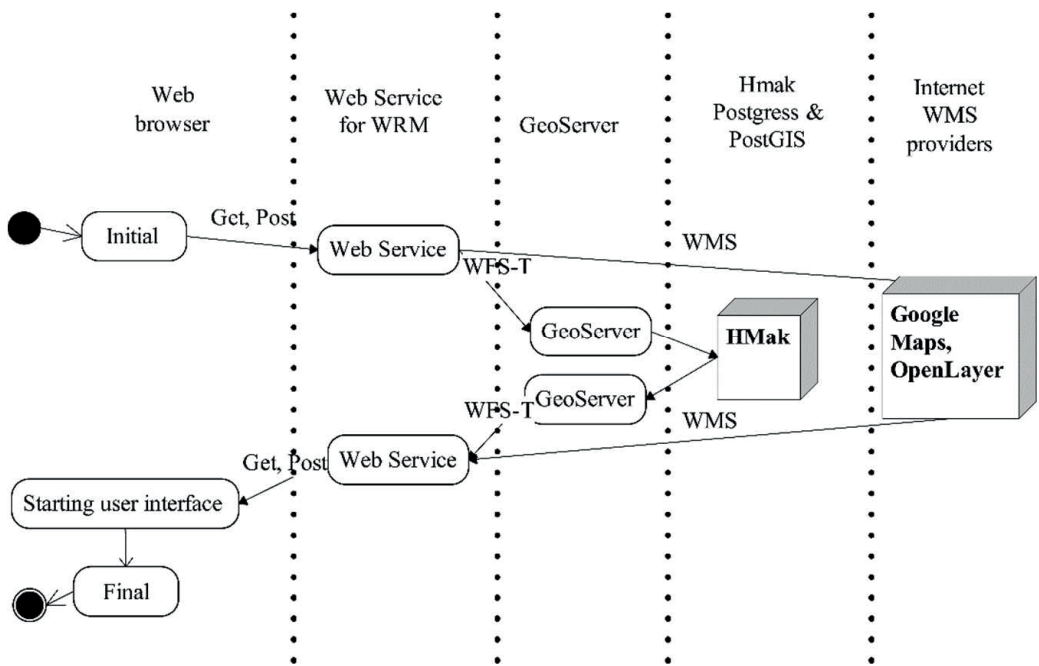

Figure 7.3 Activity diagram of web browser refresh

The WFS-T interface supports creation, deletion, and updating of geospatial data across the web. The WFS-T is an XML message. The OpenLayers library creates two way WFS-T communication between the user interface and the vector geospatial data from HMak having GeoServer are a middle tier. This is the most valuable characteristic of the presented cloud platform and it is according to SOA principles. Important milestone is that the web service for WRM runs on $VM_1$ and DI web service (GeoServer) runs on $VM_2$, which present a distributed computer system. The difference from the previous system (Delipetrev et al. 2014) is that the WFS-T communication is over the internet and not along a single VM. This clearly proves the possibilities to deploy many web services and link them accordingly.

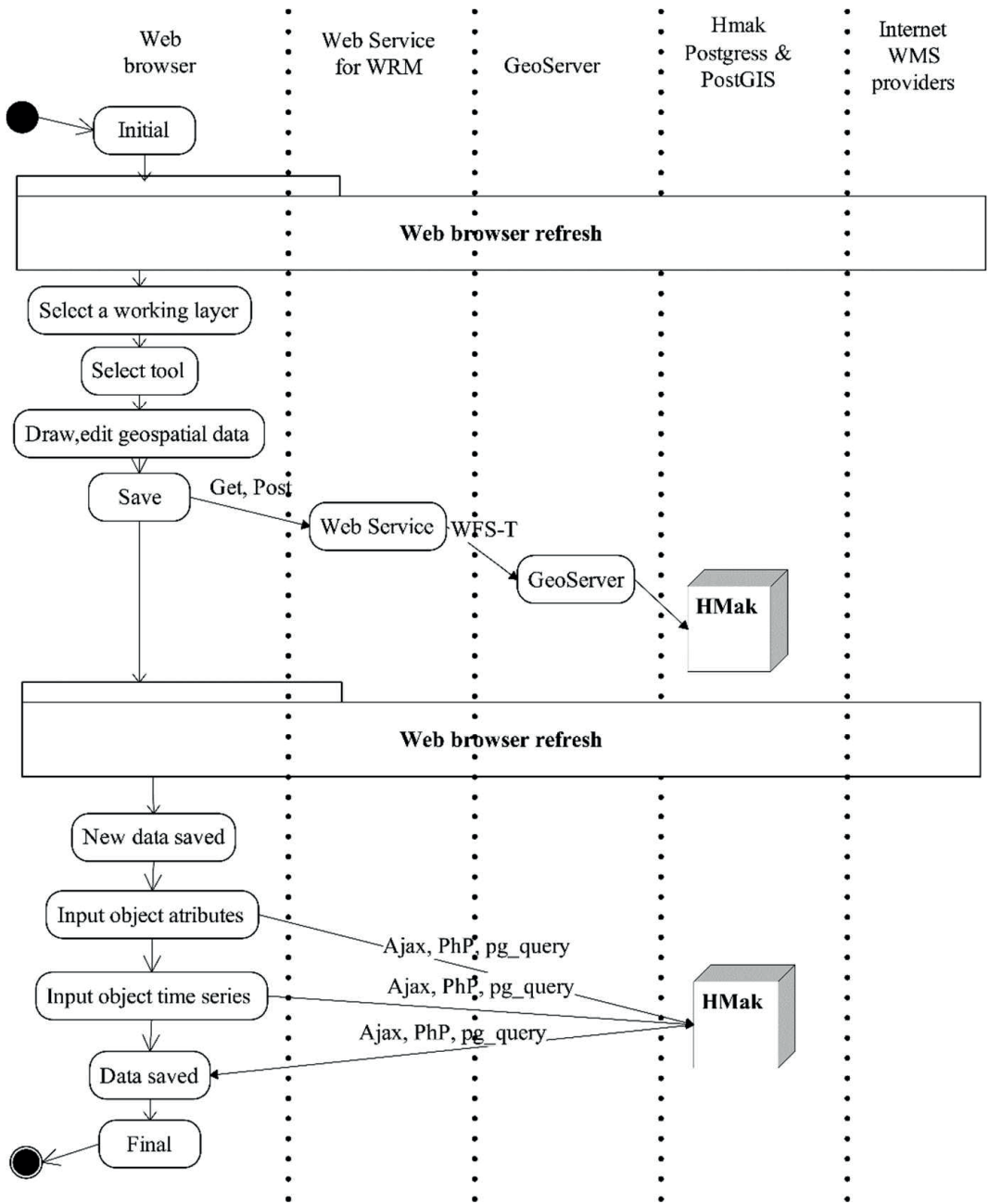

Figure 7.4 Web service for support of WRM activity diagram

When the user saves changes in the user interface, a WFS-T message is created that goes to the GeoServer, which translates the message and correspondingly changes the geospatial data stored in HMak. After each refreshing of the user browser a WFS-T request is invoked that reads data from HMak over the GeoServer and WMS from the internet providers, as shown in Figure 7.3. The "Get" and "Post" are HTTP request methods used by the web browser to get raster geospatial maps from WMS and vector geospatial maps from WFS. The activity diagram of the web service for support to WRM is shown in Figure 7.4.

As part of the web interface (see Figure 7.2), below the map screen there are three tabs entitled as: "Attribute info", "Time series data" and "Optimization". Additional PHP and Ajax scripts were developed to work with attribute data because this is not supported by WFS-T. When an element (object) from the map is selected the Ajax and PHP scripts are executed and fill the attribute information of the specified object in the "Attribute info" tab. These attributes can be modified and stored back into HMak. The second tab "Time series data" provides the capability to upload time series data associated with a selected object. In the current implementation, "Time series data" tab calls PHP and JavaScript functions that upload a Comma Separated Values (CSV) time series file into the HMak database. The "Time series data" tab demonstrates the possibility to upload additional information, except attribute tables, of any geospatial object.

### 7.2.3  Web service for water resources optimization

The web service for water resources optimization is created from several components: web form for entering data, prototype code for uploading data into HMak database coded in PHP and Ajax, nDP, nSDP, and nRL Java applications and a web page for presenting results. The input data are uploaded using "Optimization" tab shown in Figure 7.2. The input data are depending on the algorithm that is used and need to be saved in files in CSV format. Three algorithms are embedded in the web service for WRO: 1) nDP, 2) nSDP and 3) nRL. The input data are loaded and stored into the DI. The nDP, nSDP, and nRL algorithms are exported as an executable JAR file saved in the file directory on the $VM_2$.

The service is started by selecting a reservoir from the cloud platform interface and clicking on the "Optimization" tab. The optimization tab opens a new window for entering data. After data is uploaded, a new window opens for starting optimization. Depending on the algorithm selected the appropriate JAR file is started that connects to HMak and reads the input data from HMak, processes the input data and saves the results data in HMak. The resulting data are then plotted by using the JavaScript library "highcharts.js." (A screenshot of such a result plot is presented in the following section 7.3). The activity diagram of the web service for water resources optimization is shown in Figure 7.5.

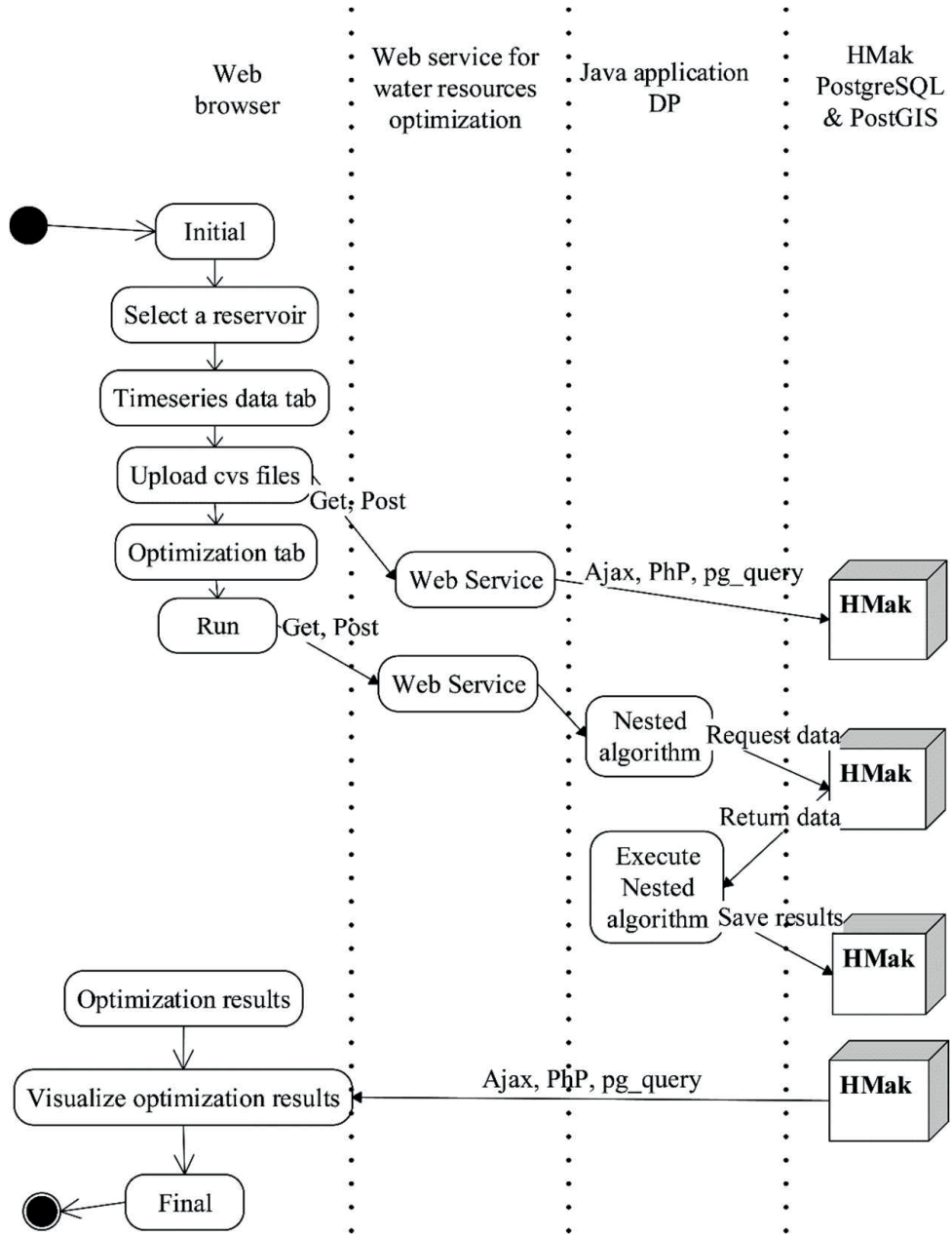

Figure 7.5 Web service for water resources optimization activity diagram

### 7.2.4 Web service for user management

The user management web service is deployed on the $VM_1$. The current implementation is simple with managing user access to the cloud platform. The user management web service contains an administrator panel for managing users, adding new users, controlling usage time of the cloud platform, deleting existing users, etc. When the user logs in the cloud platform, it activates a PHP session that measures the user usage time. The usage time is saved in the user profile. Further development of this service will include users' computer power and storage usage. Using this information, the administrator can effectively manage the cloud platform users.

## 7.3 Results and tests

The cloud-based decision support platform for water resources was deployed on $VM_1$ with operating system Ubuntu, running as a micro instance on AWS, and $VM_2$ with operating system Fedora 16 that is part of the Xen cloud platform running on IBM x3400 M3. The Xen Cloud platform is managed with the Citrix XenCenter application that provides management tools to control the server environment e.g. control the CPU usage, memory, disks and network connections on all virtual machines.

The cloud-based decision support platform was tested using the case study data. The main objective was to demonstrate that the platform works and supports the real time multi-user activity and embodies several optimization algorithms. The Zletovica river basin was modelled using the web service for support of WRM shown in Figure 7.6. (The titles of the towns, reservoir, and map legend in Figure 7.6 were added additionally and are not part of the web service).

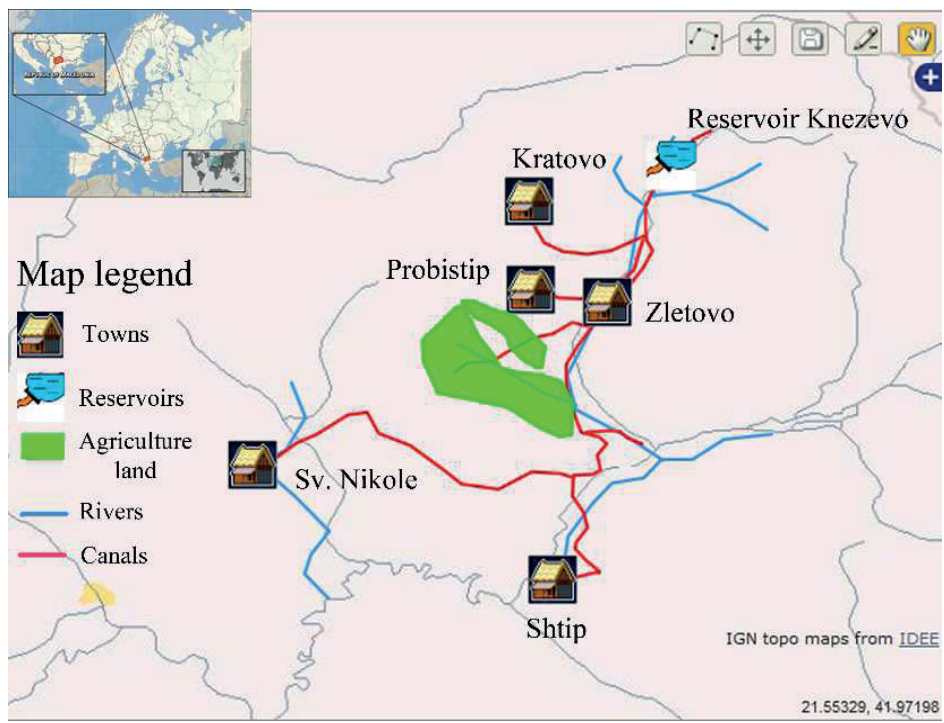

Figure 7.6 WRM of Zletovica river basin

In addition to a standard web browser, other applications using WFS-T interface can connect with the cloud platform through the DI web service, in the same manner as the web service for support of WRM. After a connection is established, these applications can work with the geospatial data. Two geospatial (mapping) applications, uDig and MapInfo, were tested and connected to the web service. This demonstrates the platform flexibility and openness to other applications.

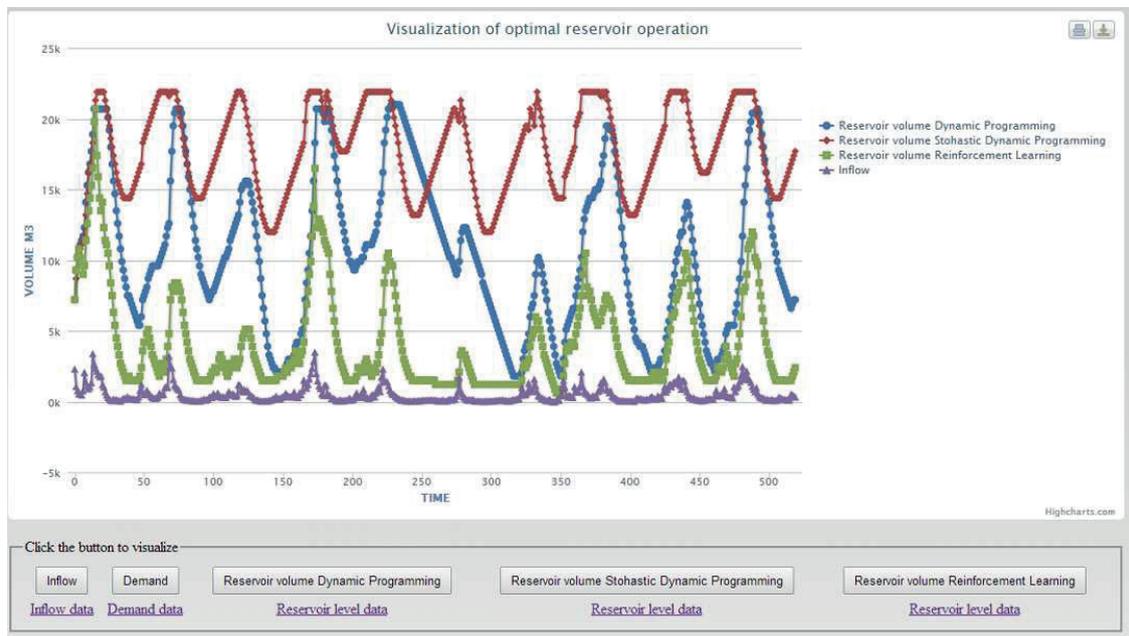

Figure 7.7 ORO results of nDP, nSDP, and nRL algorithms

The WRO web service was tested using nDP, nSDP, and nRL algorithms on the Zletovica case study. The WRO web service optimization results present the optimal reservoir volume in the testing period (10 years weekly data) of nDP, nSDP, and nRL, as shown in Figure 7.7. The tabs below present the specified algorithms results. This interface was very useful in developing and controlling of the algorithms. It was especially useful for tuning the nRL algorithm, because after each 10.000 episodes a test of the optimal reservoir policy could be performed and visualised. From this view the learning of nRL agent was directly visible, same as shown in Figure 6.11.

Eight students carried out the multi-user activity test on the cloud platform simultaneously. The test objectives were to examine the cloud platform stability and support for real time concurrent multi-users activity. All students had the same access to the four services and performed the following:

- Search and download of maps and data from DI web service.

- Draw, modify, enter, and delete geospatial objects and their attributes using the service for support of WRM.

- Optimize the reservoir operation using the service for water resources optimization

At the beginning, students were instructed how to use the cloud platform and its web services and their task was to test the system with increased workload from concurrent multi-users access. These are the main conclusions from the performed tests:

The DI web service supported concurrent multi-users access. The users performed search, view, and download of maps and data. The service was functioning normally, as expected. The web service for support of WRM provided constant service to concurrent multi-users that collaborated together in the same working environment, e.g. jointly drew rivers, canals and other objects, entered attribute data or together developed a 'model' of the hydro system. The web service for water resources optimization provided optimization results to all the users. Multiple users uploaded CSV files, executed optimization, and jointly viewed the results. The only limitation was that each optimization could be executed one by one. The main conclusion from this test is that the cloud platform is stable and functional supporting multi-users and an increased workload.

## 7.4 Discussion

The developed application demonstrated that the cloud paradigm "only a web browser is needed to use the application" could be accomplished. Of course, the presented prototype cloud platform is far from being operational for solving water resources modelling and optimization problems. Another important objective that was realised is the embedding of the nested optimization algorithms in the cloud platform. This is a proof of concept that optimization algorithm can be included in the cloud platforms. This cloud platform presents what are current possibilities for development of the next generation of software product and services.

The reason why WRO and DI web services were deployed on $VM_2$ is related to their high computational demands. The most computationally demanding part of the cloud system is the WRO service with its nested algorithms, especially nRL. The other two services, the service for support of WRM and users' management need relatively low computational power.

Further, we want to consider the NIST definition (Mell and Grance 2011) of cloud computing with the proposed cloud platform and explain the advantages, disadvantages and further development of the system. The first two essential characteristics of the cloud application are "on-demand self- service" and "broad network access." The cloud platform is available and accessible all the time and from anywhere and it only requires a web browser to be accessed and used. Moreover, the interaction with the application is on-demand and driven by user needs. The cloud platform is also available on mobile phones and laptops through the browser.

The third and fourth essential characteristics of the cloud are its capability for "resources pooling" and "rapid elasticity." The cloud platform is deployed on two physical servers running two separate VMs. The basic adjustment concerning the workload can be performed by modifying the current server's computational power (adding extra memory and computational power). The VMs workload can be monitored over AWS console and XenCenter and adjusted appropriately. The cloud application components, standards, and programming languages are interoperable and can be deployed on an unlimited number of servers. Issues concerning scalability and resource pooling can be resolved by adding and connecting additional VMs.

Additionally, the issues concerning scalability and resource pooling can be resolved by creating many data repositories similar to the DI that can be deployed anywhere on the internet storing large quantity of geospatial and other types of data. Furthermore, the GeoServer application can connect to multiple data repositories that are geographically dispersed; the web service for support of WRM can connect to many instances of the GeoServer and finally, the web services themselves can be distributed on several servers. Depending on the number of users, workload, storage capacities, processing power, number of servers available, etc., the optimal cloud environment can be adapted. Obviously, these operations in the future should be automated. The deployment experiments clearly demonstrate that the cloud platform can run on different physical and virtual servers, having many data sources and GeoServer instances.

Concerning service models, the presented cloud computing application belong to SaaS. Users with a web browser access the cloud application and do not care about underlying cloud infrastructure. The current deployment model is hybrid of public - private cloud because the $VM_1$ is running on the Amazon web services, which is a public cloud, while $VM_2$ is on the Xen cloud platform, which is a private cloud.

The last essential characteristic of cloud computing is "measured service" which is rudimentary supported by measuring the time of each user's usage of the system. This satisfies cloud computing criteria, but needs to be vastly improved (e.g. with measuring processing power consumption, storage capacity utilization, etc.).

After affirming the "cloud" platform, the next discussion is about the web services capabilities and their further improvements.

The web service for support of WRM can be enhanced by the introduction of new layers and improvement of attribute information. Further improvement can be in the connectivity between objects e.g. one river flowing into another, a canal supplying a water user, etc. OpenLayers provides tools and capabilities for snapping and connecting objects, which can be used for this purpose. Additionally, the service can include development of the computational engine for performing actual simulations.

The future development of the WRO web service can be in supporting diverse water resources problems, MO optimization, and better results presentation. Many possible directions can be envisaged here, also, depending on whether one aims for developing generic web-based simulation / optimization system that can be used for different cases, or for a fully developed system for a particular case.

The cloud platform most valuable characteristic is the real time collaboration platform capabilities. Multiple users from all around the world, using only a web browser can work jointly using the web services and collaborate in real time on the same working environment. An example of this is when a user saves the current work in the web service for support of WRM, after which all other distributed users with just refreshing the web browser window can see this newly updated data (new/modified rivers, water users etc.). Another example is when one user enters the time series files for inflow, demand and other input data and runs the optimization, and other users can jointly view the results. All of the data and models are stored on the Internet and users do not have to be concerned with software version or the hardware and software support infrastructure.

Another valuable characteristic is the cloud platform flexibility for creating new services and components or upgrading the existing ones. Implementation of the OGC standards and SOA allows even greater flexibility in connecting other software to the cloud platform that was demonstrated when connecting uDig and MapInfo with the web service for managing, storing, and presenting geospatial data. The combination where a desktop application is directly working with a cloud platform can provide additional benefits, e.g. storage of data locally.

The presented cloud platform is created using open source software, which is increasing its value because many software companies could use similar technologies and components to create various solutions without license fees. This does not exclude the possibility of adding additional commercial software components (data repositories, libraries, applications etc.).

The major drawback of the presented application is that without internet access or due to a potential server downtime, it cannot be used. The first and second drawback can be solved by additional internet connection and backup servers, respectively. Another important consideration is the cost. The development of the presented cloud platform to the stage of operational solution would involve significant human and material resources. Resolving the operational and service cost and the continuous user support cost of a fully functional web solution will be challenging.

Another potential development of the cloud decision support platform is acquisition of real time meteorological information. This data can be input to a forecasting model to predict future reservoir inflows, tributary inflows, and other data needed for the forecast. The nDP, nSDP, and nRL can be further developed to include these real time data and provide short and long term optimal reservoir policy. Importantly, this holds the capacity to produce a collaborative platform that can be accessed by all users and provide fresh ways of communicating important information, potentially leading to more participative decision making processes involving water resources management and planning.

## 7.5   **Conclusion**

Future applications, software and services will increasingly be cloud oriented. The presented cloud-based decision support system demonstrates that there are existing open source software and technologies to develop robust and complex cloud platforms for water resources. The solution was successfully tested on a case study and with concurrent multi-users activity. The cloud-based decision support system embedded the previously described nested optimization algorithms nDP, nSDP and nRL. The nested algorithms can be accessed over the web interface, their case study data can be uploaded, optimization executed and results presented. The cloud platform was tested in a distributed computer environment running on two VMs.

Further development of the cloud-based platform can be in creating and connecting new data repositories by including more diverse water related data, such as population growth, urbanization, meteorological and climate data, etc. Additional modelling, optimization, and other decision support services are envisioned to be added to the existing platform, so that it will evolve in a fully cloud-based decision support platform for water resources modelling and optimisation.

.

# Chapter 8     Conclusions and recommendations

*"Stay hungry, stay foolish"*

*~Steve Jobs*

This chapter presents the conclusions and recommendations for further research.

___________________________________________________________________

## 8.1     Summary

The PhD thesis presented the novel optimal reservoir operation algorithms nDP, nSDP, nRL and their corresponding multi-objective versions created by MOSS, the MOnDP, MOnSDP and MOnRL. The novel idea is to include a nested optimization at each state transition of classical DP, SDP and RL that lowers the starting problem dimension and alleviates the curse of dimensionality. The nDP algorithm research is published in the Journal of hydroinformatics (Delipetrev et al. 2015), while upcoming publications will present the application of the nesting idea with the other algorithms.

The case study of the Zletovica hydro system is researched and analysed in details, concerning all facets of the system (reservoirs, irrigation channels, irrigation studies, water resources, water demands, water distribution, hydropower, etc.). The system is represented in a corresponding water allocation model, used for defining the optimization problem formulation, decision variables, constraints and the OF (OFs).

The nDP, nSDP and nRL implementations in the Zletovica hydro system demonstrated their capabilities and limitations. The added scientific value is that due to the complexity of the Zletovica hydro system, many possibilities and alternatives were investigated, especially in the case of nRL implementation. The nRL implementation is explained in details and can be used as a starting point for further ORO RL research.

The nDP was compared to AWD DP and classical DP, indicating that it has better modelling capabilities than AWD DP and classical DP.

Identification of optimal solutions in multi-objective settings was performed with MOnDP, MOnSDP and MOnRL algorithms. These algorithms identified the Pareto ORO solutions. From the Pareto solutions it is possible to select an ORO policy that is the most suitable concerning various objectives, like irrigation deficits, water supply deficits, minimum critical level deviations, maximum critical level deviations, and total deficits.

The nSD, nSDP and nRL algorithms are embedded in a cloud decision support platform. The cloud decision support platform presents the latest ICT capabilities in building applications with a) cloud computing b) SOA and c) web GIS, using open source software and open standards (OGC). The cloud decision support platform has a flexible architecture and currently includes four web services for 1) data infrastructure, 2) for support of water resources modelling 3) for water resources optimization and 4) users' management.

This development acts as a proof of concept, which demonstrates that a complex platform can be developed with several cloud computing centres (in this case AWS and Xen Server) that can support multiple geographically dispersed users. The system demonstrated the cloud platform scalability, distributed computer environment, availability, accessibility, real-time multiuser collaboration environment, and its advantages over classical desktop application.

The specific conclusions concerning the two main parts of the PhD thesis, 1) ORO algorithms and 2) cloud decision support platform are presented in the following sections.

## 8.2   Conclusions

### 8.2.1   Conclusions concerning the algorithms

Regarding the developed ORO algorithms (nDP, nSDP, nRL, MOnDP, MOnSDP and MOnRL) presented in Chapter 2, 3, 4, 5 and 6, the following conclusions can be drawn:

1. Several ORO algorithms were designed and developed nDP, nSDP, nRL as well as their MO variants, MOnDP, MOnSDP and MOnRL. Due to their specific design using nesting, these algorithms can solve ORO problems with multiple decision variables, successfully alleviating the curse of dimensionality. These algorithms were implemented and tested in the case of the Zletovica hydro system with eight objectives and six decision variables.

2. The optimization problem formulation of the Zletovica hydro system has been solved by the nDP, nSDP and nRL. The nDP allowed for full representation and solution of the optimization problem formulation of the Zletovica hydro system, including all decision variables, constraints, and OF. The nSDP has issues in implementing several state variables without provoking the curse of dimensionality, so adjustments were needed to fit nSDP to the case study optimization problem requirement. The nRL showed its true power with including all four stochastic variables implementing the complete optimization problem formulation, but its implementation and tuning requires additional effort. The main conclusion from the implementation of the algorithms is that nDP can implement complex optimization problem formulations without significant problems. The nSDP has limitations when additional optimization problem variables are included. The nRL is very powerful in implementing complex optimization problems, but needs tuning concerning its design, parameters, action list, convergence criteria, etc.

3. The nDP was compared with classical DP and the AWD DP algorithm. It was confirmed that the classical DP cannot be used for complex problems as the one considered due to its huge computational demands while the AWD DP solves a simplified problem so cannot provide solutions as accurate as nDP provides.

4. The nDP was employed in optimization experiments on monthly and weekly time-series data of the case study. The nDP monthly optimization experiments demonstrated that the weights given to each objective directly influence the ORO, and thus the modeller or a decision maker can choose the objectives to be mostly satisfied. The experiments with the nDP on monthly data demonstrated the applicability of the nested algorithms for the variable storage discretization and variable weights at each time step. With respect to the case study requirements, the nDP optimization on weekly data gave results where the ecological flow and water supply users are satisfied over 99% of the time, and the irrigation – over 75% of the time, for the time period of 55 years (1951-2005).

5. The nSDP and nRL were used to derive one-year weekly optimal reservoir policy. The available weekly data (1951-2004) were divided into a training (1951-1994) and testing part (1994-2004). The nSDP and nRL optimized/learned the optimal reservoir policy on training data, and their policy was examined on the testing data. The nDP solved the ORO problem in the testing period (1994-2004) and this solution was used as a target for both nSDP and nRL. Interesting results were to observe how the nRL agent learns with the increase of the number of episodes. The nRL optimal reservoir policy is the best between 80,000-160,000 learning episodes. The nSDP and nRL policies were benchmarked against the nDP results and it was found that the nRL performs better than nSDP overall and for all objectives separately. The main conclusion is that the nRL is a better choice than the nSDP, at least for the considered case study.

6. The case study problem was also solved by using the multi-objective nested optimization algorithms MOnDP, MOnSDP, and MOnRL. The found solutions form the Pareto optimal set in eight dimensional objective function space (since the eight different objectives were considered). The MOnDP was used as a

    scanning algorithm with 10 sets of varying weights that can identify the most desirable MO solutions. The MOnDP was selected because it is much quicker than MOnSDP and MOnRL. From the 10 sets of weights and their MOnDP results, three sets were selected to be used by MOnSDP and MOnRL. (The results also confirmed the previous conclusions about the relative performance of various algorithms.) The solutions generated by the MOnRL were found to be much better than those of the MOnSDP.

7. The developed nested algorithms are computationally efficient and can be run on standard personal computers. For the considered case study, on a standard PC, nDP executes in 1-3 min, nSDP in 2-5 min, while nRL 8-20 min (the longest is nRL-Q). The MOnDP, MOnSDP, and MOnRL execution time depends on the number of weight sets used in the MOSS approach.

8. The developed and tested nested versions of DP, SDP and RL proved the effectiveness and efficiency of such an approach, which allows for including additional objectives in optimization (characteristic of multi-purpose reservoirs) without a substantial modification in the source code or a considerable increase in computational complexity.

### 8.2.2 Conclusions concerning the decision support platform

The second part of the PhD thesis has been devoted to the decision support platform. The conclusions can be drawn:

1. The cloud platform was tested on distributed computer systems with two virtual machines ($VM_1$ and $VM_2$). The cloud platform is available and accessible from everywhere and at any time, and only a web browser is needed to use it. It is interoperable and can run on different operating systems and platforms. The platform is flexible for adding additional services, and can connect with desktop software, which was demonstrated by linking to MapInfo and uDig. The cloud platform was validated with the NIST (Mell and Grance 2009) definition of a cloud.

2. The cloud platform uses the mechanism of web services, which in this study proved to be an effective method for flexible integration of various components. For examples, the nDP, nSDP, and nRL are part of the web service for optimization, which links to other services for storing their input data and results, and visualization.

3. The cloud platform provides a multiuser collaboration environment where different users from all around the world can access a shared application and collaborate in real-time. The collaboration environment was successfully tested with a group of students imitating the decision procedures in water resources.

## 8.3    **Recommendations**

The recommendations for further research are summarized in the following paragraphs.

1. It is recommended to carry out an investigation into the possibilities of using the ideas of nested optimization for multi-reservoir systems. Optimization of such systems is by an order of magnitude computationally more demanding than of single-reservoir ones.

2. A full-fledged comparison of MOSS and multi-objective algorithms is recommended. The use of MOSS in this study was motivated by earlier experiments that showed that for a number of practical problems MOSS is a more efficient alternative to a full MO optimization that provides a small Pareto set which however for many problems is sufficient. MOSS approach needs further research, tuning and testing on more examples.

The most interesting field that deserves much more attention and research in the future is nRL and MO RL.

3. The nRL agent exploration / exploitation strategy is a field requiring further research. In the current setting, the exploration is random and the agent picks a random action from all possible actions. In ORO, the number of possible actions is limited, which means that we can include additional intelligence in selecting the exploration action. Currently the nRL agent can choose an exploration action from full to empty reservoir, but with additional intelligence the agent would be restricted in its choice of actions. In the present case, these actions (from full to empty) are usually far from optimal and do not become part of the derived policy. On the other hand, an agent with additional intelligence for selecting the best set of exploration action may increase the complexity of nRL, and its computation cost. Moreover, the kind of intelligence needed may be case-specific. There will be trade-offs between the implementation of a more intelligent agent that chooses a set of possible exploration actions, and the random selections currently implemented in this thesis.

4. Multi-agent and MO RL definitely need more research. The idea and concept of employing multiple agents in searching for an optimal solution is powerful. Including multiple agents naturally opens up the possibilities for parallel computation that is highly desired in any problem. The previously described MO nested algorithms, can be parallelized and supported by a high performance computer or cloud infrastructure. Most of the nested algorithms are quite fast on a standard computer, however for more complex problems parallelization in MO and multi-agent optimization would be a way to keep the computational time within the reasonable limits.

5. It can be recommended to carry out research into the ways of including real time and forecasted meteorological and hydrological data in nRL. The nRL has to be

>upgraded to include additional variables and the incoming data streams to be able to continuously and reasonably fast learn the optimal reservoir policy.

Having researched the algorithms and systems for optimal reservoir operation, it is also realized that the most significant challenge is the actual implementation of the presented tools in the form of an operating decision support system to be used in the Zletovica hydro system. The research, knowledge, and prototypes are here, but without their real implementation and usage, they may remain just research results from another PhD thesis, without their use in real practice. This will be definitely the next goal.

# Abbreviations

ANN Artificial neural network

API Application programming interface

AWD Aggregated water demand

AWS Amazon web services

DI Data infrastructure

DP Dynamic programming

DSS Decision support system

FAO U.N. Food and Agriculture Organization\

FQI Fitted Q-iteration

IaaS Infrastructure as a service

ICT Information and communication technologies

GIS Geographic information systems

GPS Global positioning system

MO Multi-objective

MOFQI Multi-objective fitted Q-iteration

MOnDP Multi-objective nested dynamic programming

MOnSDP Multi-objective nested stochastic dynamic programming

MOnRL Multi-objective nested reinforcement learning

MOPSO Multi-objective parameterization-simulation-optimization

MOSS Multi-objective optimization by a sequence of single-objective optimization searches

OECD Organization for Economic Cooperation and Development

OGC Open geospatial consortium

OF Objective function

ORO Optimal reservoir operation

SaaS Software as a service

SDP stochastic dynamic programming

SO Single objective

SOA Service oriented architecture

SOAWS Single-objective aggregated weighted sum

PaaS Platform as a service

RL reinforcement learning

nDP Dynamic programming

NIST National institute of standards and technology

nSDP stochastic dynamic programming

nRL reinforcement learning

WEF World Economic Forum

WFS Web feature service

WFS-T Web feature service – transactional

WMS Web map service

WRI World Resources Institute

WRM Water resources modelling

WRO Water resources optimization

XML Extended markup language

# References

Links to the nested algorithms Java source code:

nDP: https://github.com/deblagoj/DP-3Objectives.git

nSDP: https://github.com/deblagoj/SDP-3Objectives-improved.git

nRL: https://github.com/deblagoj/RL-4states.git

Cloud computing platform for water resources: https://github.com/deblagoj/IWRM.git

# Samenvating

Bevolkingsgroei, verbeterde standaard, en klimaatverandering zetten extra druk op de reservoirs wereldwijd. Het reservoir purpouse is om water te leveren voor de bevolking, industrie, landbouw, waterkracht, enz. Het reservoir operatie is een multi-objective probleem voor optimale waterverdeling tussen verschillende vaak tegenstrijdige gebruikers en functie. Een optimale werking reservoir (beleid) algoritmen nodig die voldoet aan de gebruiker en functioneert zo veel mogelijk. Deze optimale reservoir operatie algoritmes zijn vaak onderdeel van een beslissingsondersteunend systeem dat de tools voor het waterbeheer biedt.

De vraag naar schoon water in de Republiek Macedonië wordt voortdurend groeit met de toename van de standaard van de bevolking, de ontwikkeling van nieuwe industrieën en landbouw. De regio Macedonië in de zone van continentaal klimaat, gekenmerkt met natte en koude winter en de lange droge zomerseizoen. Het verstrekken water voor gebruikers in de zomerperiode is zeer belangrijk. Het beheer van deze complexe vraagstukken vraagt ontwikkelen van reservoir optimalisatie algoritmes en beslissingsondersteunende systemen die potentiële water conflicten zullen evenwicht tussen bevolking, industrie, landbouw, energie, enz, en te helpen bij de planning en ontwikkeling van nieuwe infrastructuur.

Dit zijn de belangrijkste drijfveren voor dit promotieonderzoek. Het proefschrift onderzoek is in twee belangrijke wetenschappelijke onderwerpen in Hydroinformatics 1) optimale reservoir werking en 2) beslissingsondersteunende systemen.

De klassieke oplossing voor de optimale reservoir operatie probleem wordt berekend door het dynamisch programmeren (DP), stochastisch dynamisch programmeren (SDP) en recent reinforcement leren (RL). De DP en SDP methoden zijn bekend en gevestigde, maar last van dubbele vloek 1) vloek van de dimensionaliteit en 2) vloek van modellering. De toename van het aantal variabelen in de stand-actie ruimte van de multi-objective optimale reservoir operatie probleem provoceert de vloek van de dimensionaliteit. Dit is vooral merkbaar wanneer meerdere vraag waterbescherming (steden, landbouw, industrie enz.) Betreft, hetgeen vaak in veel optimale reservoir gebruik tegenkomt. Dit brengt de eerste centrale onderzoeksvraag in dit werk aan bod: Hoe om meerdere gebruikers van water naar de optimale reservoir operatie probleem omvatten zonder provoceren de vloek van de dimensionaliteit?

Het proefschrift legt de ontwikkeling van een nieuw idee, genaamd "genest", die kunnen bestaan uit extra variabelen in de DP, SDP en RL zonder provoceren de vloek van de dimensionaliteit. De "genest" idee werd in de DP, SDP en RL geïmplementeerd en navenant drie nieuwe algoritmen ontwikkeld genoemde genest DP (NDP), genesteld SDP (NSDP) en geneste RL (NRL). De geneste algoritmen bestaan uit twee algoritmen: 1) DP, SDP of RL en 2) genest optimalisatie algoritme geïmplementeerd Simplex en kwadratische Knapsack. Het idee is om de geneste optimalisatie-algoritme in de staat overgang die de startende probleem dimensie verlaagt en verlicht de vloek van de dimensionaliteit bevatten.

De meervoudige doel optimale reservoir Bedieningsprobleem wordt beschreven door een enkele geaggregeerde gewogen doelfunctie waarbij elk van de afzonderlijke doelstellingen gewichten toegekend. Door het gebruik van verschillende sets van gewichten, de geneste algoritmes verwerven multi-objectieve eigenschappen, aangeduid als multi-objective NDP (MOnDP), multi-objective NSDP (MOnSDP) en multi-objectieve nRL (MOnRL). Deze algoritmen kunnen meerdere doelstellingen optimale oplossingen te identificeren.

Deze algoritmen werden in het Zletovica hydro systeem casestudy, gelegen in de Republiek Macedonië geïmplementeerd. De Zletovica hydro systeem heeft zes beslissing variabelen en acht doelstellingen, waaruit twee gerelateerd zijn aan het hoofd, vijf voor de vraag naar water gebruikers reservoir, en één voor waterkracht. Omdat het Zletovica hydro systeem case study is ingewikkelder dan een klassieke enkele reservoir geval, de NPD, NSDP en nRL algoritmes werden gewijzigd om de case study te passen. De modificatie toonde de geneste algoritmen beperkingen en mogelijkheden.

Na de uitvoering ervan, werden de geneste algoritmes getest met behulp van 55 jaar (1951-2005), maandelijkse en wekelijkse gegevens van de Zletovica hydro systeem. De NDP werd getest op 55 jaar de maandelijkse gegevens die aantonen dat het om een merk nieuw algoritme, en het is beter in staat dan de klassieke DP en geaggregeerde vraag naar water DP-algoritme. De NSDP en nRL opgeleid / leerde de optimale reservoir beleid met behulp van 45 jaar (1951-1995) trainingsgegevens. De NSDP en RL optimale reservoir beleid werden getest met behulp van 10 jaar (1995-2005) testgegevens. De NDP berekende de optimale reservoir operatie in dezelfde 10 (1995-2005) jaar periode en werd ingesteld als een doelwit voor de nSPD en nRL beleid. De resultaten tonen dat de nRL produceert betere optimale reservoir beleid de NSDP. De NDP, NSDP en nRL algoritmes kunnen multi-objective optimalisatie problemen op te lossen, zonder significante toename van de complexiteit algoritme en de computationele kosten. Computationeel, de algoritmen zijn zeer efficiënt en kan dicht en onregelmatige variabele discretisatie behandelen. De NDP kan overweg met meerdere model en besluitvorming variabelen, terwijl NSDP is beperkt in het aanvaarden van bijkomende modelvariabelen. NRL beter in staat dan de NSDP bij de behandeling extra modelvariabelen maar een complexe en moeilijke implementatie

De MO "geneste" algoritmes werden getest met meerdere gewicht sets. De MOnDP werd uitgevoerd met 10 gewichten sets te scannen naar geschikte MO oplossingen. Daarna werden vier gewicht sets geselecteerd en werkzaam in MOnSDP en MOnRL. Drie clusters van resultaten werden gemaakt: MOnDP, MOnRL en MOnSDP cluster. De MOnDP cluster resultaten waren het doelwit. De MOnRL cluster beter dan MOnSDP in alle vier gewichten sets, die de vorige resultaat bevestigd.

De "genest" algoritmen moeten worden opgenomen in een platform (applicatie), dus ze zijn toegankelijk en beschikbaar aan meerdere gebruikers. De klassieke oplossing is om een desktop applicatie die u zal voorzien van deze algoritmen (en eventueel bieden andere functionaliteiten) te ontwikkelen. De klassieke desktop applicatie is ontworpen om te werken op één computer en missen vaak het ondersteunen van samenwerking van gebruikers. De samenwerking ondersteuning betekent dat meerdere gebruikers de toepassing tegelijkertijd gebruiken met dezelfde werkomgeving en gereedschappen. Daarnaast wordt de klassieke desktop applicatie data en modellen draagbaarheid beperkt en ingetogen binnen de software-versie, en er zijn rigide beperking van schaalbaarheid van de software. Dit zijn de belangrijkste redenen, die de tweede onderzoeksvraag gedefinieerd: Hoe kan ik een toepassing die 24/7 beschikbaar, overal toegankelijk is ontwikkelen, dat is schaalbaar en interoperabel zijn, en kan de samenwerking van gelijktijdige meerdere gebruikers te ondersteunen?

De tweede onderzoeksvraag heeft geleid tot de ontwikkeling van een wolk decision support platform voor watervoorraden. De cloud-platform ingebed de eerder ontwikkelde algoritmes NDP, NSDP en nRL. De cloud-platform is gemaakt met de nieuwste ontwikkelingen in de informatie- en communicatietechnologie (ICT), open source software en web-GIS. De cloud-platform heeft vier web services voor: (1) data-infrastructuur, (2) ondersteuning van de watervoorraden modellering (3) watervoorraden optimalisatie en (4) gebruikersbeheer. De cloud-platform is ontwikkeld met behulp van meerdere programmeertalen (PHP, Ajax, JavaScript en Java), bibliotheken (OpenLayers, JQuery), en open source software componenten (GeoServer, PostgreSQL, PostGIS).

De wolk decision support platform werd getest op het Zletovica hydro systeem. Met behulp van de web service voor watervoorraden modellering, werd het Zletovica hydro systeem met succes gemodelleerd met ruimtelijke objecten voor de reservoir Knezevo, Zletovica rivier en haar zijrivieren, agrarische grachten, watergebruikers, zijrivier instroom punten en landbouw gronden. De webdienst voor watervoorraden modelleren presenteert aangepaste web GIS-applicatie in de waterhuishouding, het verstrekken van online GIS-mogelijkheden. De webdienst voor watervoorraden optimalisatie biedt web-interface voor de NDP, NSDP en nRL algoritmen. De webdienst voor watervoorraden optimalisatie biedt webformulieren voor het invoeren van algoritmen invoergegevens, knoppen om de geneste algoritmen uit te voeren, en de presentatie van optimalisatie resultaten. Een groep studenten testte de webservices, waaruit blijkt dat de gebruikers samen kunnen werken en het model van de watervoorraden, uitvoeren optimalisaties en de resultaten bekijken. De gepresenteerde cloud-platform heeft de volgende voordelen: het is beschikbaar de hele tijd, het is overal toegankelijk, het creëert real time samenwerking met meerdere gebruikers platform, het werkt in een gedistribueerde computer-omgeving draait op twee afzonderlijke virtuele machines (VM), de programmeertalen code en componenten zijn interoperabel, het is flexibel voor het toevoegen van extra componenten en diensten en het is schaalbaar, afhankelijk van de werklast.

Beide onderzoeksthema's worden geïmplementeerd in de Zletovica hydro-systeem, gelegen in het northeaster deel van de Republiek Macedonië. Dit onderzoek kan de basis voor de ontwikkeling van verbeterde waterbronnen optimalisering van het systeem, de planning en het beheer in de Republiek Macedonië worden.

# Acknowledgments

This PhD thesis could not have been possible without the support of UNESCO-IHE, Hydroinformatics chair group, my family, colleagues and friends. I am deeply thankful to all of them for their contributions to my research as well as for the good times we spent together during my four years as a PhD student.

My sincere gratitude goes to my supervisor Dr. Andreja Jonoski. He helped me from the start, by providing me assistance in applying for a PhD position at UNESCO-IHE, guiding me through the PhD studies, improving my writing, and helping to finalize this thesis. We had long discussions about many topics, including science, economics, and politics. I appreciate all his contributions of time and patience, and I thank him for making my PhD experience productive and stimulating.

I want to thank my promotor prof. Dr. Dimitri Solomatine for the continuous support of my PhD study and research, for his patience, motivation, and enthusiasm. His guidance helped me continuously during my research and in the period of the writing of this thesis. His advice on both research as well as on my career has been priceless.

I want to thank the following members of the Hydroinformatics chair group of UNESCO-IHE: Dr. Ioana Popescu, Dr. Biswa Bhattacharya, Dr. Gerald Corzo, Dr. Schalk Jan van Andel, Dr. Leonardo Alfonso and Jos Bult.

My deepest gratitude goes to the Netherlands Fellowship Programme (NFP) for providing funding for my PhD studies.

I am thankful to my colleagues and friends, Nikola Stanic, Ljijana Zlatanovic, Mirjana Vemic, Amila Hodzic, Aleksandar Pavlov, Milan Jaksic, Ruzica Jacimovic, Girma Yimer, Guy Beaujot, Katerina Gjorgjievska, Kun Yan, Maurizio Mazzoleni, Neiler Medina, Yaredo Abayneh, Quan Pan, Endalkachew Bogale, Aleksandra Bogdan, Kirill Gorshkov, Zahrah Musa, Oscar Marquez Calvo, Micah Mukolwe, and Mario Castro Gama.

I want to thank the ex-Yugoslav basketball team for the good times together: Slavco, Djuro, Damir, Zoka, Sale, Nemanja, Hansko, Bogdan, Misa, Nikolas, and all other enthusiasts who joined our Sunday morning basketball sessions.

Last but not least, I want to thank my family, my wife Ilinka and my sons Toshe and Tome, for their patience and support during my PhD studies. This was an extraordinary experience for all of us. Toshe attended Dutch school without knowing a word of Dutch, Tome started to be raised the Netherlands, and my wife Ilinka left her job in Macedonia in order to support me in my PhD studies. I also want to thank my father Todor, my mother Zorica, my brother Marjan, and his wife Katarina.

Lastly, I would like to thank all members of the doctoral committee for their review and the provided comments. Their comments and review were very valuable for improving the quality of this thesis

# ABOUT THE AUTHOR

Blagoj Delipetrev was born on the 4th of March 1980 in Shtip, Republic of Macedonia. He graduated from the Faculty of Electrical Engineering and Information Technologies, at University Ss. Cyril and Methodius in Skopje in 2003.

Because of his passion for science instilled by university professors in his family, especially his father, he continued his education for Master degree, while in the meantime he obtained his first employment as a lecture assistant at Faculty of Mining, Geology, and Technology in Shtip. Blagoj started his Master studies in 2004 at the Faculty of Electrical Engineering and Information Technologies, University Ss. Cyril and Methodius in Skopje, working on a thesis entitled "Geo-model of the Republic of Macedonia", focused on information systems technologies, Geographical Information Systems (GIS) and Spatial Data Infrastructures (SDI) and their potential applications in Macedonia. He successfully defended his Master thesis in June 2007.

During 2008, Blagoj applied for a PhD research position at UNESCO-IHE, supported by the Netherlands Fellowship Programme of the Government of the Netherlands. His PhD proposal on the topic of decision support systems for water resources management in the Republic of Macedonia was accepted and in January 2010 Blagoj started his PhD research at UNESCO-IHE. This publication presents his PhD thesis entitled "Nested algorithms for optimal reservoir operation and their embedding in a decision support platform", focused on novel algorithms for Optimal Reservoir Operation (ORO) and development of cloud decision support systems.

Currently Blagoj is working as an assistant professor at Faculty of Computer Science, University Goce Delcev in Shtip, Republic of Macedonia.

# Scientific publications

## Peer Reviewed International Journals

Delipetrev, B., Jonoski A., and Solomatine D. A novel nested stochastic dynamic programming (nSDP) and nested reinforcement learning (nRL) algorithm for multipurpose reservoir optimization. *Journal of Hydroinformatics.* Submitted paper.

Delipetrev, B., Jonoski A., and Solomatine D. (2015) A novel nested dynamic programming (nDP) algorithm for multipurpose reservoir optimization. *Journal of Hydroinformatics.* Vol 17, No 4, IWA Publishing. 570–583.doi:10.2166/hydro.2015.066.

Delipetrev, B., Jonoski A., and Solomatine D. (2014) Development of a web application for water resources based on open source software. *Computers & Geosciences* 62: 35-42.

Delipetrov B., Mihajlov D., Delipetrov M., and Delipetrev T. (2010) Model of the Hydro-Information System of the Republic of Macedonia. *Journal of Computing and Information Technology CIT*, Vol 18, No 2, p.201-204

Panovska S., Delipetrov T., Delipetrov M., and Delipetrev B. (2008) Analysis of geophysical models on the territory of the Republic of Macedonia. *Physica Macedonica,* Vol 57-58, 155-163.

Delipetrev B., Delipetrov M., Panovska S., and Delipetrev T. (2007) Basic model of the geo-database of the Republic of Macedonia. *Geologica Macedonica* Vol. 21, 63-67.

Delipetrev B., Panovska S., Delipetrev M., and Dimov G. (2005) Digital model of the Basic geological map of the Republic of Macedonia. *Geologica Macedonica v. 19*

## Book Chapters

Delipetrev, B., Stojanova, A., Ljubotenska, A., Kocaleva, M., Delipetrev, M. and Manevski, V. (2016) Collaborative Cloud Computing Application for Water Resources Based on Open Source Software. in *ICT Innovations 2015*. Loshkovska, S. and Koceski, S. (eds), pp. 69-78, Springer International Publishing.

## Conference Proceedings

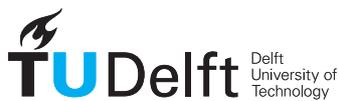

Reservoir operation is a multi-objective optimization problem traditionally solved with dynamic programming (DP) and stochastic dynamic programming (SDP) algorithms. The thesis presents novel algorithms for optimal reservoir operation named nested DP (nDP), nested SDP (nSDP), nested reinforcement learning (nRL) and their multi-objective (MO) variants correspondingly MOnDP, MOnSDP and MOnRL.

The novel idea is to include a nested optimization algorithm into each state transition that reduces the initial problem dimension and alleviates the curse of dimensionality. These algorithms can solve multi-objective optimization problems, without significantly increasing the algorithm complexity, the computational expenses and can handle dense and irregular variable discretization. All algorithms are coded in Java and tested on the case study of Knezevo reservoir in the Republic of Macedonia.

Nested optimization algorithms are embedded in a cloud application platform for water resources modeling and optimization. The platform is available 24X7, accessible from everywhere, scalable, distributed, interoperable, and it creates a real-time multiuser collaboration platform.

This thesis contributes with new and more powerful algorithms for optimal reservoir operation and cloud application platform. All source code is available for public use and can be used by researchers and practitioners to advance the mentioned areas further.

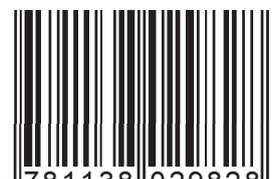